\documentclass[11pt]{article}
\usepackage{amsfonts,amsmath,amssymb,mathtools}
\usepackage[left=2cm,right=2cm,top=2cm,bottom=2cm,bindingoffset=0cm]{geometry}
\usepackage{graphicx}
\usepackage[hidelinks, unicode, pdftex]{hyperref}
\usepackage{cite}
\usepackage{authblk}
\usepackage{slashed}
\usepackage{braket}
\usepackage{comment}
\usepackage{xcolor}

\definecolor{linkcolor}{rgb}{0,0,0}
\definecolor{urlcolor}{rgb}{0,0,1}

\hypersetup{pdfstartview=FitH,  linkcolor=linkcolor,urlcolor=urlcolor,citecolor=urlcolor, colorlinks=true}

\newcommand\bem{\begin{pmatrix}}
\newcommand\eem{\end{pmatrix}}
\newcommand\beq{\begin{equation}}
\newcommand\eeq{\end{equation}}
\newcommand\beqs{\begin{equation*}}
\newcommand\eeqs{\end{equation*}}
\newcommand\nonum{\nonumber\\}
\newcommand{\tr}{\text{tr}}
\newcommand{\sgn}{\,\text{sgn}}

\title{\bf On quantization in background scalar fields}
\author[1,2]{E.~T.~Akhmedov\thanks{\href{mailto:akhmedov@itep.ru}{akhmedov@itep.ru}}}
\author[1,2]{E.~N.~Lanina\thanks{\href{mailto:lanina.en@phystech.edu}{lanina.en@phystech.edu}}}
\author[1,2]{D.~A.~Trunin\thanks{\href{mailto:dmitriy.trunin@phystech.edu}{dmitriy.trunin@phystech.edu}}}
\affil[1]{Institutskii per, 9, Moscow Institute of Physics and Technology, 141700, Dolgoprudny, Russia}
\affil[2]{B. Cheremushkinskaya, 25, Institute for Theoretical and Experimental Physics, 117218, Moscow, Russia}
\date{\today}

\begin{document}

\maketitle

\begin{abstract}
    We consider (0+1) and (1+1) dimensional Yukawa theory in various scalar field backgrounds, which are solving classical equations of motion: $\ddot{\phi}_{cl} = 0$ or $\Box \phi_{cl} = 0$, correspondingly. The (0+1)--dimensional theory we solve exactly. In (1+1)--dimensions we consider background fields of the form $\phi_{cl} = E\, t$ and $\phi_{cl} = E\, x$, which are inspired by the constant electric field. Here $E$ is a constant. We study the backreaction problem by various methods, including the dynamics of a coherent state. We also calculate loop corrections to the correlation functions in the theory using the Schwinger--Keldysh diagrammatic technique. 
\end{abstract}

\newpage
\tableofcontents
\newpage

\section{Introduction}\label{sec:intro}

The main goal of quantum field theory is to find the response of the fields to external perturbations, i.e. to find correlation functions or, more generically, correlations between an external influence on the system and its backreaction on it. In classical field theory correlation functions are solutions of equations of motion. In quantum field theory one also should take into account quantum fluctuations, i.e. calculate loop corrections to the tree-level correlation functions. Usually one treats quantum fluctuations using Feynman diagrammatic technique. It implicitly assumes that external perturbations do not change the initial state of the theory, i.e. the system remains stationary.

However, strong background fields usually take the state of the quantum field theory out of equilibrium; in this situation standard (stationary or Feynman) technique incorrectly describes the dynamics of the fields. For instance, stationary approximation is violated in an expanding space--time (see, e.g., \cite{Krotov, Akhmedov:dS, Bascone, Akhmedov:2019cfd, Akhmedov:2017ooy}), in strong electric fields~\cite{Akhmedov:Et, Akhmedov:Ex}, during the gravitational collapse \cite{Akhmedov:H} and in a number of other non-trivial physical situations~\cite{Alexeev, Astrakhantsev, Trunin, Diatlyk}. In such situations loop corrections to the tree-level correlation functions grow with time. This indicates the breakdown of the perturbation theory. Namely, every power of the small coupling constant is accompanied by a large (growing with evolution time) factor. This raises the question of the loop resummation. 

Such a resummation was performed only in a limited number of cases \cite{Akhmedov:dS, Bascone, Akhmedov:2019cfd, Akhmedov:2017ooy, Akhmedov:Et, Akhmedov:Ex}. Moreover, even in these cases one can catch only the leading qualitative effects in the limit of long evolution period and small coupling constant. In this respect it would be nice to find a simple but nontrivial example of a non-equilibrium field theory, in which calculations and dynamics itself are more transparent than in complex gravitational and electromagnetic analogs.

As an example of such a non-equilibrium situation we propose to consider the Yukawa theory of interacting fermions and massless bosons in $(D+1)$-dimensional Minkowski spacetime:
\begin{equation}
    S = \int d^{D+1}x\left[\frac{1}{2}(\partial_\mu\phi)^2+i\bar{\psi}\slashed{\partial}\psi-\lambda\phi\bar{\psi}\psi\right].
\end{equation}
We start with $D=0,1$. Usually one quantizes this theory on the trivial background $\phi_{cl} = 0$, $\psi_{cl} = 0$ and uses the standard equilibrium approach to find scattering amplitudes~\cite{Peskin}. This approach is not applicable in the presence of a strong background scalar field $\phi_{cl}$, at least if there is a pumping of energy into the system, which may generate an increase of the higher level populations and anomalous quantum averages. To study such an out of equilibrium situation, we quantize the fields on a non-zero classical background and then calculate correlation functions using non-equilibrium Schwinger--Keldysh diagrammatic technique~\cite{Kamenev,Berges,Rammer,Calzetta,Landau:vol10,Schwinger,Keldysh}.

Namely, in this paper we rely on the following program. First, we assume that there is a strong scalar field, i.e. a classical solution $\phi_{cl}(x) \gg 1$ for some values of $(D+1)$-dimensional $x$ and $\psi_{cl} = 0$. For instance, we separately study linearly growing in two dimensions background fields of the form $\phi_{cl} = \frac{m}{\lambda} + Et$ and $\phi_{cl} = \frac{m}{\lambda} + Ex$ inspired by the strong background electric field in QED~\cite{Akhmedov:Et, Akhmedov:Ex}. Whereas the separate paper~\cite{Diatlyk} considers the case of the strong scalar wave background of the form $\phi_{cl} = \frac{1}{\lambda} \Phi\left(\frac{t-x}{\sqrt{2}}\right)$. Second, we split each field into the sum of the ``classical background'' and ``quantum fluctuations'': $\phi = \phi_{cl} + \phi_q$, $\psi = \psi_q$, quantize the ``quantum'' part and find tree-level correlation functions. We use the exact fermion modes instead of plane waves; thereby we explicitly find the response of the fermion field (at least at the tree--level in such backgrounds). Then we find at tree--level the response of the scalar field itself on the background. 

Finally, we calculate loop corrections to the correlation functions using non-equilibrium Schwinger--Keldysh diagrammatic technique. In particular, we are interested in the loop corrections to the Keldysh propagators for scalar and fermion fields, because these propagators reflect the change of the state of the theory. Namely, at the loop level they show the time dependence of the corresponding level populations and anomalous quantum averages. The usual equilibrium technique is not applicable if these quantities are non-zero. For instance, this is the case of strong electric~\cite{Akhmedov:Et, Akhmedov:Ex} and gravitational~\cite{Akhmedov:dS,Akhmedov:H} fields, where loop corrections to the Keldysh propagator grow with time. 

Feynman technique takes into account only contributions of the zero point fluctuations into correlation functions. To take into account the change of the initial state of the theory (change in the anomalous averages) and of the excitation of higher than zero point levels (for the exact modes in background fields) one has to apply the Schwinger--Keldysh technique. 

However, in this paper we show that strong scalar fields under consideration do not share the properties of the background electric and gravitational fields: even in the limit of indefinitely long evolution period loop corrections to the level population and anomalous quantum average remain finite in the first loop level. Which means that while in the strong electric and gravitational fields to understand the dynamics one has to resum the leading contributions from all loops (see e.g. \cite{Akhmedov:dS} for a review), in the background scalar fields under consideration one does not need to do that.  

Let us also emphasize the other two apparent important differences between strong scalar field and strong electric and gravitational fields. The equations of motion for a point like relativistic particle in the $\phi_{cl} = \frac{m}{\lambda} + Et$, $\phi_{cl} = \frac{m}{\lambda} + Ex$ or $\phi_{cl} = \frac{1}{\lambda} \Phi \left(\frac{t-x}{\sqrt{2}}\right)$ backgrounds does not have Euclidean world--line instanton solutions and the effective actions in the scalar background fields are real~\cite{Diatlyk}. Therefore, there is no particle tunneling in the strong scalar fields under consideration. This distinguishes strong scalar field from the strong electric~\cite{Narozhnyi, Nikishov, Grib} or gravitational~\cite{Birrel} ones. However, the situation with the particle creation in the scalar field background $\phi_{cl} = \frac{1}{\lambda} \Phi \left(\frac{t-x}{\sqrt{2}}\right)$ is not that trivial as is shown in \cite{Diatlyk} on the tree--level. This can signal that in the latter background field loop corrections also may grow with time, but that is a subject for a separate paper and is not considered here.

The paper is organized as follows. 
In section~\ref{sec:1D} we discuss the one-dimensional problem. This is the simplest case to our knowledge, because in $(0+1)$ dimensions the scalar current $\lambda \langle \bar{\psi} \psi \rangle$ can be calculated exactly. Moreover, the theory can be solved exactly. Using operator formalism we show that first loop corrections to the scalar two-point functions are fully determined by corrections to one-point functions. Then we reproduce this result in Schwinger--Keldysh diagrammatic technique and extend it to all orders of perturbation theory.

In sections~\ref{sec:2D-Et} and~\ref{sec:2D-Ex} we consider the case of linearly growing in time, $\phi_{cl} = \frac{m}{\lambda} + Et$, or in space, $\phi_{cl} = \frac{m}{\lambda} + Ex$,  scalar field in $(1+1)$ dimension. We discuss the subtleties of choosing the correct fermion modes and quantize the fermion field. Using these modes we calculate the tree-level scalar current and first loop corrections to the scalar and fermion propagators. We find that in both cases these corrections remain finite in the limit of infinitely long evolution periods.

In section~\ref{sec:coherent} we consider another approach to the scalar field background: we examine the time evolution of the ``coherent state'' corresponding to the initial value of the field $\phi_{cl}(x) = \frac{m}{\lambda} + Ex$: 

$$
\left\langle \phi_{cl}\left| \hat{\phi}(t=0,x)\right|\phi_{cl}\right\rangle = \phi_{cl}(x).
$$
Such an approach corresponds to a different setup for the background field, which at first sight seems to be the same. On one side, if we consider the background field $\phi_{cl} = \frac{m}{\lambda} + Ex$ for all times and find the exact fermion modes in it, this should correspond to the situation that the background field is maintained somehow for all times in its fixed form under consideration. Or this approach is applicable when the backreaction on the background is very week. On the other hand, if we consider a background field set up by the initial coherent state $|\phi_{cl}\rangle$, which is then released to evolve freely, such an approach can be used for the case when the backreaction is strong.  

To the best of our knowledge, the last approach has not yet been considered for other non-equilibrium systems. However, we find that the behavior of the scalar field in different setups are qualitatively the same, which seems to be a peculiarity of the scalar background fields under consideration.   

Finally, we discuss the results and conclude in section~\ref{sec:discussion}. In addition, we discuss the asymptotic expansion for the parabolic cylinder functions in appendix~\ref{sec:asymptotics}, review textbook derivation of the Feynman effective action and renormalizations for the scalar field in appendix~\ref{sec:effective}, discuss some subtleties in the mode decomposition in appendix~\ref{sec:definite} and derive the coherent state in appendix~\ref{pa}.

\section{Strong scalar field in one dimension}
\label{sec:1D}
\setcounter{equation}{0}

To start with we consider the most simple situation --- the $(0+1)$--dimensional quantum field theory of interacting fermions and real scalar field. In considering this simplest $(0+1)$--dimensional situation we will show many technical details for pedagogical reasons to introduce the non--stationary technique and set up the notations.

There are two options to describe fermions in one dimension. First one is determined by the following action:

\begin{equation}\label{1s2f1}
    S=\int dt\left[\frac{1}{2}\dot{\phi}^2+i\bar{\psi}\dot{\psi}-\lambda\phi\bar{\psi}\psi\right],
\end{equation}
where we denoted the conjugated fermion as $\bar{\psi} = \psi^\dagger$. The fermions become Grassmanian upon quantization.  Another option is the theory with two-component spinors:

\begin{equation}\label{1s3f1}
    S=\int dt\left[\frac{1}{2}\dot{\phi}^2+i\bar{\psi}\gamma_0\dot{\psi}-\lambda\phi\bar{\psi}\psi\right],
\end{equation}
where $\gamma_0=\begin{pmatrix} 1 & 0 \\ 0 & -1
\end{pmatrix}$, $\bar{\psi} = \psi^\dagger \gamma^0$. 

It can be shown that the situation in the latter theory is just a bit more complicated than in the former one. Essentially the dynamics is the same. The main complication of (\ref{1s3f1}) in comparison with the theory (\ref{1s2f1}) is that in (\ref{1s3f1}) upon quantization we have four fermion Fock space states, $|0,0\rangle$, $|0,1\rangle$, $|1,0\rangle$ and $|1,1\rangle$, rather than two, $|0\rangle$ and $|1\rangle$ as it is the case for \eqref{1s2f1}. In what follows we consider only the theory \eqref{1s2f1}. We address the theory under consideration as if it is the simplest one dimensional quantum field theory. Namely instead of calculating quantum mechanical transition amplitudes we calculate correlation functions. Our main goal is to find the backreaction on a strong scalar field, to be described below, in these very simple settings.

The equations of motion for the action~\eqref{1s2f1} are as follows:
\begin{equation}\label{1s2f2}
    \begin{cases}
        \ddot{\phi}=-\lambda\bar{\psi}\psi,\\
        i\dot{\psi}=\lambda\phi\psi.
    \end{cases}
\end{equation}
These equations have the following classical solution:

\begin{equation}\label{1s2f3}
    \phi_{cl}(t)=\frac{m}{\lambda}+\frac{\alpha}{\lambda}t, \quad \psi_{cl}=\bar{\psi}_{cl}=0,
\end{equation}
which we will consider as a background.

Then we consider mode decomposition for quantum parts of these fields over the classical background~\eqref{1s2f3}:
\begin{align}
\hat{\psi}(t) = \hat{a} p(t), \quad \hat{\bar{\psi}}(t) = \hat{a}^\dagger p^*(t),\label{eq:1D-fermions} \\
\hat{\phi}(t) = \hat{\alpha} f(t) + \hat{\alpha}^\dagger f^*(t), \nonumber 
\end{align}
where operators $\hat{a}$ and $\hat{\alpha}$ obey the standard (anti)commutation relations:

\beq \{\hat{a}, \hat{a}^\dagger\} = 1, \quad [\hat{\alpha}, \hat{\alpha}^\dagger] = 1. \eeq
The equations for the modes on this background are as follows:

\begin{equation}\label{1eq:anti}
    \begin{cases}
        \ddot{f}=0,\\
        \left(i\frac{d}{dt}-m-\alpha t\right)p=0.
    \end{cases}
\end{equation}
Thus, we have the first order differential equation for the fermion modes, hence, their form is

\begin{equation}\label{1s2f5}
p(t) = e^{-i\int\limits^t(m+\alpha t')dt'}.
\end{equation}
As a result, the tree--level expectation value of the equal-time product of two fermion operators does not depend on time:

\begin{equation}\label{1s2f6}
    \left\langle 0\right|\bar{\psi}\psi\left|0\right\rangle =0 \quad \text{ and } \quad \left\langle 1\right|\bar{\psi}\psi\left|1\right\rangle =1,
\end{equation}
where $\hat{a} |0\rangle = \hat{a}^\dagger | 1\rangle = 0$.
To find $\langle \bar{\psi}\psi\rangle $ exactly, note that the full Hamiltonian of the theory is as follows:

\begin{equation}\label{1s2f7}
    H_{full}=\lambda\phi\bar{\psi}\psi+\frac{\pi^2}{2},
\end{equation}
where $\pi$ is the momentum conjugate to the scalar field, $[\phi,\pi]=i$, $\{\psi,\bar{\psi}\}=1$. Using such a Hamiltonian one can find that:

\begin{equation}\label{1s2f8}
    [\bar{\psi}\psi,H_{full}] = 0, \quad \text{hence}, \quad \langle 0| \bar{\psi} \psi | 0\rangle _{exact}(t) = 0 \quad \text{and} \quad \langle 1| \bar{\psi}\psi |1\rangle _{exact}(t) = 1.
\end{equation}
Thus, we have two options for the backreaction problem:

\beq \label{cl_eqs} \begin{aligned}
    \ddot{\langle \phi\rangle} &\equiv -\lambda \langle 0| \bar{\psi}\psi | 0\rangle = 0,  \quad {\rm and}\\
    \ddot{\langle \phi\rangle} &\equiv -\lambda \langle 1| \bar{\psi}\psi | 1\rangle = - \lambda, 
\end{aligned} \eeq
i.e. either the background force is zero or non--zero, but constant. 

It should be stressed at this point that the result under consideration does not depend whether we quantize in the background scalar field (\ref{1s2f3}) or we put the background field to zero. However, to complete the solution of the problem, we also have to calculate the scalar and fermion two--point functions, when the points do not coincide.

To do that let us point out one important issue. Consider one--dimensional scalar with a non--zero mass:

\begin{equation}
    S_0=\frac{1}{2}\int dt\left[\dot{\phi}^2-\omega^2\phi^2\right].
\end{equation}
The standard mode in this case is $f(t) = \frac{1}{\sqrt{2 \omega}} e^{-i \omega t}$.

Consider the two--point Wightman functions in this theory in the limit $\omega\rightarrow 0$:

\begin{equation}\label{1s2f9}
\begin{aligned}
    \phi(t)=\frac{1}{\sqrt{2\omega}}(\alpha e^{-i\omega t}+\alpha^\dagger e^{i\omega t}) &\xrightarrow{\omega \rightarrow 0}\frac{\alpha+\alpha^\dagger}{\sqrt{2\omega}}+i\sqrt{\frac{\omega}{2}}(\alpha^\dagger-\alpha)t,\\
    {\rm and} \quad \langle \phi(t)\phi(t')\rangle =\frac{e^{-i\omega(t-t')}}{2\omega} &\xrightarrow{\omega \rightarrow 0}\frac{1}{2\omega}-\frac{i}{2}(t-t').
\end{aligned}
\end{equation}
Note that if we just omit the term $\frac{1}{2\omega}$ in the propagator it can be used as the tree--level Wightman scalar function in the theory (\ref{1s2f1}). In fact, the latter one does solve the appropriate differential equation:

$$
\left(\frac{d^2}{dt^2} + \omega^2\right)\, G(t-t') = 0,
$$
and can be used as a basis for the construction of other propagators. Such as e.g. Feynman, retarded and Keldysh two--point functions.

On the other hand, consider the direct quantization of the scalar part of the theory (\ref{1s2f1}).  Then the mode is $f(t) = \frac{1 - i t}{\sqrt{2}}$ and the expansion of the field operator is:

\begin{equation}\label{1s2f10}
    \phi(t)=\frac{1}{\sqrt{2}}\left[(\alpha+\alpha^\dagger)+i(\alpha^\dagger-\alpha)t\right].
\end{equation}
It is easy to check that such $\phi$ satisfies the equation of motion and $[\phi,\pi]=i$.

Now we can calculate the tree--level boson Wightman propagator:

\begin{equation}\label{1s2f11}
    \langle \phi(t)\phi(t')\rangle_0 =\frac{1}{2}\left\langle 0\right|\left[(\alpha+\alpha^\dagger)+i(\alpha^\dagger-\alpha)t\right]\left[(\alpha+\alpha^\dagger)+i(\alpha^\dagger-\alpha)t'\right]\left|0\right\rangle =\frac{1-i(t-t')+tt'}{2}.
\end{equation}
This provides another option for the two--point function in the theory.
The two choices of the Wightman propagators in the theory correspond to two different choices of states. While the second choice corresponds to a ground state in the Fock space, the first one is a sort of a coherent state. A somewhat similar situation appears for the massless scalar field in two--dimensional flat space or in de Sitter space \cite{Bertola:2006df}.

Please also note that while the first choice of the propagator respects the time translational invariance, but does not respect so called positivity, $\langle \phi^2(t)\rangle > 0$ (in the present case $\langle \phi^2(t)\rangle$ is just vanishing, while in two--dimensions similar Wightman function can become negative), the second choice does respect positivity, but violates the time translational invariance.

What remains to be done now is to calculate the exact two--point Wightman function for the scalars and fermions. In the next two subsections we will do that in two different, but related, ways. But before doing this let us explain the resulting solution of the problem in simple terms. Consider a solution of the second equation in~\eqref{cl_eqs}:

\begin{equation}\label{totsol}
    \ddot{\langle \phi\rangle} = - \lambda. 
\end{equation}
It is given by

\begin{equation}
    \langle \phi\rangle = -\frac{\lambda}{2}t^2+c_1 t+c_2,
\end{equation}
where $c_{1,2}$ are integration constants. Hence, the field operator $\hat{\phi}(t)$ can be written in the following form:
\begin{equation}\label{phi11}
    \hat{\phi}(t) = \frac{m}{\lambda}+\frac{\alpha}{\lambda}t + \frac{1}{\sqrt{2}}\left[\left(\hat{\alpha} + \hat{\alpha}^\dagger\right) + i \left(\hat{\alpha}^\dagger - \hat{\alpha}\right)t\right]-\frac{\lambda}{2}t^2+c_1 t+c_2.
\end{equation}
Then, the boson propagator has the following form:

\begin{equation}\label{phi}
    \Delta \langle \phi(t_1)\phi(t_2) \rangle = \langle \phi(t_1) \rangle \langle \phi(t_2) \rangle = \frac{\lambda^2}{4}t_1^2 t_2^2 -\frac{\lambda}{2}c_1(t_1^2 t_2 + t_2^2 t_1)-\frac{\lambda}{2}c_2(t_1^2+t_2^2)+c_1 c_2(t_1+t_2)+c_1^2 t_1 t_2+c_2^2.
\end{equation}
This expression coincides with the exact result shown e.g. in eq.~\eqref{1s2f25} if we set 

\begin{eqnarray}
    c_1=\lambda t_0, \\
    c_2=-\frac{\lambda}{2}t_0^2,
\end{eqnarray}
That is true because the exact expression follows from the ``tadpole'' diagram, which corresponds to the solution of the equation \eqref{totsol}.

\subsection{Two-point functions and perturbative corrections}\label{1s2.1}

Let us make the field $\phi$ dynamical and calculate corrections to the tree--level propagators. The potential operator in the interaction picture is as follows:
\begin{equation}\label{1s2f12}
    V(t)=U_0^\dagger(t,t_0)\left(\lambda\phi(t_0)\bar{\psi}\psi\right)U_0(t,t_0)=\lambda\phi(t)\bar{\psi}\psi=\lambda\left(\hat{\alpha} f(t)+\hat{\alpha}^\dagger f^*(t)\right)\hat{a}^\dagger \hat{a},
\end{equation}
where $t_0$ is the time after which the self-interaction $\lambda\phi\bar{\psi}\psi$ is adiabatically turned on. We recall that $\bar{\psi} \psi$ does not depend on time and $f(t)=\frac{1-it}{\sqrt{2}}$. Evolution operator in the interaction picture is as follows:
\begin{equation}\label{1s2f13}
\begin{aligned} U\left(t_{b}, t_{a}\right) &=T \exp \left[-i \int_{t_{a}}^{t_{b}} d \eta V(\eta)\right]=1-i \int_{t_{a}}^{t_{b}} d \eta V(\eta)+(-i)^{2} \int_{t_{a}}^{t_{b}} d \eta V(\eta) \int_{t_{a}}^{\eta} d \xi V(\xi)+\cdots \equiv \\ & \equiv 1+U_{1}\left(t_{b}, t_{a}\right)+U_{2}\left(t_{b}, t_{a}\right)+\cdots \end{aligned}
\end{equation}
One can explicitly calculate the first and second order corrections to the evolution operator:

\begin{equation}\label{1s2f14}
\begin{aligned}
    U_1(t_b,t_a)&=-\frac{i\lambda}{\sqrt{2}}\hat{a}^\dagger \hat{a}\left[(t_a-t_b)\left(-1+\frac{i}{2}(t_a+t_b)\right)\hat{\alpha} + h.c. \right],\\
    U_2(t_b,t_a)&=-\frac{\lambda^2}{2}\hat{a}^\dagger \hat{a} \Bigg[\frac{1}{24}(t_a-t_b)^2\left(12+3t_a^2+t_b(3t_b+4i)+t_a(6t_b-4i)\right) \hat{\alpha}^\dagger \hat{\alpha} - \\ &\phantom{=-\frac{\lambda^2}{2}\hat{a}^\dagger \hat{a} \Bigg(}-\frac{1}{8}(t_a-t_b)^2(2i+t_a+t_b)^2 \hat{\alpha} \hat{\alpha} + h.c.\Bigg],
\end{aligned}
\end{equation}
where we have used the identity $\hat{a}^\dagger \hat{a} \hat{a}^\dagger \hat{a} = \hat{a}^\dagger \hat{a}$. Now let us calculate the Wightman function of two boson fields in the vacuum state of the scalar field, $\hat{\alpha} |0\rangle = 0$:

\begin{equation}\label{1s2f16}
\begin{aligned}
    D_{exact}(t_1,t_2)&=\left\langle \phi(t_1)\phi(t_2)\right\rangle =\left\langle U^\dagger(t_1,t_0)\phi(t_1)U(t_1,t_2)\phi(t_2)U(t_2,t_0)\right\rangle = \\ &=\big\langle \left[1+U_1(t_0,t_1)+U_2(t_0,t_1)+\dots\right]\phi_1\left[1+U_1(t_1,t_2)+U_2(t_1,t_2)+\dots\right]\phi_2\times\\
    &\phantom{=}\times\left[1+U_1(t_2,t_0)+U_2(t_2,t_0)+\dots\right]\big\rangle=D_0(t_1,t_2)+\Delta D(t_1,t_2)+\dots,
\end{aligned}
\end{equation}
where we denote $\phi(t_a)\equiv\phi_a$ for short. 

Note that if we average over the vacuum for fermions, $a |0\rangle = 0$, all contributions except the bare boson propagator vanish because they always contain the combination $\psi\left|0\right\rangle = 0$. So in this case the tree-level expression for the boson propagator is exact:

\beq
\label{eq:D-zero}
D_{exact}(t_1,t_2)=D_{0}(t_1,t_2).
\eeq
Now consider the averaging over the state $\hat{a}^\dagger\left|1\right\rangle = 0$ for fermions, which gives a less trivial result. 
Using the decomposition of the evolution operator, one finds that the correction to the tree--level propagator grows with time:

\begin{equation}\label{1s2f19}
\begin{aligned}
    \Delta D(t_1,t_2)&=\frac{\lambda^2}{8}(t_0-t_1)(t_0-t_2)\big\{(t_0+t_1-2i)(t_0+t_2+2i)f(t_1)f^*(t_2)+\\
    &+(t_0+t_1-2i)(t_0+t_2-2i)f(t_1)f(t_2)+h.c\big\}=\frac{\lambda^2}{4}(t_1-t_0)^2(t_2-t_0)^2.
\end{aligned}
\end{equation}
To calculate $\left\langle \phi(t_2)\phi(t_1)\right\rangle $ we should simply change $t_1\leftrightarrow t_2$. For the future reference we show here expressions for the Keldysh and retarded/advanced (R/A) propagators~\cite{Kamenev,Rammer,Berges,Calzetta,Landau:vol10}:

\begin{equation}\label{1s2f20}
\begin{aligned} D^{K}\left(t_{1}, t_{2}\right) &=\frac{1}{2}\left\langle \left\{\phi\left(t_{1}\right), \phi\left(t_{2}\right)\right\}\right\rangle, \\ D^{R / A}\left(t_{1}, t_{2}\right) &=\pm \theta\left( \pm t_{1} \mp t_{2}\right)\left\langle \left[\phi\left(t_{1}\right), \phi\left(t_{2}\right)\right]\right\rangle. \end{aligned}
\end{equation}
Note that
\begin{equation}\label{1s2f21}
\begin{aligned}
    D^A(t_1,t_2)=D^R(t_2,t_1).
\end{aligned}
\end{equation}
This means that advanced and retarded propagators behave similarly and we need to calculate only the retarded one. Thus, it follows that

\begin{equation}\label{1s2f22}
\begin{aligned}
    D_0^K&=\frac{1}{2}\left[f(t_1)f^*(t_2)+f^*(t_1)f(t_2)\right]=\frac{1+t_1t_2}{2}, \\
    \Delta D^K&=\frac{\lambda^2}{4}(t_1-t_0)^2(t_2-t_0)^2, \\
    D_0^{R}&=\theta(t_1 - t_2)\left[f(t_1)f^*(t_2)-f^*(t_1)f(t_2)\right]=i\theta(t_1-t_2)(t_2-t_1), \\
    \Delta D^R&=0.
\end{aligned}
\end{equation}
Here subscript $0$ denotes tree--level propagators, while $\Delta D$ --- perturbative corrections which we calculate here. To understand the obtained result let us calculate the expectation value of the single operator:

\begin{equation}\label{1s2f23}
    \left\langle \phi_1\right\rangle =\left\langle U^\dagger(t_1,t_0)\phi_1 U(t_1,t_0)\right\rangle.
\end{equation}
Up to the first order in $\lambda$ the correction looks as follows:
\begin{equation}\label{1s2f24}
    \Delta\left\langle \phi_1\right\rangle =-i\lambda\int\limits_{t_0}^{t_1}dt_2\left(\left\langle \phi_1\phi_2\right\rangle -\left\langle \phi_2\phi_1\right\rangle \right)=\lambda\int\limits_{t_0}^{t_1}dt_2(t_2-t_1)=-\frac{\lambda}{2}(t_1-t_0)^2.
\end{equation}
Hence, we see that $\Delta D$ is completely determined by the correction to the one-point correlation function:

\begin{equation}\label{1s2f25}
    \Delta D(t_1,t_2)=\Delta D^K(t_1,t_2)=\Delta\left\langle \phi_1\right\rangle \Delta\left\langle \phi_2\right\rangle =\frac{\lambda^2}{4}(t_1-t_0)^2(t_2-t_0)^2.
\end{equation}
In the following subsection we will see that this contribution corresponds to the so-called ``tadpole'' diagrams. And, thus, although we have obtained the result under consideration only at the order $\lambda^2$ in the expansion of (\ref{1s2f16}) it is actually the exact expression.

Apart from other things the observations that we have made in this section indicate that the growth of the two--point function with times $t_{1,2}$ has no connection to the change of the state in the theory unlike the case of non--stationary situations in higher dimensional quantum field theories. Namely, the time evolution in the theory does not lead to a generation of the anomalous quantum averages and level populations neither for fermions nor for the boson. In other words, the initial state does not change despite the non-stationarity of the theory.

\subsection{Schwinger-Keldysh diagrammatic technique}
\label{sec:technique}

In this subsection we recalculate the results of the previous subsection with the use of the diagrammatic technique. We use Schwinger-Keldysh diagrammatic technique~\cite{Schwinger, Keldysh, Kamenev, Landau:vol10, Berges, Rammer, Calzetta}. This technique uses the following fermionic propagators:

\beq
\label{eq:fermion-correators-def}
\begin{aligned}
i G^{--}(x_1, x_2) &\equiv \langle T \psi(x_1) \bar{\psi}(x_2) \rangle = \theta(t_1 - t_2) i G^{+-}(x_1, x_2) + \theta(t_2 - t_1) i G^{-+}(x_1, x_2), \\
i G^{++}(x_1, x_2) &\equiv \langle \tilde{T} \psi(x_1) \bar{\psi}(x_2) \rangle = \theta(t_1 - t_2) i G^{-+}(x_1, x_2) + \theta(t_2 - t_1) i G^{+-}(x_1, x_2), \\
i G^{+-}(x_1, x_2) &\equiv \langle \psi(x_1) \bar{\psi}(x_2) \rangle, \\
i G^{-+}(x_1, x_2) &\equiv - \langle \bar{\psi}(x_2) \psi(x_1) \rangle,
\end{aligned}
\eeq
where $\langle \, \cdots \rangle$ denotes averaging over an appropriate initial state, $T$ stands for the time ordering and $\tilde{T}$ --- for the anti-time ordering. Corresponding bosonic correlation functions are as follows:

\beq
\label{eq:boson-correators-def}
\begin{aligned}
i D^{--}(x_1, x_2) &\equiv \langle T \phi(x_1) \phi(x_2) \rangle = \theta(t_1 - t_2) i D^{+-}(x_1, x_2) + \theta(t_2 - t_1) i D^{-+}(x_1, x_2), \\
i D^{++}(x_1, x_2) &\equiv \langle \tilde{T} \phi(x_1) \phi(x_2) \rangle = \theta(t_1 - t_2) i D^{-+}(x_1, x_2) + \theta(t_2 - t_1) i D^{+-}(x_1, x_2), \\
i D^{+-}(x_1, x_2) &\equiv \langle \phi(x_1) \phi(x_2) \rangle, \\
i D^{-+}(x_1, x_2) &\equiv \langle \phi(x_2) \phi(x_1) \rangle.
\end{aligned}
\eeq
In what follows we will include the imaginary unit into the definition of the correlation functions~\eqref{eq:fermion-correators-def} and~\eqref{eq:boson-correators-def} for short.

One can also define these correlation functions using the Keldysh time contour, which starts at the moment $t_0$, goes to $t \rightarrow +\infty$ and then returns back to the starting point~\cite{Keldysh}. The contour appears due to the simultaneous presence of time ordered $U$ and anti--time ordered $U^\dagger$ in \eqref{1s2f16}. Ordering along this contour corresponds to the time-ordering on the ``forward'' part and to the anti-time-ordering on the ``backward'' part. Hence, one can assign ``$\mp$'' signs to the fields sitting on the forward and backward parts of the contour, correspondingly, and define correlation functions $G^{\pm \pm} \equiv \langle \psi_{\pm} \bar{\psi}_{\pm} \rangle$, $D^{\pm \pm} \equiv \langle \phi_{\pm} \phi_{\pm} \rangle$. This definition is equivalent to the definition~\eqref{eq:fermion-correators-def} and~\eqref{eq:boson-correators-def}. More details can be found in \cite{Kamenev, Berges, Rammer}.

Also note that functions $G^{\pm \pm}$ and $D^{\pm \pm}$ are not independent due to the relations:

\beq G^{++} + G^{--} = G^{+-} + G^{-+}, \quad D^{++} + D^{--} = D^{+-} + D^{-+}. \eeq
It is convenient to do the Keldysh rotation from the forward and backward (``$\pm$'') components of the fields to the so called classical and quantum components\footnote{In general, matrices which rotate fields $\phi$, $\psi$, and $\bar{\psi}$ are independent, but here we choose them to be equal to each other.} \cite{Keldysh,Kamenev,Berges}:

\beq \label{1s2f28} \bem \phi_{cl} \\ \phi_q \eem = \hat{R} \bem \phi_+ \\ \phi_- \eem, \quad \bem \psi_{cl} \\ \psi_q \eem = \hat{R} \bem \psi_+ \\ \psi_- \eem, \quad \bem \bar{\psi}_{cl} \\ \bar{\psi}_q \eem = \hat{R} \bem \bar{\psi}_+ \\ \bar{\psi}_- \eem, \quad \hat{R} = \bem \frac{1}{2} & \frac{1}{2} \\ -1 & 1 \eem, \eeq
and introduce the Keldysh and retarded/andvanced propagators:

\beq \begin{aligned}
G^K &\equiv \langle \psi_{cl} \bar{\psi}_{cl} \rangle = \frac{1}{2} \left(  G^{++} + G^{--} \right), & D^K &\equiv \langle \phi_{cl} \phi_{cl} \rangle = \frac{1}{2} \left(  D^{++} + D^{--} \right),\\
G^R &\equiv \langle \psi_{cl} \bar{\psi}_q \rangle = G^{--} - G^{-+}, & D^R &\equiv \langle \phi_{cl} \phi_q \rangle = D^{--} - D^{-+}, \\
G^A &\equiv \langle \psi_q \bar{\psi}_{cl} \rangle = G^{--} - G^{+-}, & D^A &\equiv \langle \phi_q \phi_{cl} \rangle = D^{--} - D^{+-}.
\end{aligned} \eeq
This definition is equivalent to the one used in~\eqref{1s2f20}.

Note that in $(0+1)$--dimensions diagrammatic technique works only for correlation functions averaged over the vacuum or thermal (stationary) state, because diagrammatics is based on Wick's theorem~\cite{Trunin, Evans}, which is applicable only in stationary situations in one dimension. However, in our case this restriction does not bother us, because the state of the fields does not change in time. In higher dimensional quantum field theory this restriction disappears because of the infinite space volume which kills unsuitable operator averages~\cite{Landau:vol9,Landau:vol10}.

\begin{figure}[t]
\begin{center}
\begin{minipage}[t]{0.3\linewidth}
\includegraphics[width=1\linewidth]{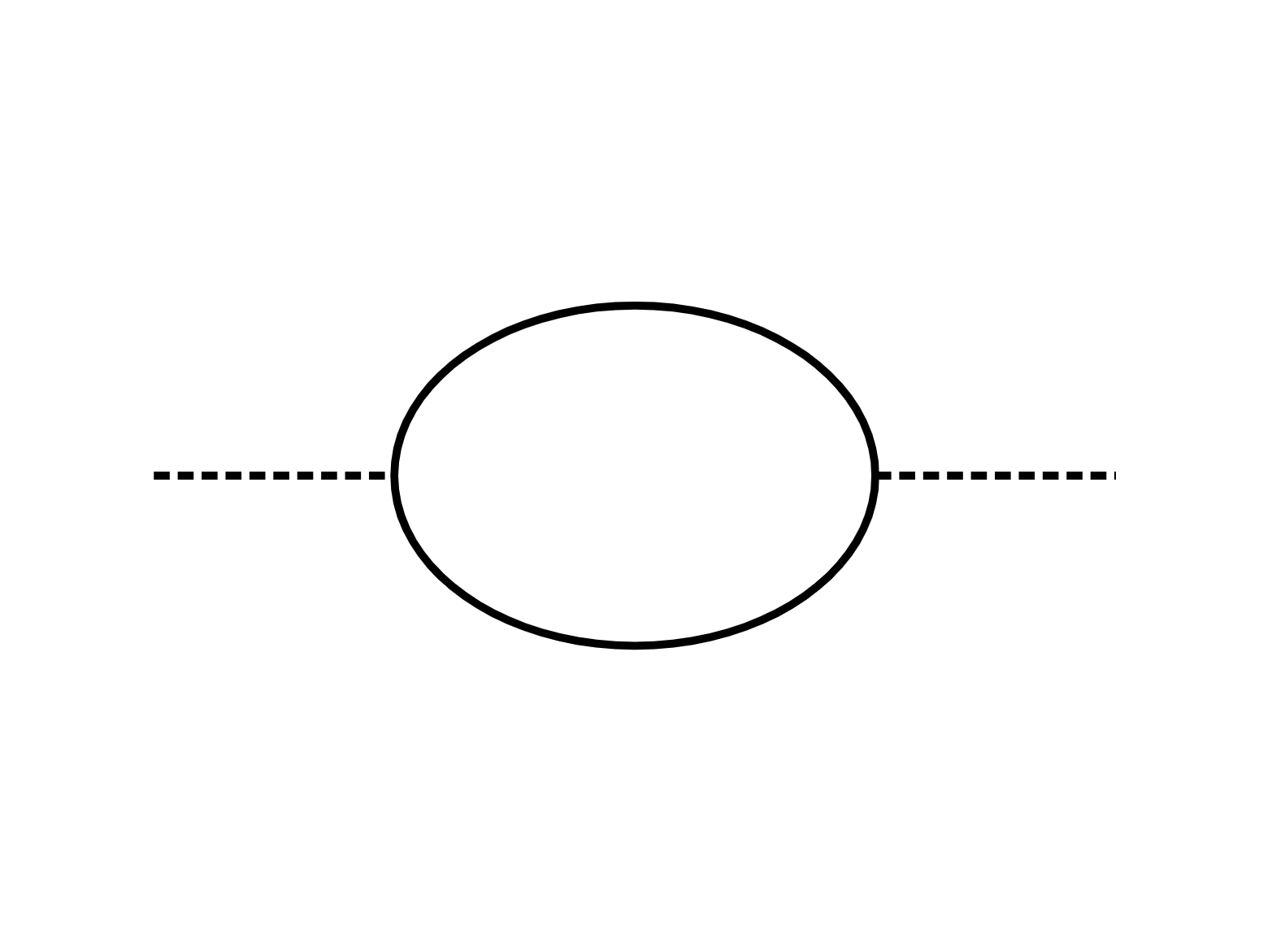}
\caption{One-loop correction} 
\label{fig:one-loop-1}
\end{minipage}
\hfill 
\begin{minipage}[t]{0.3\linewidth}
\includegraphics[width=1\linewidth]{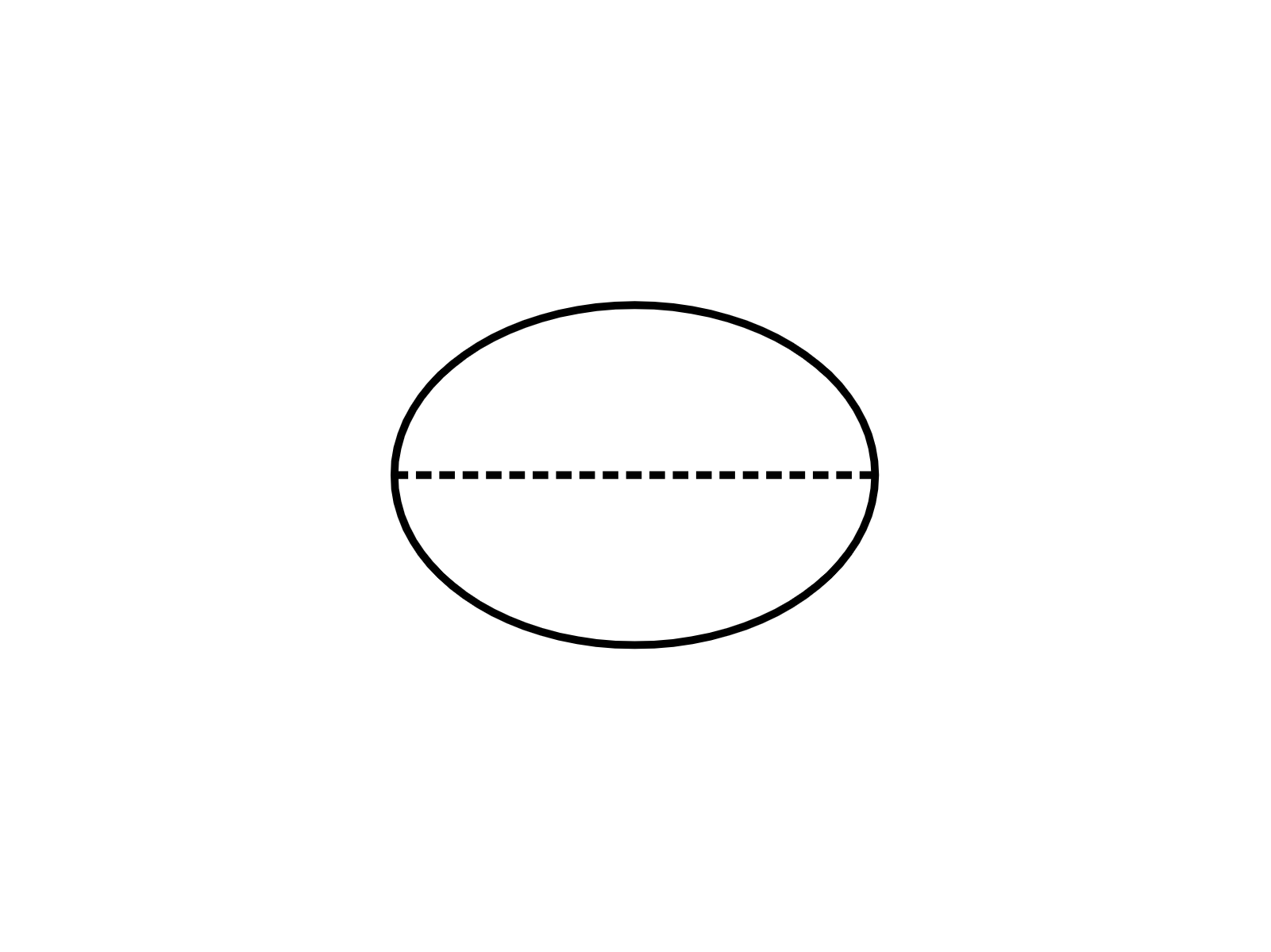}
\caption{Bubble diagram}
\label{fig:bubble-diagramm}
\end{minipage}
\hfill 
\begin{minipage}[t]{0.3\linewidth}
\includegraphics[width=1\linewidth]{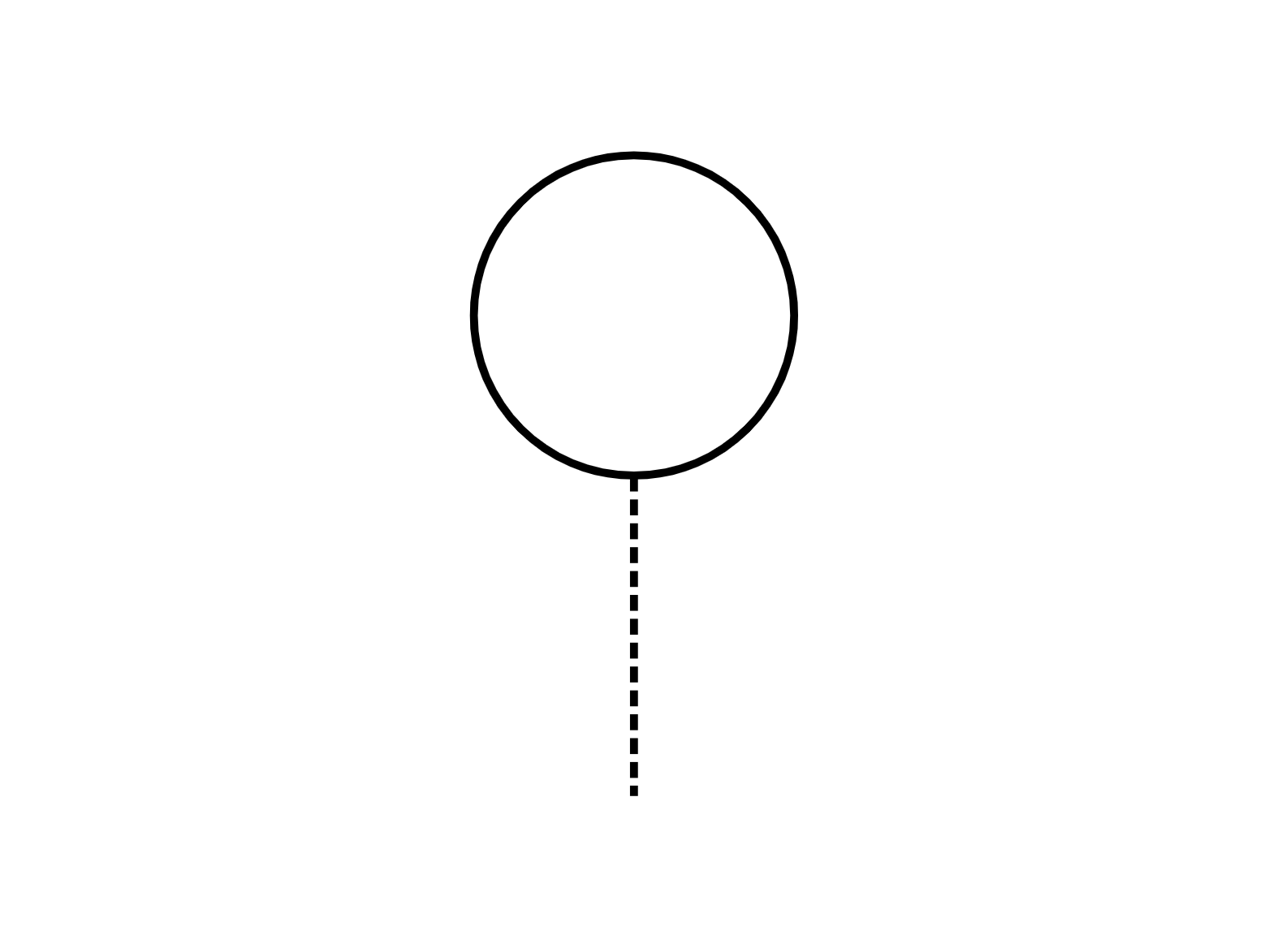}
\caption{Tadpole diagram}
\label{fig:tadpole-diagramms}
\end{minipage}
\end{center}
\end{figure}

Let us calculate the first loop correction to the boson two-point correlation function using Schwinger-Keldysh diagrammatic technique. First, we consider the averaging over the state $\left|0\right\rangle _\psi\left|0\right\rangle _\phi$. In this case tree--level propagators have the following form (it is easy to restore the remaining four correlators using definitions~\eqref{eq:fermion-correators-def} and~\eqref{eq:boson-correators-def}):

\beq \label{1s2f26} \begin{aligned}
    G_0^{+-}(t_1,t_2)&=\exp\left\{-i\int\limits^{t_1}_{t_2}(m+\alpha t')dt'\right\}, \\
    G_0^{-+}(t_1,t_2)&=0, \\
    D_0^{+-}(t_1,t_2)&=f(t_1)f^*(t_2)=\frac{(1-it_1)(1+it_2)}{2}, \\
    D_0^{-+}(t_1,t_2)&=\left(D_0^{+-}(t_1,t_2)\right)^*=\frac{(1+it_1)(1-it_2)}{2}.
\end{aligned} \eeq
The one-loop corrections to the scalar field propagators (Fig.~\ref{fig:one-loop-1}) is vanishing:

\begin{equation}\label{1s2f31}
    \Delta D^{+-}(t_1, t_2)=-\lambda^2 \int dt_3 dt_4 \sum_{\sigma_{3,4} = \{+,-\}} D^{+\sigma_3}(t_1,t_3) G^{\sigma_3 \sigma_4}(t_3,t_4) G^{\sigma_4 \sigma_3}(t_4,t_3) D^{\sigma_4 -}(t_4,t_2) \, \text{sgn}(\sigma_3 \sigma_4)=0,
\end{equation}
because $G^{-+}=0$ and $\theta_{34}\theta_{43}=0$, where for short we denote $\theta_{34}\equiv\theta(t_3-t_4)$. Thus, $\Delta D^K(t_1,t_2)=\Delta D^{R/A}(t_1,t_2)=0$. Due to the same reason the so-called ``bubble''
diagram (Fig.~\ref{fig:bubble-diagramm}) is also equal to zero\footnote{In the Schwinger--Keldysh diagrammatic technique vacuum bubbles always cancel out.}. Finally, the tadpole diagrams (Fig.~\ref{fig:tadpole-diagramms}) are zero because they contain the free fermion propagators in coincident points: $\left\langle 0\right|\psi\bar{\psi}\left|0\right\rangle =0$.
Thus, one-loop corrections to the boson propagator is zero for the case of averaging over the state $|0\rangle_\psi |0\rangle_\phi$. This is exactly what we have seen in the previous subsection (see eq.~\eqref{eq:D-zero}).

Now let us take the average over the state $\left|1\right\rangle_\psi\left|0\right\rangle _\phi$. In this case tree--level boson propagators do not change, whereas tree--level fermion propagators acquire the following form:
\begin{equation} \label{1s2f32}
\begin{aligned}
G_0^{+-}(t_1,t_2)&=0, \\ G_0^{-+}(t_1,t_2)&=-\exp\left\{-i\int\limits^{t_1}_{t_2}(m+\alpha t')dt'\right\}.
\end{aligned}
\end{equation}
The diagrams (Fig.~\ref{fig:one-loop-1}) and (Fig.~\ref{fig:bubble-diagramm}) in this case are zero again for the same reasons. Hence, we recalculate only the tadpole diagrams (Fig.~\ref{fig:tadpole-diagramms}):
\begin{equation}\label{1s2f34}
\begin{aligned}
    \Delta\left\langle \phi_1^+\right\rangle &=-i\lambda\int dt_2\sum\limits_{\sigma=\{+,-\}}D^{+\sigma}(t_1,t_2)G_{aa}^{\sigma\sigma}(t_2,t_2)\text{sgn}(-\sigma)=\\
    &=-i\lambda\int\limits_{t_0}^{+\infty}dt_2 D^{R}(t_1,t_2)=\lambda\int_{t_0}^{t_1}dt_2(t_2-t_1)=-\frac{\lambda}{2}(t_1-t_0)^2,\\
    \Delta\left\langle \phi_1^-\right\rangle &=-i\lambda\int dt_2\sum\limits_{\sigma=\{+,-\}}D^{-\sigma}(t_1,t_2)G_{aa}^{\sigma\sigma}(t_2,t_2)\text{sgn}(-\sigma)=\\
    &=-i\lambda\int\limits_{t_0}^{+\infty}dt_2 D^{R}(t_1,t_2)=\lambda\int_{t_0}^{t_1}dt_2(t_2-t_1)=-\frac{\lambda}{2}(t_1-t_0)^2=\Delta\left\langle \phi_1^+\right\rangle .
\end{aligned}
\end{equation}
Hence, the correction to the boson correlation function looks as follows:
\begin{equation}\label{1s2f35}
   \Delta D^{+-}(t_1,t_2)=\Delta D^K(t_1,t_2)=\Delta\left\langle \phi_1^+\right\rangle \Delta\left\langle \phi_2^-\right\rangle =\frac{\lambda^2}{4}(t_1-t_0)^2(t_2-t_0)^2,
\end{equation}
which coincides with the result~\eqref{1s2f25} from the previous subsection.

Note that if we choose the bare scalar Wightman propagator as follows:

\begin{equation}\label{1s2f36}
    \left\langle \phi_1\phi_2\right\rangle_0 =-\frac{i}{2}(t_1-t_2),
\end{equation}
which, as we have discuss around eq. \eqref{1s2f9}, respects the time translational invariance, we will get the same answer for the tadpole diagram:
\begin{equation}\label{1s2f37}
    \Delta\left\langle \phi_1^-\right\rangle =\Delta\left\langle \phi_1^+\right\rangle =-i\lambda\int\limits_{t_0}^{+\infty}dt_2 D^{R}(t_1,t_2)=\lambda\int\limits_{t_0}^{t_1}dt_2(t_2-t_1)=-\frac{\lambda}{2}(t_1-t_0)^2,
\end{equation}
because retarded propagators do not depend on the state. Thus, the diagrammatic technique gives the correct combinatoric factors and reproduces the result of the direct calculation performed above in the subsection~\ref{1s2.1}.

\subsection{Exact boson propagators}\label{1s2.3}

As we have already pointed out in the subsection~\ref{1s2.1}, the tree-level expression for the boson propagator is exact if we average over the fermion vacuum $\hat{a}|0\rangle_\psi = 0$: $D_{exact}(t_1,t_2)=D_0(t_1,t_2)$. So in this subsection we consider averaging over the state $\hat{a}^\dagger\left|1\right\rangle_\psi = 0$. We will see that in this case the situation is nearly the same.

\begin{figure}[t]
\begin{minipage}[t]{0.45\linewidth}
\center{\includegraphics[width=1\linewidth]{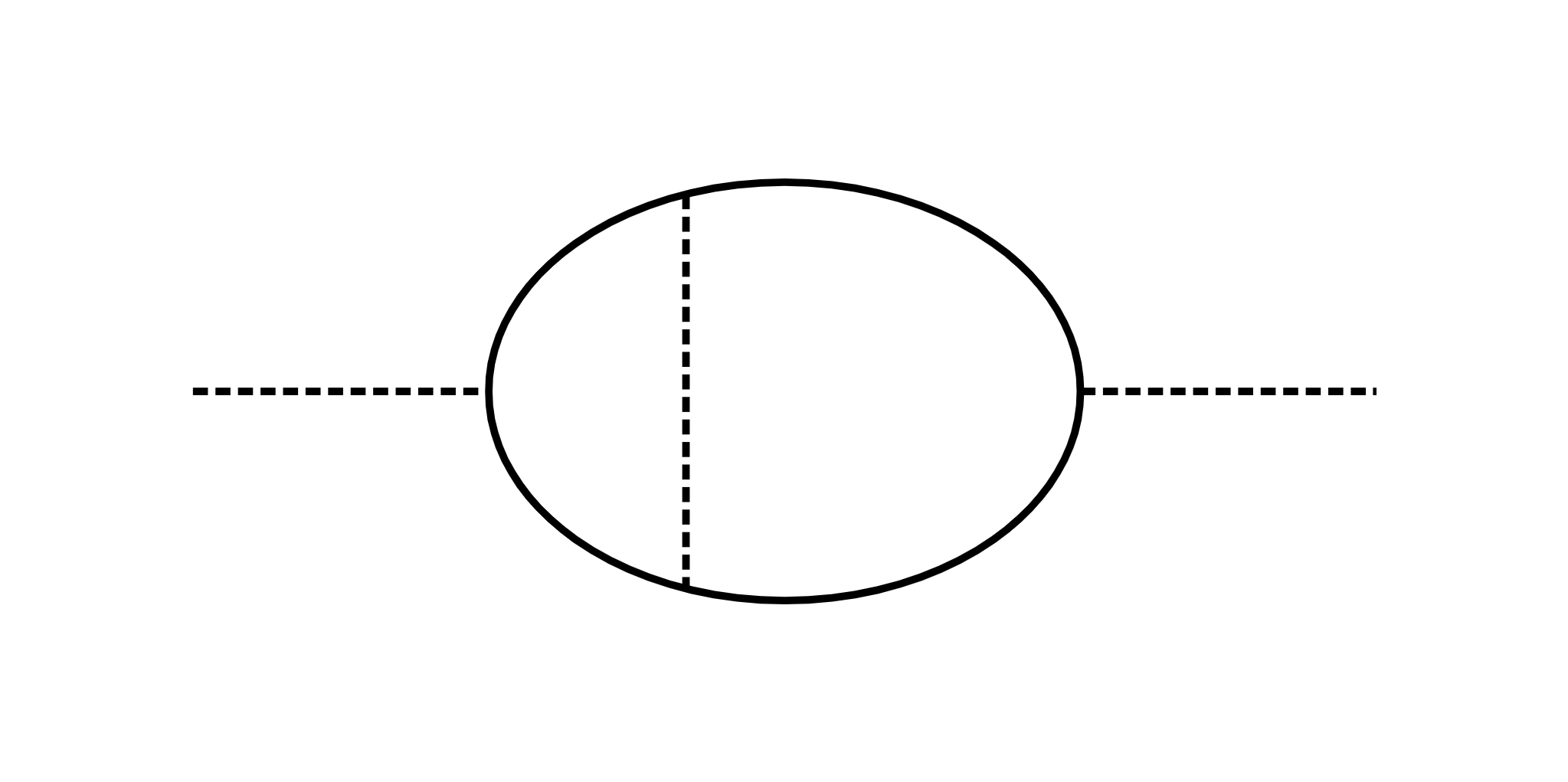}}
\caption{Diagram with corrected vertex} \label{fig:digram+vertex}
\end{minipage}
\hfill 
\begin{minipage}[t]{0.45\linewidth}
\center{\includegraphics[width=1\linewidth]{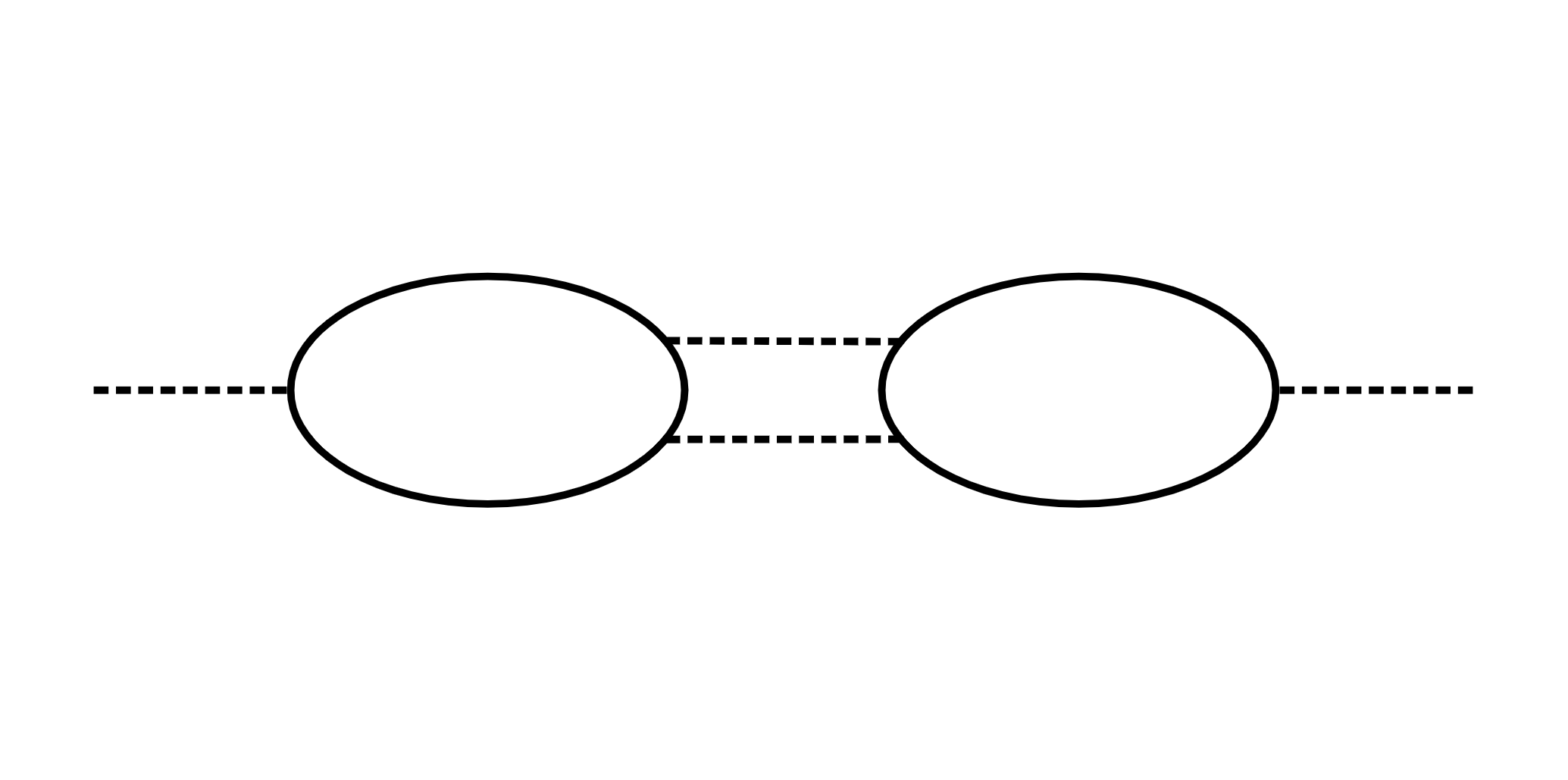}}\caption{Two loops connected with double boson propagator} \label{fig:two-loop}
\end{minipage}
\end{figure}
Let us classify what sort of diagrams can provide contributions to the exact boson propagator $\left\langle \phi_1\phi_2\right\rangle $. First, note that corrections to fermion propagators vanish due to the fact that they come from the interaction vertex $V$, which contains fields in coincident points, and $\left\langle \bar{\psi}\psi\right\rangle _{exact}=\left\langle \bar{\psi}\psi\right\rangle _{0}$, as we have already shown above\footnote{This observation means that we know the exact value of the fermionic two--point functions in the theory under consideration.}. Many-loop diagrams containing (Fig.~\ref{fig:one-loop-1}) and (Fig.~\ref{fig:bubble-diagramm}) and even such diagrams with corrected vertexes, for example, (Fig.~\ref{fig:digram+vertex}) vanish for the same reasons as have been discussed in the previous subsection.

Consider loops connected with more than one boson propagator, for example, (Fig.~\ref{fig:two-loop}). 
\begin{figure}[t]
\begin{minipage}[b]{0.45\linewidth}
\center{\includegraphics[width=1\linewidth]{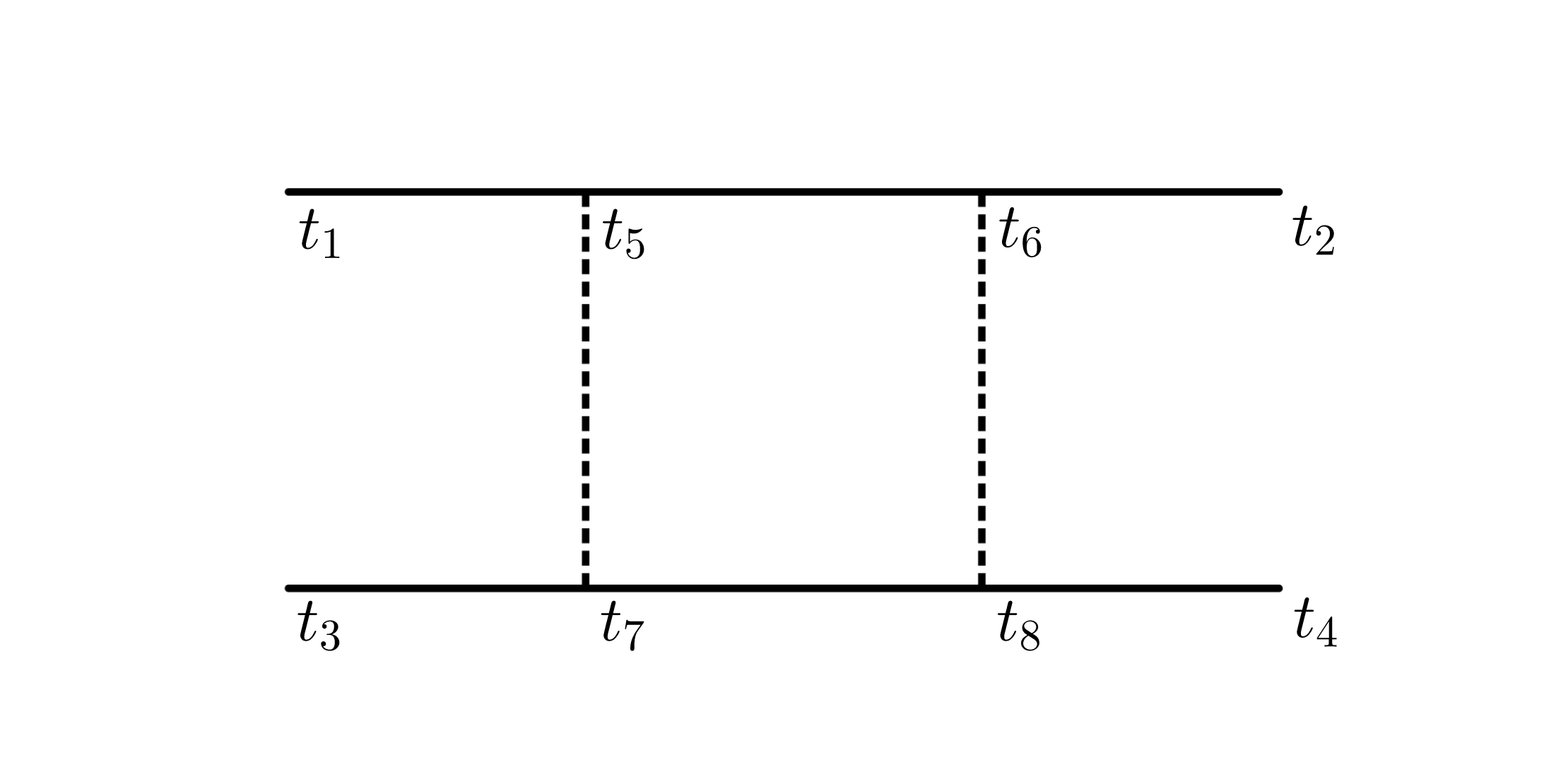}}
\caption{Ladder parallel diagram} 
\label{fig:ladder-1}
\end{minipage}
\hfill 
\begin{minipage}[b]{0.45\linewidth}
\center{\includegraphics[width=1\linewidth]{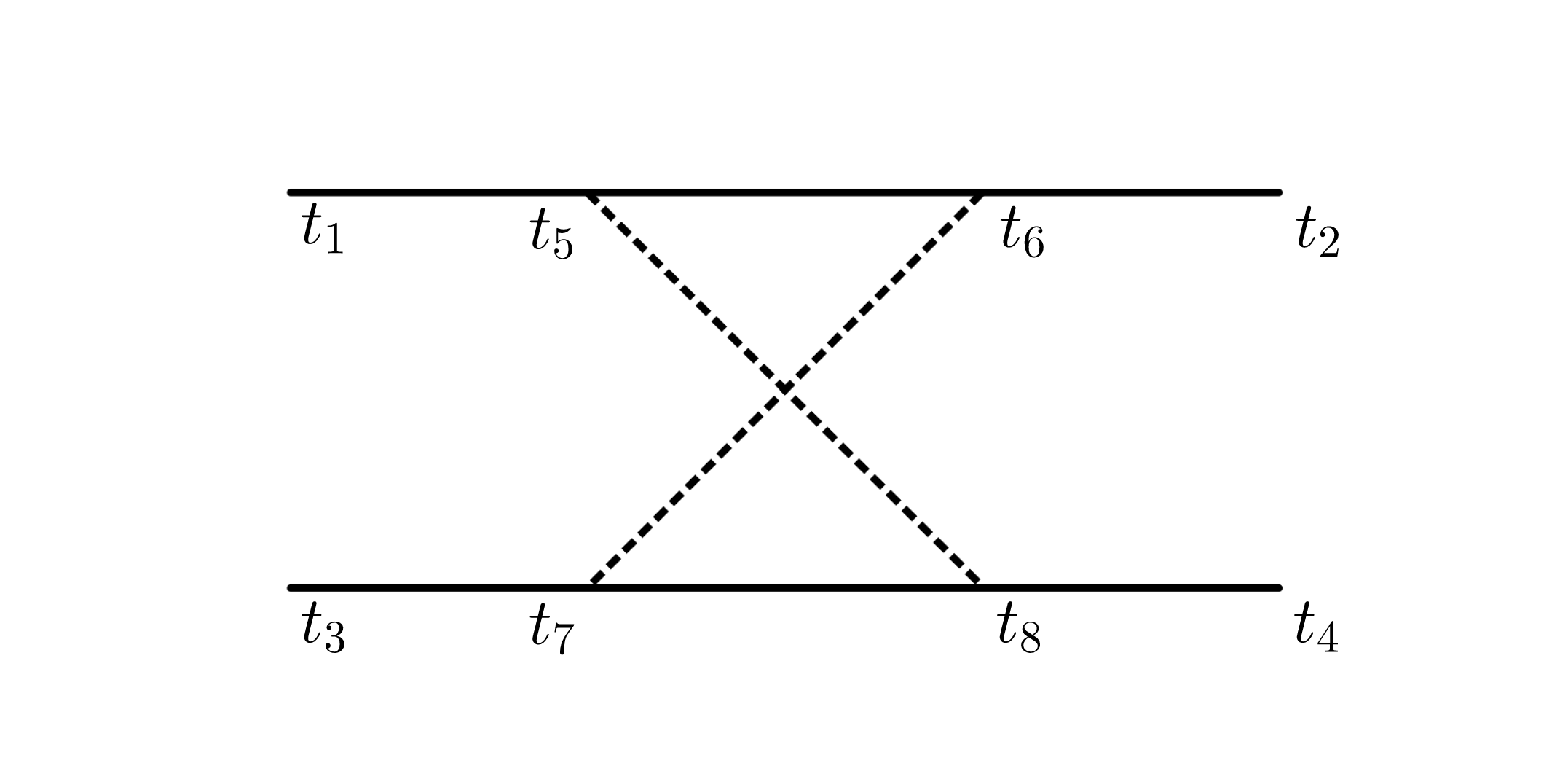}}
\caption{Ladder cross diagram}
\label{fig:ladder-2}
\end{minipage}
\end{figure}
To prove that this diagram also vanishes, we consider such diagrams as depicted on the Figs.~\ref{fig:ladder-1} and~\ref{fig:ladder-2}.
These two diagrams are described by the following expression:

\begin{multline}\label{1s2f38}
    \Delta G^{\sigma_1\sigma_2\sigma_3\sigma_4}(t_1,t_2,t_3,t_4)=\frac{\lambda^4}{2}\int dt_5dt_6dt_7dt_8\sum\limits_{\sigma_{5,6,7,8}=\{+,-\}}G^{\sigma_1\sigma_5}(t_1,t_5)G^{\sigma_5\sigma_6}(t_5,t_6)G^{\sigma_6\sigma_2}(t_6,t_2)\times \\
    \times G^{\sigma_4\sigma_7}(t_4,t_7)G^{\sigma_7\sigma_8}(t_7,t_8)G^{\sigma_8\sigma_3}(t_8,t_3)D^{\sigma_5\sigma_7}(t_5,t_7)D^{\sigma_6\sigma_8}(t_6,t_8)\text{sgn}(\sigma_5\sigma_6)\text{sgn}(\sigma_7\sigma_8).
\end{multline}
Note that $G^{+-}=0$, so only expressions of the following form:
\begin{equation}\label{1s2f39}
\begin{aligned}
    G^{\sigma_1+}(t_1,t_5)G^{++}(t_5,t_6)G^{+\sigma_2}(t_6,t_2)G^{\sigma_4+}(t_4,t_7)G^{++}(t_7,t_8)G^{+\sigma_3}(t_8,t_3)D^{++}(t_5,t_7)D^{++}(t_6,t_8), \\
    G^{\sigma_1+}(t_1,t_5)G^{++}(t_5,t_6)G^{+\sigma_2}(t_6,t_2)G^{\sigma_4-}(t_4,t_7)G^{--}(t_7,t_8)G^{-\sigma_3}(t_8,t_3)D^{+-}(t_5,t_7)D^{+-}(t_6,t_8), \\
    G^{\sigma_1-}(t_1,t_5)G^{--}(t_5,t_6)G^{-\sigma_2}(t_6,t_2)G^{\sigma_4+}(t_4,t_7)G^{++}(t_7,t_8)G^{+\sigma_3}(t_8,t_3)D^{-+}(t_5,t_7)D^{-+}(t_6,t_8), \\
    G^{\sigma_1-}(t_1,t_5)G^{--}(t_5,t_6)G^{-\sigma_2}(t_6,t_2)G^{\sigma_4-}(t_4,t_7)G^{--}(t_7,t_8)G^{-\sigma_3}(t_8,t_3)D^{--}(t_5,t_7)D^{--}(t_6,t_8)
\end{aligned}
\end{equation}
may give non-zero contributions. But due to the presence of theta-functions they are proportional to $(t_1-t_2)^2$ and, as we remember, such diagrams come from the interaction vertex $V$ which contains fields in the coincident points, where $\bar{\psi}(t)\psi(t)$ does not depend on time $t$ due to the form of modes~\eqref{1s2f5}. Hence, for example, in the diagram of the Fig.~\ref{fig:two-loop} we have, as a part, the four point correlation function $G^{\sigma_1\sigma_2\sigma_3\sigma_4}(t,t,t,t)$, in which we can set all its arguments equal to $t$. Hence, that enforces $t_1=t_2$ and contributions from the Fig.~\ref{fig:ladder-1} and Fig.~\ref{fig:ladder-2} vanish in the case under consideration. The contributions of higher-loop diagrams are also zero for the same reason.

As a result, only the remaining tadpole diagrams (Fig.~\ref{fig:tadpole-diagramms}) can give the non--vanishing contribution. So the exact propagators are as follows:

\begin{equation}\label{1eq:anti0}
\begin{aligned}
    D^{+-}_{exact}(t_1,t_2)&=D_0^{+-}(t_1,t_2)+\Delta\left\langle \phi_1\right\rangle \Delta\left\langle \phi_2\right\rangle =\frac{(1-it_1)(1+it_2)}{2}+\frac{\lambda^2}{4}(t_1-t_0)^2(t_2-t_0)^2,\\
    D^{K}_{exact}(t_1,t_2)&=D_0^{K}(t_1,t_2)+\Delta\left\langle \phi_1\right\rangle \Delta\left\langle \phi_2\right\rangle =\frac{1+t_1 t_2}{2}+\frac{\lambda^2}{4}(t_1-t_0)^2(t_2-t_0)^2,\\
    D^{R/A}_{exact}(t_1,t_2)&=D^{R/A}_0(t_1,t_2)=\pm i\theta(\pm t_1 \mp t_2)(t_2-t_1).
\end{aligned}
\end{equation}
This generalizes the result of the subsection~\ref{1s2.1} to the arbitrary order in $\lambda$ and, as we have explained above, comes from the solution of the eq.(\ref{totsol}) with the tadpole appearing due to non--zero right hand side $\langle 1 | \bar{\psi} \psi | 1\rangle$.

\section{Linearly growing in time background scalar field in two dimensions}
\label{sec:2D-Et}
\setcounter{equation}{0}

In this section we consider the Yukawa model of interacting fermions and real scalar field in $(1+1)$-dimensional Minkowski space-time with $(+, -)$ signature of the metric:
\beq
\label{eq:action}
S = \int d^2 x \left[ \frac{1}{2} \partial_{\mu} \phi \partial^{\mu} \phi + i \bar{\psi} \slashed{\partial} \psi - \lambda \phi \bar{\psi} \psi \right],
\eeq
where we denote $\slashed{\partial} \equiv \gamma^\mu \partial_\mu$, $\bar{\psi} = \psi^\dagger \gamma^0$ and assume that the coupling parameter is $\lambda > 0$. In this section we use the Dirac-Pauli representation for the Clifford algebra:
\beq
\label{eq:gamma-1}
\gamma^0 \equiv \bem 1 & 0 \\ 0 & -1 \eem, \quad \gamma^1 \equiv \bem 0 & 1 \\ -1 & 0 \eem.
\eeq
The equations of motion for the action~\eqref{eq:action} are as follows:
\beq
\label{eq:system}
\begin{cases}
  \partial^2 \phi + \lambda \bar{\psi} \psi = 0, \\
  \left(i \slashed{\partial} - \lambda \phi \right) \psi = 0.
\end{cases}  
\eeq
Their classical solutions can be taken as $\psi_{cl} = 0, \phi_{cl} = \mathcal{F}(t - x) + \tilde{\mathcal{F}}(t + x)$, where $\mathcal{F}$ and $\tilde{\mathcal{F}}$ are arbitrary smooth functions. In what follows we consider such classical solutions as external backgrounds and split the classical and quantum parts of the fields: $\phi = \phi_{cl} + \phi_q$, $\psi = \psi_{cl} + \psi_q$. Our goal is to calculate correlation functions.

Concretely, in this section we consider the background field which linearly grows with time: $\phi_{cl} = E t$, where $E$ is some real positive constant\footnote{Of course, one can obtain classical solutions with other values and signs of this constant via time shifts: $t \rightarrow t + \delta t \Longrightarrow \phi_{cl} = E \delta t + E t$, or reversals: $t \rightarrow - t \Longrightarrow \phi_{cl} = \frac{m}{\lambda} - E t$. E.g. one can give a mass $m_\psi$ to the fermion field by the time shift $\delta t = \frac{m_\psi}{\lambda E}$. However, these transformations do not bring anything substantially new into our discussion. So, we consider positive $E$ and zero mass without loss of generality.}. Specifically, in the limit $E \rightarrow 0$ this background reproduces free massless fermion field. When $E \neq 0$ the Hamiltonian depends on time, i.e. the situation is not stationary. Hence, one may expect the particle creation that is similar to the one in strong electric~\cite{Schwinger, Nikishov, Narozhnyi, Grib} or gravitational fields~\cite{Birrel}. 

However, let us emphasize the difference between e.g. the pair creation in the electric field background (the well-known Schwinger effect~\cite{Schwinger}) and processes in the scalar field background. On the one hand, at tree--level the particle creation in the electric field can be attributed to the quantum tunneling through the classically forbidden region. The rate of such a process is described by an imaginary part of the effective action; moreover, the expression for the rate is not an analytic function of the background field~\cite{Grib}. On the other hand, as we will see below the imaginary part of the Feynman effective action on the scalar field background is zero (see Appendix~\ref{sec:effective}). Hence, non-zero quantum expectation values indicate rather vacuum polarization than particle creation. But, as is discussed in \cite{Diatlyk} the situation with the particle creation in the background scalar fields is not that trivial.

Furthermore, note that the background scalar field $\phi_{cl} = E t$ is rather unrealistic, since an indefinitely growing field requires infinite amount of energy. However, it allows one to grasp the main properties of the model. It would be more appropriate to consider the pulse background $\phi_{cl} = ET \tanh \frac{t}{T}$, which becomes constant at the past and future infinities and reproduces the linear growth for $|t| \ll T$. Such a configuration does not solve the equations of motion without a source in~\eqref{eq:system}. Another possibility is to consider a strong wave, i.e. $\mathcal{F}(t-x)$ which has compact support. The latter classical background was considered in~\cite{Diatlyk}.

\subsection{Modes}
\label{sec:modes}

To set up the notations let us start with the consideration of the free massive fermion field without a background scalar field. This field can be decomposed into the modes as follows:
\beq
\label{eq:modes-decomp}
\psi(t,x) = \int \frac{dp}{2 \pi} \left[ a_p \psi_p^{(+)}(t,x) + b_p^\dagger \psi_p^{(-)}(t,x) \right].
\eeq
The functions $\psi_p^{(+)}(t,x) \equiv u_p e^{-i p x}$ and $\psi_p^{(-)}(t,x) \equiv v_p e^{i p x}$, which are positive and negative frequency modes, solve the free equations of motion:
\beq
\label{eq:free}
(i \slashed{\partial} - m)\psi = 0,
\eeq
and creation and annihilation operators obey the standard anticommutation relations:

\beq
\label{eq:anti-a}
\left\{ a_p, a_q^\dagger \right\} = \left\{ b_p, b_q^\dagger \right\} = 2 \pi \delta(p-q).
\eeq
This fixes the equal-time anticommutation relations for $\psi$ and $\psi^\dagger$:
\beq
\label{eq:anti}
\left\{\psi_a(t,x), \psi_b^{\dagger}(t,y) \right\} = \delta(x-y) \delta_{ab},
\eeq
where we restored the spinor indices $a,b=1,2$. The form of $u_p$ and $v_p$ spinors is as follows:
\beq
\label{eq:free-modes}
u_p = \bem u_{p,1} \\ u_{p,2} \eem =  \frac{\text{sgn}(p)}{\sqrt{2 \omega (\omega - m)}} \bem p \\ \omega - m \eem, \quad v_p  = \bem v_{p,1} \\ v_{p,2} \eem = \frac{\text{sgn}(p)}{\sqrt{2 \omega (\omega - m)}} \bem \omega - m \\ p \eem,
\eeq
where $\omega = \sqrt{p^2 + m^2}$ and we have used the Dirac representation for gamma-matrices~\eqref{eq:gamma-1}. For further purposes (see footnote~\ref{footnote:phase}) we introduced the phase factor $\text{sgn}(p)$ which does not affect the conditions~\eqref{eq:free} and~\eqref{eq:anti}. In what follows we omit the index $p$ of $u_p$, $v_p$ and $\psi_p$ where it can be easily restored.

The fermion field in the time-dependent background can be decomposed in the way similar to~\eqref{eq:modes-decomp}, except that functions $\psi^{(\pm)}$ solve the equations of motion~\eqref{eq:system} with $\phi = \phi_{cl}$ instead of the free equations~\eqref{eq:free}:

\begin{equation}
\label{eq:Et-main}
\left[ i \gamma^{\mu} \partial_{\mu} - M(t) \right] \psi = 0,
\end{equation}
where we define for short:
\beq
\label{eq:M-definition}
M(t) = \alpha t, \quad \alpha = \lambda E. \eeq
Because of the spatial homogeneity it is convenient to represent the modes in the following form:

\beq 
\psi(t,x) = \psi_p(t) e^{ipx}.\eeq
Substituting this factorized solution into~\eqref{eq:Et-main}, one obtains the equation for the time dependent part of the modes:

\beq
\label{eq:Et-time}
\left[i \gamma^0 \partial_t - \gamma^1 p - M(t) \right] \psi_p(t) = 0.
\eeq
One can decouple this system applying the operator $\left[-i \gamma^0 \partial_t - \gamma^1 p - M(t) \right]$ to its left hand side and keeping in mind that the eigenvalues of $\gamma^0$ are $\pm 1$. Hence, the equation reduces to:
\beq \begin{gathered}
\left[ \partial_t^2 + \left(\omega_p^{(1,2)}\right)^2(t)\right] \psi_{1,2}(t) = 0, \quad \text{where} \\
\left(\omega_p^{(1,2)}\right)^2(t) \equiv p^2 + \alpha^2 t^2 \pm i \alpha.
\end{gathered} \eeq
Note that this resembles the equation for the massive charged scalar field on the constant electric field background~\cite{Krotov, Akhmedov:Et, Akhmedov:Ex}. Its exact solution is the sum of linearly independent parabolic cylinder functions $D_{\nu}(z)$:
\beq
\label{eq:mode-options} \begin{gathered}
\psi_1\left[ z(t) \right] = A_1 D_{\nu}\left[ z(t) \right] + B_1 D_{-\nu - 1}\left[ i z(t) \right], \\ \psi_2\left[ z(t) \right] = A_2 D_{\nu - 1}\left[ z(t) \right] + B_2 D_{-\nu}\left[ i z(t) \right],
\end{gathered} \eeq
where $A_{1,2}$, $B_{1,2}$ are complex constants which we fix below, and we define for convenience:
\beq \label{eq:variables-def} z \equiv \frac{1+i}{\sqrt{\alpha}} M(t), \quad \nu \equiv -\frac{i p^2}{2 \alpha}. \eeq
It is not possible to define usual in-- and out-- modes as well as positive and negative frequency solutions in our case due to the fact that the external field is never switched off. Indeed, parabolic cylinder function has the following asymptotic behavior~\cite{Bateman-2,Whittaker}:
\beq \begin{aligned}
\label{eq:D-bigz}
& D_{\nu}(z) = z^{\nu} e^{-\frac{z^2}{4}} \left[\sum_{n = 0}^N \frac{\left(-\frac{\nu}{2} \right)_n \left(\frac{1}{2} - \frac{\nu}{2} \right)_n}{n! \left(-\frac{z^2}{2} \right)^n} + \mathcal{O}\left|z^2\right|^{-N-1} \right], \\
& \left(\gamma\right)_0 = 1, \quad \left(\gamma\right)_{n \neq 0} = \gamma\left(\gamma + 1 \right) \cdots \left(\gamma + n - 1 \right),
\end{aligned} \eeq
for $|z| \gg |\nu|$ and $\left|\text{Arg}(z)\right| < \frac{\pi}{2}$. In our case $\text{Arg}(z) = \pm \frac{\pi}{4}$ and the condition $|z| \gg |\nu|$ is satisfied for sufficiently large times $|t| \gg \frac{p^2}{\alpha^{3/2}}$. So, in the leading order as $t \rightarrow +\infty$ one obtains:
\beq
\psi_{1,2}(z(t)) \sim A_{1,2}(p) \exp\left(-\frac{i}{2} \alpha t^2 - \frac{i p^2}{2 \alpha} \log t\right) + B_{1,2}(p) \exp\left(\frac{i}{2} \alpha t^2 + \frac{i p^2}{2 \alpha} \log t\right),
\eeq
where $A_{1,2}(p)$ and $B_{1,2}(p)$ are some constants that do not depend on time (but depend on the momentum). Thus, the modes $\psi_{1,2}(t,x)$ cannot be reduced to the sum of positive and negative frequency plane waves, and the interpretation in terms of particles is meaningless. Please keep in mind that in non-stationary situations it is more appropriate to calculate correlation functions rather than amplitudes, at least because there are no asymptotic particle states \cite{Akhmedov:2009be, Akhmedov:2009vh, Akhmedov:2008pu}.

However, let us check the other limit --- the ultraviolet region, where $|p| \gg \sqrt{\alpha}$ for a fixed $t$. In such a limit we expect that the modes in the strong scalar background and in the free theory have similar behavior. In fact, in this case the parabolic cylinder function has the following asymptotic expansion (see Appendix~\ref{sec:asymptotics} for details):

\beq
\label{eq:D-bignu}
D_\nu\left[z(t)\right] \simeq \frac{e^{\frac{\pi p^2}{8 \alpha}}}{\sqrt{2}} \left(\frac{M}{\sqrt{M^2 + p^2}} + 1\right)^{\frac{1}{2}} e^{\frac{i p^2}{4 \alpha} - \frac{i p^2}{4 \alpha} \log \frac{\left(\sqrt{M^2 + p^2} + M\right)^2}{2 \alpha} - \frac{i M \sqrt{M^2 + p^2}}{2 \alpha}} \left[1 + \mathcal{O}\left(\frac{\alpha}{M^2 + p^2}\right) \right].
\eeq
Hence, for times $|t| \ll \frac{|p|}{\alpha}$ the exact modes behave as follows:

\beq \begin{gathered} 
\psi_{1,2}(t, x) \sim A'_{1,2}(p) e^{-i |p| t + i p x} + B'_{1,2}(p) e^{i |p| t + i p x},
\end{gathered} \eeq
which means that for fixed time and large momenta one obtains the standard flat space plane waves. Now it is clear that functions $D_\nu\left[z(t)\right]$ and $D_{\nu - 1}\left[z(t)\right]$ correspond to ``positive frequency'' modes, i.e. the exact harmonics should be as follows:

\beq \psi^{(+)}(t) \equiv \bem \psi_1^{(+)}(t) \\ \psi_2^{(+)}(t) \eem = A^{(+)} \bem D_\nu\left[z(t)\right] \\ \frac{\left(i \partial_t - M(t)\right)}{p} D_\nu\left[z(t)\right] \eem, \eeq
where we used the system~\eqref{eq:Et-time} to relate the first and second components of the spinor. One can simplify this expression using the following relations for parabolic cylinder functions~\cite{Bateman-2,Whittaker}:

\begin{equation}\label{s2f19}
\begin{gathered}
\partial_zD_\nu(z)+\frac{1}{2}zD_\nu(z) - \nu D_{\nu-1}(z) = 0, \\
\partial_zD_\nu(z)-\frac{1}{2}zD_\nu(z) + D_{\nu+1}(z) = 0,
\end{gathered}
\end{equation}
and represent the ``positive frequency'' modes in the form:

\beq
\label{eq:mode+}
\psi_p^{(+)}(t,x) = A^{(+)} \bem D_\nu\left[z(t)\right] \\ \frac{1 + i}{\sqrt{2}} \frac{p}{\sqrt{2 \alpha}} D_{\nu - 1}\left[z(t)\right] \eem e^{i p x}.
\eeq
They behave as $\psi \sim e^{-i |p| t + i p x}$ for sufficiently large momenta. We choose to consider such modes out of all options present in eq.~\eqref{eq:mode-options} because they have proper UV behavior, i.e. tend to the free fermion field modes in the limit $ p \rightarrow \infty$. Propagators expanded in such modes possess the proper Hadamard behaviour. Which means that they lead to the same UV renormalization as in the absence of the background field. On general grounds we think that this is the appropriate physical picture. We come back to the discussion of other options below at the end of this subsection.

In the same way one obtains the ``negative frequency'' modes:
\beq
\label{eq:mode-}
\psi_p^{(-)}(t,x) = A^{(-)} \bem \frac{1 - i}{\sqrt{2}} \frac{p}{\sqrt{2 \alpha}} D^*_{\nu - 1}\left[z(t)\right] \\ D^*_\nu \left[z(t)\right] \eem e^{-i p x},
\eeq
which behave as $\psi \sim e^{i |p| t - i p x}$ for sufficiently large momenta.

Let us fix the coefficients $A^{(+)}$ and $A^{(-)}$ using the equal-time anticommutation relations~\eqref{eq:anti}:

\beq \begin{aligned}
& \left\{\psi_a(t,x), \psi_b^{\dagger}(t,y) \right\} = \\ & = \iint \frac{dp}{2\pi} \frac{dq}{2\pi} \left[ \left\{a_p, a_q^+\right\} \psi_{a,p}^{(+)}(t) \psi_{b,q}^{(+)}(t)^{*} e^{i(p x - q y)} + \left\{b_p^+,b_q \right\} \psi_{a,p}^{(-)}(t) \psi_{b,q}^{(-)}(t)^{*} e^{-i\left(px-qy \right)} \right] = \\ &
= \int \frac{dp}{2\pi} \left[\psi_{a,p}^{(+)}(t) \psi_{b,p}^{(+)}(t)^{*} + \psi_{a,-p}^{(-)}(t) \psi_{b,-p}^{(-)}(t)^{*} \right] e^{ip(x-y)} = \delta(x-y) \delta_{ab},
\end{aligned} \eeq
where we use the canonical anticommutation relations~\eqref{eq:anti-a}. This condition is satisfied if

\beq
\label{eq:Et-normalization}
\psi_{a,p}^{(+)}(t) \psi_{b,p}^{(+)}(t)^* + \psi_{a,-p}^{(-)}(t) \psi_{b,-p}^{(-)}(t)^{*} = \delta_{ab} \iff \begin{cases} |A^{(+)}|^2 \left| D_\nu(z) \right|^2 + |A^{(-)}|^2 \frac{p^2}{2 \alpha} \left| D_{\nu-1}(z) \right|^2 = 1, \\ |A^{(-)}|^2 \left| D_\nu(z) \right|^2 + |A^{(+)}|^2 \frac{p^2}{2 \alpha} \left| D_{\nu-1}(z) \right|^2 = 1, \\ \left( |A^{(+)}|^2 - |A^{(-)}|^2 \right) \frac{1-i}{\sqrt{2}} \frac{p}{\sqrt{2 \alpha}} D_\nu(z) D_{\nu-1}^*(z) = 0, \end{cases}
\eeq
for arbitrary times $z(t)$. Note that this condition is time-independent due to the equations of motion and the relation $\psi_{a,-p}^{(-)}(t) = - \gamma_{ab}^1 \left( \psi_{b,p}^{(+)}(t) \right)^*$ which follows from the symmetry of the system~\eqref{eq:system}:

\beq \partial_t \left( \psi_{a,p}^{(+)}(t) \psi_{b,p}^{(+)}(t)^* + \psi_{a,-p}^{(-)}(t) \psi_{b,-p}^{(-)}(t)^* \right) = 0. \eeq
First, \eqref{eq:Et-normalization} implies that $|A^{(+)}|^2 = |A^{(-)}|^2 = |A|^2$. Second, it allows one to find the constant $|A|^2$ by setting the argument of parabolic cylinder functions equal to any convenient value, e.g. to zero:

\beqs |A|^2 \left[\frac{\pi}{\left|\Gamma\left(\frac{1}{2} + \frac{i p^2}{4 \alpha} \right) \right|^2} + \frac{p^2}{4 \alpha} \frac{\pi}{\left|\Gamma \left(1 + \frac{ip^2}{4\alpha} \right) \right|^2} \right] = 1. \eeqs
Using the properties of the Gamma function:

\beqs \left|\Gamma(iy)\right|^2 = \frac{\pi}{y \sinh(\pi y)}, \quad \left|\Gamma \left(\frac{1}{2} + iy \right) \right|^2 = \frac{\pi}{\cosh(\pi y)}, \eeqs
we find that

\beq |A|^2 = e^{-\frac{\pi p^2}{4 \alpha}}. \eeq
Let us sum up the main results of this subsection, i.e. write down the asymptotic expressions for the modes.

For $t > 0$, $\alpha |t| \ll |p|$, $|p| \gg \sqrt{\alpha}$ one obtains up to a $\mathcal{O} \left(\frac{M^2}{p^2}\right)$ that the modes behave as:

\beq
\label{eq:modes-bigp}
\psi^{(+)}(t,x) \simeq \frac{1}{\sqrt{2}} \bem 1 + \frac{|M|}{2|p|} \\ \text{sgn}(p) \left( 1 - \frac{|M|}{2|p|} \right) \eem e^{-i |p| t + i p x + \frac{i p^2}{4 \alpha} - \frac{i p^2}{4 \alpha} \log \frac{p^2}{2 \alpha} + i \tilde{\varphi}},
\eeq
where $\tilde{\varphi}$ is an arbitrary constant phase independent of $t$ and $p$. Up to an irrelevant phase this asymptotic behaviour coincides\footnote{\label{footnote:phase} For this reason we have introduced the phase factor $\text{sgn}(p)$ in eq.~\eqref{eq:free-modes}.} with the free modes~\eqref{eq:free-modes}. 

At the same time, for $t > 0$, $\alpha |t| \gg |p|$, $|t| \gg \frac{1}{\sqrt{\alpha}}$ one obtains up to a $\mathcal{O} \left(\frac{p^2}{M^2}\right)$ that the modes behave as
\beq
\label{eq:modes-smallp}
\psi^{(+)}(t,x) \simeq \bem 1 \\ \frac{p}{2 |M|} \eem \left( 2 \alpha t^2 \right)^{\frac{i p^2}{4 \alpha}} e^{-\frac{i \alpha t^2}{2} + i p x + \frac{i p^2}{4 \alpha} \log \frac{p^2}{2 \alpha} + i \tilde{\varphi}}.
\eeq
The ``negative frequency'' modes are obtained from the ``positive frequency'' ones by the charge conjugation operation: 

\beq 
\psi_p^{(-)}(t,x) = \gamma^5 \psi_p^{(+)*}(t,x), 
\eeq
where $\gamma^5 = \gamma^0 \gamma^1$. Also one can check that the modes obey the following relation:
\beq
\label{eq:mirror}
\psi_p^{(+)}(-t,x) = \sgn p \, \gamma^5 \psi_p^{(+)*}(t,x).
\eeq
Finally let us point out the following important issue. In this subsection we have found a complete basis of modes solving the classical equations of motion. But there is an ambiguity in the choice of such a basis. Depending on this choice, there are different ``ground'' Fock space states in the theory. In fact, instead of (\ref{eq:mode+}) and (\ref{eq:mode-}) one could consider canonically transformed basis of modes:

\begin{equation}
    \label{bt02}
    \widetilde{\psi}^{(+)}_{p}(t,x) = \int \dfrac{dq}{2\pi}\bigg[a_{pq}\psi^{(+)}_{q}(t,x) + b_{pq} \psi^{(-)}_{q}(t,x)\bigg], \quad
    \widetilde{\psi}^{(-)}_{p}(t,x) = \int \dfrac{dq}{2\pi} \bigg[c_{pq}\psi^{(+)}_{q}(t,x) + d_{pq}\psi^{(-)}_{q}(t,x)\bigg].
\end{equation}
To respect the canonical anti--commutation relations for the fermionic fields and for the corresponding creation and annihilation operators the Bogoliubov coefficients, $a_{pq}, b_{pq}, c_{pq}$ and $d_{pq}$, should satisfy certain relations which are listed in \cite{Diatlyk}.

On physical grounds one also should demand that

\begin{eqnarray} \label{alphapq}
    a_{pq} \approx d_{pq} \approx \delta(p-q), \quad b_{pq} \approx c_{pq} \approx 0,
\end{eqnarray}
as $p$ is taken to infinity. That is necessary for the propagators to have the proper Hadamard behaviour.

Thus, there is no unique way to choose the basis of modes and all possibilities in (\ref{bt02}) are in principle allowed and may lead to different physical situations. This fact is apparent when there is no preferable basis of special functions found in XIX century and listed in the standard text books.

For a given choice of modes one can define a new Fock space ``ground'' state:

\begin{equation}
    \hat{\widetilde{a}}_p |a,b,c,d\rangle = \hat{\widetilde{b}}_p |a, b,c,d\rangle = 0,\label{abgest}
\end{equation}
where $\hat{\widetilde{a}}_p$ and $\hat{\widetilde{b}}_p$ are canonically transformed annihilation operators. For this new state certain physical quantities will be different from those for the original state \cite{Diatlyk}. However, we will argue, as it was also done in \cite{Diatlyk}, that the scalar current, $\langle \bar{\psi} \psi \rangle$, at leading approximation for large and slowly changing background scalar field does not depend on the choice of the initial state. 

\subsection{Tree-level scalar current}
\label{sec:current}

In the previous subsection we derived the exact modes for the fermion field, which in a sense describes the fermion response to the strong scalar field background. In this subsection we find the response of the scalar field itself due to the presence of the non--trivial fermion zero--point fluctuations in the scalar field background under consideration. 

Quantizing the Hamiltonian of the theory~\eqref{eq:action}:
\beq \hat{H} = \int dx \left[ \frac{1}{2} \left( \partial_t \hat{\phi} \right)^2 + \left( \partial_x \hat{\phi} \right)^2 - i \hat{\bar{\psi}} \gamma^1 \partial_x \hat{\psi} + \lambda \hat{\phi} \hat{\bar{\psi}} \hat{\psi} \right], \eeq
and using Hamilton's equations:
\beq \dot{\hat{\phi}}(x) = i \left[ \hat{H}, \hat{\phi}(x) \right], \quad \dot{\hat{\psi}}(x) = i \left[ \hat{H}, \hat{\psi}(x) \right], \eeq
one obtains the following operator equation for the scalar field:
\beq 
\label{eq:opEOM}
\partial^2 \hat{\phi} + \lambda \hat{\bar{\psi}} \hat{\psi} = 0, 
\eeq
which reproduces one of the classical equations of motion~\eqref{eq:system}. Hence, one needs to calculate the scalar current $j_{cl}(t) \equiv \langle \hat{\bar{\psi}} \hat{\psi} \rangle$ to find the response of the classical field $\phi_{cl} = \langle \hat{\phi} \rangle$. This current has the following form:
\beq
\label{eq:current}
\begin{aligned}
\langle \bar{\psi} \psi \rangle (t) &= \iint \frac{dp}{2\pi} \frac{dq}{2\pi} \left[ \langle b_p b_q^\dagger \rangle \left( \psi_{1,p}^{(-)}(t) \psi_{1,q}^{(-)}(t)^{*} - \psi_{2,p}^{(-)}(t) \psi_{2,q}^{(-)}(t)^{*} \right) e^{i(p-q)x} \right] = \\ &= \int \frac{dp}{2\pi} \left( \left|\psi_{1,p}^{(-)}(t)\right|^2 - \left|\psi_{2,p}^{(-)}(t)\right|^2 \right)  = \int \frac{dp}{2\pi} \left( 1 - 2 e^{-\frac{\pi p^2}{4\alpha}} \left| D_\nu\left[z(t)\right] \right|^2 \right),
\end{aligned}
\eeq
where we have used the notations of Sec.~\ref{sec:modes} for short, and in the last line also we have used one of the relations~\eqref{eq:Et-normalization}. Note that in principle the equation under consideration provides an implicit expression for the current. However, this form of the current is hard to interpret in physical terms. To obtain physically tractable equations we will consider only the leading contribution in the limit $t \rightarrow \infty$ for small $\alpha$.

Before evaluating the integral~\eqref{eq:current}, consider the case of a free fermion field with a mass $m$. Using the free modes~\eqref{eq:free-modes} one obtains the following free current:
\beq
\label{eq:free-current}
\langle \bar{\psi} \psi \rangle_{free} = - \int_{- \Lambda}^{\Lambda} \frac{dp}{2\pi} \frac{m}{\sqrt{m^2 + p^2}} \approx \frac{m}{\pi} \log \frac{m}{2 \Lambda},
\eeq
where we have introduced the ultraviolet cutoff at the scale $\Lambda$. Note that the constant classical background $\phi_{cl} = \frac{m}{\lambda}$, substituted into the system~\eqref{eq:system}, reproduces this case. The analog of the mass parameter $m$ in the theory~\eqref{eq:action} is $M(t) = \lambda \phi_{cl} = \lambda E t$. Thus one expects the following behavior for the current~\eqref{eq:current}:

\beq
\label{eq:current-proposal}
\langle \bar{\psi} \psi \rangle (t) \simeq \frac{\lambda \phi_{cl}}{\pi} \log \frac{\lambda \phi_{cl}}{2 \Lambda}.
\eeq
Let us check this conjecture by calculating the integral~\eqref{eq:current} in such an approximation when $\phi_{cl}$ is large and slowly changing function. Note that $M(t) = \alpha t$ grows indefinitely with time, so it can overcome an arbitrarily large fixed scale $\Lambda$. Due to this fact we consider cases $M < \Lambda$ and $M > \Lambda$ separately. In both cases we assume $M^2 \gg \alpha$ to single out the leading contributions. The case $M > \Lambda$ is rather unphysical as we have already mentioned. However, we consider it for integrity.

In the case $M \ll \Lambda$ we divide the region of integration into two segments: $[0, \Lambda] = [0, \sqrt{\alpha}] + [\sqrt{\alpha}, \Lambda]$, and estimate integrals over these segments using expansions~\eqref{eq:D-bigz} and~\eqref{eq:D-bignu} correspondingly:

\begin{align}
\label{eq:current-calc-1}
\int_0^{\sqrt{\alpha}} dp \left( 1 - 2 e^{-\frac{\pi p^2}{4\alpha}} \left| D_\nu\left[z(t)\right] \right|^2 \right) &\simeq \int_0^{\sqrt{\alpha}} dp \left[1 - 2 + \frac{p^2}{2M^2} + \mathcal{O}\left(\frac{\alpha^2}{M^4}\right) \right] \simeq \nonum &\simeq - \sqrt{\alpha} \left[ 1 - \frac{1}{6} \frac{\alpha}{M^2} + \mathcal{O}\left( \frac{\alpha^2}{M^4} \right) \right]; \\
\label{eq:current-calc-2}
\int_{\sqrt{\alpha}}^\Lambda dp \left( 1 - 2 e^{-\frac{\pi p^2}{4\alpha}} \left| D_\nu\left[z(t)\right] \right|^2 \right) &\simeq \int_{\sqrt{\alpha}}^\Lambda dp \left[1 - \left(1 + \frac{M}{\sqrt{M^2 + p^2}}\right) \left(1 + O\left(\frac{\alpha}{p^2}\right) \right) \right] \simeq \nonum &\simeq M \left[ \log \frac{M}{2 \Lambda} + \mathcal{O}\left( \frac{M^2}{\Lambda^2}, \frac{\sqrt{\alpha}}{M} \right) \right].
\end{align}
Hence, in the limit $t \to \infty$ we obtain that:

\beq
\label{eq:Et-current}
\langle \bar{\psi} \psi \rangle (t) \simeq \frac{\alpha t}{\pi} \log \frac{\alpha t}{2 \Lambda} + \cdots \, ,
\eeq
where we denoted the subleading contribution as ``$\cdots$''. This expression coincides with~\eqref{eq:current-proposal} in the approximation under consideration. It also reproduces the behaviour of the scalar current found in \cite{Diatlyk}.

In the case $M \gg \Lambda$ one can use the decomposition~\eqref{eq:D-bigz} in the entire domain $[0, \Lambda]$:
\beq
\label{eq:Et-current-2}
\langle \bar{\psi} \psi \rangle (t) \sim \int_0^\Lambda dp \left[ - 1 + \frac{p^2}{2 M^2} + \mathcal{O}\left(\frac{\alpha^2}{M^4}\right) \right] \simeq - \Lambda + \frac{1}{6} \frac{\Lambda^3}{M^2} + \cdots,
\eeq
i.e., in the leading order the current does not depend on time and linearly diverges as $\Lambda \rightarrow \infty$. We think that this behavior has no physical sense, e.g., it does not allow us to treat UV divergences properly. This means that an indefinitely growing scalar field is not self-consistent because it is not realistic, as we have mentioned already.

However, this problem can be avoided if one considers a pulse background $\phi_{cl} = ET \tanh \frac{t}{T}$ instead of the $\phi_{cl} = E t$ one. On the one hand, for times $t \ll T$ these backgrounds coincide, hence, the result~\eqref{eq:Et-current} is valid. On the other hand, for times $t \gg T$ the pulse background reproduces the free Dirac field with constant mass $m = \pm \lambda E T$. Hence, if one chooses the UV cutoff $\Lambda \gg  M(T)$, the condition $\lambda \phi_{cl} \ll \Lambda$ is always satisfied, and the equality~\eqref{eq:current-proposal} holds.

Thus, the effective equation of motion for the boson field gets modified in the following way:

\beq \label{eq:current-eom} \partial^2 \langle \phi \rangle + \frac{\lambda^2 \langle \phi\rangle}{\pi} \log\frac{\lambda \langle \phi\rangle}{\Lambda} \approx 0. \eeq
This identity is valid for the fields from the interval $\sqrt{\lambda E} \ll \lambda \phi_{cl} \ll \Lambda$ and $\langle \phi \rangle = \phi_{cl} + \dots$. Note that $\phi_{cl} = Et$ does not solve this equation, i.e. the classical field must restructure itself to satisfy the corrected equation. We discuss the origin of such a behavior in the concluding section and in the Appendix B.

Also note that the true equation of motion cannot depend on the artificial UV cutoff $\Lambda$. This problem can be solved by renormalization of the bare mass of scalar field. It turns out that quantum fluctuations break the symmetry of the problem and bring to the scalar field constant non-zero value $\phi = \langle \phi \rangle_{GS}$ (see Appendix B). First, this means that the UV cutoff in the expression~\eqref{eq:current-eom} is replaced by the vacuum value $\lambda \langle \phi \rangle_{GS}$. Second, excitations of the scalar field near the new vacuum have the mass $\mu \sim \lambda$. We review the derivation of these statements in the Appendix~\ref{sec:effective}.

\subsection{Loop corrections}
\label{sec:loops}
\label{sec:technique-2}

The tree-level calculation of the subsection~\ref{sec:current} indicates the decay of the strong scalar field $\phi = E t$. Usually this means that loop corrections significantly perturb the ground state of the system. Which means that the background field excites population of higher levels and anomalous averages \cite{Krotov, Akhmedov:Et, Akhmedov:Ex, Akhmedov:H, Akhmedov:dS, Bascone, Kamenev}. In this subsection we calculate loop corrections to the correlation functions and find that loop corrections actually do not grow with time, unlike the case of strong electric and gravitational fields.

Due to the non-stationarity of the theory in question we use the Schwinger-Keldysh diagrammatic technique discussed in Sec.~\ref{sec:technique}. Note that the definition~\eqref{eq:fermion-correators-def} should be corrected to take into account spinor indices of the fermions in two dimensions. For convenience we do the spatial Fourier transformation:
\beq G_{ab}^{\pm \pm}(x_1, x_2) = \int \frac{dp}{2\pi} G_{ab}^{\pm \pm}(t_1, t_2; p) e^{i p (x_1 - x_2)}, \eeq
which gives the following expressions for the fermionic propagators:
\beq
\label{eq:fermion-propagator}
\begin{aligned}
i G_{ab}^{+-}(t_1, t_2; p) &= \psi_{p1}^a \psi_{p2}^{c*} \left( \gamma^0 \right)_{cb} = \bem \psi_{p1}^1 \psi_{p2}^{1*} & - \psi_{p1}^1 \psi_{p2}^{2*}  \\ \psi_{p1}^2 \psi_{p2}^{1*} & - \psi_{p1}^2 \psi_{p2}^{2*} \eem, \\
i G_{ab}^{-+}(t_1, t_2; p) &= - \tilde{\psi}_{p1}^a \tilde{\psi}_{p2}^{c*} \left( \gamma^0 \right)_{cb} = \bem - \tilde{\psi}_{p1}^1 \tilde{\psi}_{p2}^{1*} & \tilde{\psi}_{p1}^{1} \tilde{\psi}_{p2}^{2*} \\ - \tilde{\psi}_{p1}^{2} \tilde{\psi}_{p2}^{1*} & \tilde{\psi}_{p1}^{2} \tilde{\psi}_{p2}^{2*} \eem = \bem - \psi_{p1}^{2*} \psi_{p2}^2 & - \psi_{p1}^{2*} \psi_{p2}^1  \\ \psi_{p1}^{1*} \psi_{p2}^2 & \psi_{p1}^{1*} \psi_{p2}^1 \eem,
\end{aligned}
\eeq
where $a$, $b$ enumerate spinor indices and we denoted for short $\psi_{p,a}^{(+)}(t_\alpha) = \psi_{p\alpha}^a$, $\psi_{-p,a}^{(-)}(t_\alpha) = \tilde{\psi}_{p\alpha}^a$. Here we also used the representation~\eqref{eq:gamma-1} for gamma-matrices, decomposition~\eqref{eq:modes-decomp} and the relation $\psi_{-p}^{(-)}(t) = -\gamma^1 \left(\psi_p^{(+)}(t)\right)^*$. Let us emphasize that we use the exact modes~\eqref{eq:mode+} and~\eqref{eq:mode-} rather than the plane waves~\eqref{eq:free-modes}.

Corresponding bosonic propagators are as follows:

\beq
\label{eq:boson-propagator}
\begin{aligned}
i D^{+-}(t_1, t_2; p) &= f_p(t_1) f_p^*(t_2) = \frac{1}{2 |p|} e^{-i |p| (t_1 - t_2)}, \\
i D^{-+}(t_1, t_2; p) &= f_p^*(t_1) f_p(t_2) = \frac{1}{2 |p|} e^{i |p| (t_1 - t_2)}, \\
\end{aligned}
\eeq
where functions $f_p(t)$ are nothing but the free modes of the scalar field:

\beq \phi(t,x) = \int \frac{dp}{2\pi} \left[ \alpha_p f_p(t) e^{i p x} + \alpha_p^\dagger f_p(t)^* e^{- i p x} \right]. \eeq
Operators $\alpha_p$ and $\alpha_p^\dagger$ satisfy the standard commutation relations: $[\alpha_p, \alpha_q^\dagger] = 2 \pi \delta(p - q)$.

Using mode decompositions for fermion and boson fields, one obtains that after the Keldysh rotation~\eqref{1s2f28} the tree-level propagators have the following form:

\beq \begin{aligned}
D^K(t_1, t_2; p) &= \frac{1}{2} \left[ f_p(t_1) f_p^*(t_2) + f_p^*(t_1) f_p(t_2) \right], \\
D^{R/A}(t_1, t_2; p) &= \pm \theta(\pm t_1 \mp t_2) \left[ f_p(t_1) f_p^*(t_2) - f_p^*(t_1) f_p(t_2) \right], \\
\tr G_{ab}^K(t_1, t_2; p) &= \frac{1}{2} \left( \psi_{p1}^1 \psi_{p2}^{1*} - \psi_{p1}^2 \psi_{p2}^{2*} + \psi_{p1}^{1*} \psi_{p2}^1 - \psi_{p1}^{2*} \psi_{p2}^2 \right), \\
\tr G_{ab}^{R/A}(t_1, t_2; p) &= \pm \theta(\pm t_1 \mp t_2) \left( \psi_{p1}^1 \psi_{p2}^{1*} - \psi_{p1}^2 \psi_{p2}^{2*} - \psi_{p1}^{1*} \psi_{p2}^1 + \psi_{p1}^{2*} \psi_{p2}^2 \right).
\end{aligned} \eeq
Apart from the other advantages (e.g. less bulky formulas), these notations allow one to study the behavior of each $p$-mode separately. Namely, the retarded and advanced propagators carry information about the spectrum of quasi--particles, while the Keldysh propagators allow to specify the state of the theory. In fact, if one does the quantum average over an arbitrary state $| \chi \rangle$ which respects spatial translational invariance, the Keldysh propagators acquire the following form:

\beq
\label{eq:ansatz}
\begin{aligned}
D^K(t_1, t_2; p) &= \left(n_p + \frac{1}{2} \right) f_p(t_1) f_p^*(t_2) + \kappa_p f_p(t_1) f_{-p}(t_2) + h.c., \\
\tr G_{ab}^K(t_1, t_2; p) &= \left( \frac{1}{2}-n'_p \right) \left( \psi_{p1}^1 \psi_{p2}^{1*} - \psi_{p1}^2 \psi_{p2}^{2*} \right) - \kappa'_p \left( \psi_{p1}^1 \psi_{p2}^{2} + \psi_{p1}^2 \psi_{p2}^{1} \right) + \left( c.c, \, p.c, \, h.c. \right),
\end{aligned} \eeq
where $h.c.$ denotes Hermitian conjugation, $p.c.$ denotes the change $p \rightarrow -p$ and $c.c.$ denotes the change $\psi_p^{(+)} \rightarrow \psi_p^{(-)}$. Also we introduced the notations as follows. First, the bosonic Keldysh propagator incorporates the level population of bosons $\langle \chi | \alpha_p^\dagger \alpha_{p'} | \chi \rangle \equiv 2\pi n_p \delta(p - p')$ and anomalous quantum average $ \langle \chi | \alpha_p \alpha_{-p'} | \chi \rangle \equiv 2\pi \kappa_p \delta(p - p')$ and its complex conjugate. Second, the trace of the fermionic Keldysh propagator contains the level population of fermions $\langle \chi | a_p^\dagger a_{p'} | \chi \rangle \equiv 2\pi n'_p \delta(p - p')$, anti-fermions $\langle \chi | b_{-p}^\dagger b_{-p'} | \chi \rangle \equiv 2\pi \tilde{n}'_p \delta(p - p')$ and anomalous quantum average $\langle \chi | a_p b_{-p'} | \chi \rangle \equiv 2\pi \kappa'_p \delta(p - p')$ and its complex conjugate. Note that the tree--level retarded and advanced propagators are proportional to the commutator $[\phi, \phi]$ or anticommutator $\{ \psi, \psi^\dagger\}$, correspondingly, which are c-numbers. I.e. the latter propagators do not depend on the choice of the state $| \chi \rangle$. 

\begin{figure}[t]
\begin{minipage}[h]{0.49\linewidth}
\center{\includegraphics[width=1\linewidth]{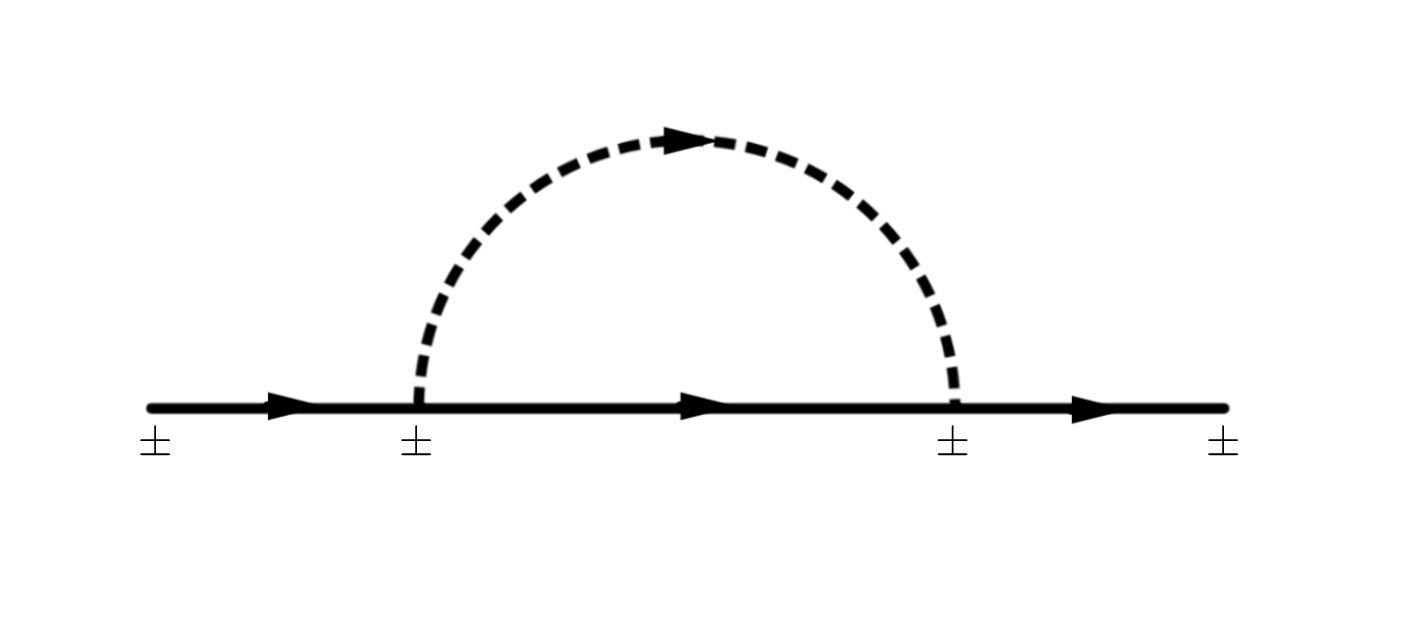} \\ a)}
\end{minipage}
\hfill
\begin{minipage}[h]{0.49\linewidth}
\center{\includegraphics[width=1\linewidth]{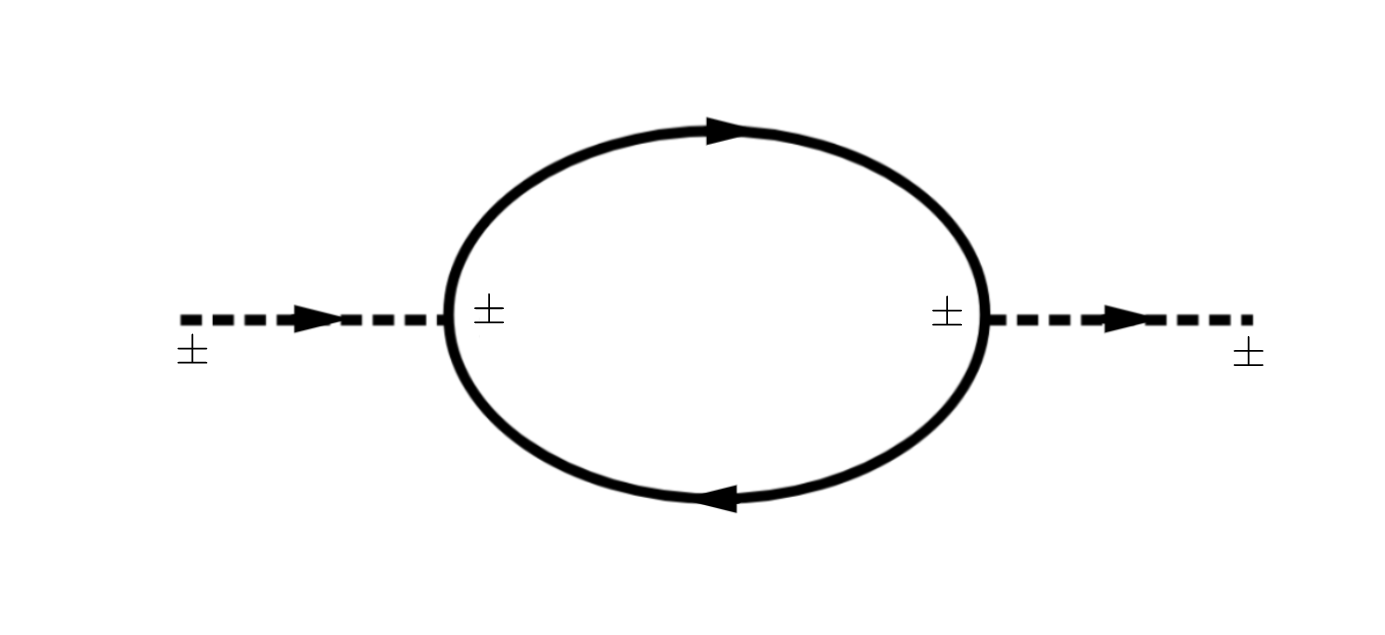} \\ b)}
\end{minipage}
\caption{One-loop corrections to the fermion (a) and boson (b) two-point functions. Solid lines correspond to the bare fermion propagators, dashed lines correspond to the bare boson propagators}
\label{fig:one-loop}
\end{figure}

Before turning on the interaction term, i.e. in the Gaussian theory, all these expectation values are exactly zero for the initial state $\hat{a}|0\rangle = \hat{\alpha} |0\rangle = 0$ that we consider. However, they can grow in time in the interacting case due to the non-stationarity of the background field. Namely, the secular growth of the level populations $n_p$, $n_p'$ or $\tilde{n}'_p$ (if present) indicates the amplification of the higher levels (than zero point fluctuations of the exact modes), whereas the growth of anomalous quantum averages (if present) means that the state of the theory at the start of the evolution is not the true vacuum state~\cite{Akhmedov:dS}. In the following sections we estimate one--loop corrections to these averages (Fig.~\ref{fig:one-loop}) and check their behavior at future infinity.  

\subsubsection{One-loop corrections to the boson propagators}
\label{sec:boson-one-loop}

In this subsubsection we calculate one-loop corrections to the boson two-point correlation functions (Fig.~\ref{fig:one-loop}). For convenience we denote $T = \frac{1}{2}(t_1 + t_2)$, $\tau = t_1 - t_2$, where $t_1$ and $t_2$ are the time arguments of the two-point functions. To simplify the expressions below we assume that the evolution of the system starts after the moment $t_0 = -T$. Note that the full evolution time is $T - t_0 = 2T$. Then we take the limit $T \rightarrow \infty$, fix $\tau \ll T$ and single out the leading contributions in this limit. Such contributions indicate the destiny of the state of the theory under consideration, because they tell about the time evolution of $n_p(T)$ and $\kappa_p(T)$ introduced in the previous subsection. For short below we use the notation 

$$\lambda E t_a = \alpha t_a = M(t_a) = M_a.$$
First, one can show that loop corrections to the retarded and advanced propagators never grow as $T\to \infty$ and $\tau = const$. In fact, due to the presence of the theta-function in these propagators one obtains the following expression for the first loop correction to the retarded propagator:

\begin{eqnarray}
\Delta D^R(t_1, t_2; p) = \nonumber \\ = -\lambda^2 \, \tr \int_{t_2}^{t_1} dt_3 \int_{t_2}^{t_3} dt_4 \int \frac{dq}{2\pi} D^R(t_1, t_3; p) G_{ab}^R\left(t_3, t_4; \frac{p+q}{2}\right) G_{ba}^K\left(t_4, t_3; \frac{p-q}{2}\right) D^R(t_4, t_2; p).
\end{eqnarray}
Due to the limits of integration over $t_3$ and $t_4$ such an expression can grow only if $\tau \rightarrow \infty$, but not when $T \rightarrow \infty$ for fixed $\tau$. The higher-order expressions posses similar behavior, because loop corrections do not change the causal properties of the retarded and advanced propagators~\cite{Akhmedov:dS, Kamenev, Berges, Rammer, Calzetta, Landau:vol10}. 

Now let us calculate the first loop correction to the Keldysh propagator:
\beq
\label{eq:boson-one-loop-pre}
\begin{aligned}
\Delta D^K(t_1, t_2; p) = \frac{1}{2} \left[ \Delta D^{++}(t_1, t_2; p) + \Delta D^{--}(t_1, t_2; p) \right] = \\ = -\frac{\lambda^2}{2} \int dt_3 dt_4 \int \frac{dq}{2\pi} \sum_{\sigma_{1,3,4} = \{+,-\}} D_{13}^{\sigma_1\sigma_3}(p) G_{34}^{\sigma_3 \sigma_4}\left(\frac{p+q}{2}\right) G_{43}^{\sigma_4 \sigma_3}\left(\frac{p-q}{2}\right) D_{42}^{\sigma_4 \sigma_1}(p) \, \text{sgn}(\sigma_3 \sigma_4),
\end{aligned}
\eeq
where we denote for short $G_{a_1 a_2}^{\pm \pm}(t_1, t_2; p) \equiv G_{12}^{\pm \pm}(p)$, $D^{\pm \pm}(t_1, t_2; p) \equiv D_{12}^{\pm \pm}(p)$ and assume the summation over the coincident spinor indices. Also we denote the one-loop corrections to the propagators $D^{++}$ and $D^{--}$ as $\Delta D^{++}$ and $\Delta D^{--}$.

Then we open the brackets in~\eqref{eq:boson-one-loop-pre} and substitute the tree--level propagators~\eqref{eq:fermion-correators-def},~\eqref{eq:boson-correators-def}. As a result, we obtain an expression of the form~\eqref{eq:ansatz}, in which leading contributions to the level population and anomalous quantum average have the following form: 

\begin{align}
n_p(T) \simeq 2 \lambda^2 \text{Re} \int_{-T}^T dt_3 \int_{-T}^{t_3} dt_4 \int \frac{dq}{2\pi} \frac{e^{i p (t_3 - t_4)}}{2p} F^*(t_3) F(t_4) = \nonumber \\ = \frac{\lambda^2}{p} \text{Re} \int_0^T dt_3 \int_0^{t_3} dt_4 \int_0^\infty \frac{dq}{\pi} \left[ e^{i p (t_3 - t_4)} F^*(t_3) F(t_4) + \sgn\left(|p|-|q|\right) e^{-i p (t_3 + t_4)} F(t_3) F(t_4) \right], \label{eq:boson-one-loop-n} \\
{\rm and} \quad \kappa_p(T) \simeq -2 \lambda^2 \int_{-T}^T dt_3 \int_{-T}^{t_3} dt_4 \int \frac{dq}{2\pi} \frac{e^{i p (t_3 + t_4)}}{2p} F(t_3) F^*(t_4) = \nonumber \\ = -\frac{2 \lambda^2}{p} \int_0^T dt_3 \int_0^{t_3} dt_4 \int_0^\infty \frac{dq}{\pi} \left[ F(t_3) F^*(t_4) \cos\left(p(t_3 + t_4)\right) +  \sgn\left(|p|-|q|\right) F(t_3) F(t_4) \sin\left(p(t_3 - t_4)\right) \right]. \label{eq:boson-one-loop-k}
\end{align}
Here we neglect the subleading (in the limit $T \rightarrow \infty$, $\tau \ll T$) contributions and introduce the function $F(t)$ to simplify the expressions:

\beq \tr\left[ G_{34}^{+-}\left(\frac{p+q}{2}\right) G_{43}^{-+}\left(\frac{p-q}{2}\right) \right] = F(t_3) F^*(t_4). \eeq
Using the expressions for the propagators~\eqref{eq:fermion-correators-def} one obtains that:
\beq
\label{eq:F}
F(t) \equiv \psi_{\frac{p+q}{2},1}^{(+)}(t)  \psi_{\frac{p-q}{2},2}^{(+)}(t) + \psi_{\frac{p+q}{2},2}^{(+)}(t)  \psi_{\frac{p-q}{2},1}^{(+)}(t),
\eeq
In both identities~\eqref{eq:boson-one-loop-n} and~\eqref{eq:boson-one-loop-k} we have divided the area of the integration over $t_3$ and $t_4$ in a specific way and then used the property~\eqref{eq:mirror} of the modes. Also we assumed that $p > 0$ and used the invariance of the function $F(t)$ under the change $q \rightarrow -q$.

It is instructive first to calculate integrals~\eqref{eq:boson-one-loop-n} and~\eqref{eq:boson-one-loop-k} in the theory without background field $\phi_{cl}$, i.e. when the fermion modes are just plane waves~\eqref{eq:free-modes}. Substituting these modes into the integrals, one finds:
\begin{align}
n_p(T) &\simeq \lambda^2 \int_{t_0}^T dt' \int_{-\infty}^{\infty} d\tau' \int_{-\infty}^{\infty} \frac{dq}{2\pi} \mathcal{N} e^{i (\omega_{\frac{p+q}{2}} + \omega_{\frac{p-q}{2}} + |p|) \tau'} = \label{eq:boson-one-loop-free-n} \\ &= \lambda^2 \int_{t_0}^T dt' \int_{-\infty}^{\infty} \frac{dq}{2\pi} \mathcal{N} \delta\left( \omega_{\frac{p+q}{2}} + \omega_{\frac{p-q}{2}} + |p| \right) \sim \mathcal{O}(\lambda^2 T^0), \nonumber \\
{\rm and} \quad \kappa_p(T=+\infty) &\simeq -2 \lambda^2 \int_{-\infty}^{\infty} dt' \int_0^{\infty} d\tau' \int_{-\infty}^{\infty} \frac{dq}{2\pi} \mathcal{N} e^{2 i |p| t' - i (\omega_{\frac{p+q}{2}} + \omega_{\frac{p-q}{2}}) \tau'} = \label{eq:boson-one-loop-free-k} \\ &= -2 \lambda^2 \int_{-\infty}^{\infty} \frac{dq}{2\pi} \mathcal{N} \delta\left( 2 |p| \right) \left( \pi \delta\left( \omega_{\frac{p+q}{2}} + \omega_{\frac{p-q}{2}} \right) - \mathcal{P} \frac{i}{\omega_{\frac{p+q}{2}} + \omega_{\frac{p-q}{2}}} \right) \sim \mathcal{O}(\lambda^2 T^0). \nonumber
\end{align}
where $\mathcal{N}$ denotes the following expression:

\beq \mathcal{N} = \frac{1}{16} \frac{(p+q) \left( \omega_{\frac{p+q}{2}} - m \right) + (p-q) \left( \omega_{\frac{p-q}{2}} - m \right)}{\omega_{\frac{p+q}{2}} \omega_{\frac{p-q}{2}} \left( \omega_{\frac{p+q}{2}} - m \right) \left( \omega_{\frac{p-q}{2}} - m \right)} , \eeq
which depends on $p$ and $q$ and does not depend on $t'=\frac{t_3+t_4}{2}$ and $\tau'=t_3-t_4$. In the second integral we used the Sokhotski--Plemelj theorem and denoted the Cauchy principal value as $\mathcal{P}$. Note that in $\kappa_p$ we put the argument $T=+\infty$ and, hence, extended the limits of integrations over times to the infinity, because we would like to show that it is not divergent as $T\to +\infty$. Thus, one obtains either finite expression\footnote{Note that the integral over $dq$ in \eqref{eq:boson-one-loop-free-k} converges.} as $T\to + \infty$ or an integration over delta-function whose argument is never zero. In other words, the one-loop correction to the free boson propagator does not grow with time $T$ due to the energy conservation which is ensured by the delta-functions. This agrees with the fact that in stationary situations correlation functions depend only on the time difference $t_1-t_2$ and do not depend on $T=(t_1+t_2)/2$.

Now let us consider the strong scalar field background, where the modes have the form~\eqref{eq:mode+} and~\eqref{eq:mode-}. Unfortunately, in this case the integrals~\eqref{eq:boson-one-loop-n} and~\eqref{eq:boson-one-loop-k} cannot be taken exactly. Hence, we will estimate them in the limit $T \rightarrow \infty$, $\tau \ll T$. Concretely, our goal here is to find if there are contributions to $n$ and $\kappa$ which survive in the limit $T \rightarrow \infty$, $\lambda \rightarrow 0$ and $\lambda^2 g(T) = \text{const}$, where $g(T)$ is some growing function of $T$ (e.g. $g(T) = T^n$ for $n\geq 1$ or $g(T) = \log T$).

Using the expansions~\eqref{eq:modes-bigp} and~\eqref{eq:modes-smallp} one can estimate the function $F(t)$:

\beq F(t) \simeq \begin{cases} \left( \frac{1 + \sgn(p - q)}{2} + \frac{1 + \sgn(q - p)}{2} \frac{2 p \alpha t}{|q^2 - p^2|} + \cdots \right) e^{- i \frac{|p+q| + |p-q|}{2} t}, \quad \text{if} \quad t < \frac{|p-q|}{2\alpha}, \\ \left(\frac{1}{\sqrt{2}} + \cdots \right) e^{-\frac{i \alpha t^2}{2} - i\left|p+q\right|t}, \quad \text{if} \quad \frac{|p-q|}{2\alpha} < t < \frac{|p+q|}{2\alpha}, \\ \left( \frac{|p+q|}{2 \alpha t} + \cdots \right) e^{-i \alpha t^2 + \frac{i(p^2 + q^2)}{2 \alpha}  \log \left(2 \alpha t^2\right)}, \quad \text{if} \quad t > \frac{|p+q|}{2\alpha}. \end{cases} \eeq
Before calculating integrals~\eqref{eq:boson-one-loop-n} and~\eqref{eq:boson-one-loop-k}, let us guess where the leading contribution may come from. First, we expect that propagators with small external momenta, $p < \alpha T$, grow faster, because corresponding low laying levels are easier to populate. Second, usually loop integrals receive leading contributions due to large virtual momenta, $q > p$ --- the main income into the lower $p$--levels comes from the higher $q$--levels. Finally, the intuition gained during the study of other background fields~\cite{Akhmedov:Et, Akhmedov:Ex, Akhmedov:H, Akhmedov:dS, Bascone, Alexeev, Astrakhantsev,Trunin} tells us that the main contribution should come from the integrands of the form $F^*(t_3) F(t_4) e^{i p (t_3 - t_4)}$, because in this case it is possible to single out the part of the integrand which does not depend on $t' = \frac{t_3 + t_4}{2}$. (Then the integral over $dt'$ may give the growing with $T$ factor.) For all other combinations of functions $F(t)$ and $e^{i p t}$ this behavior is impossible\footnote{Except the combination $F(t_3) F^*(t_4) e^{i p (t_3 - t_4)}$, which is not presented in the integrals~\eqref{eq:boson-one-loop-n} and~\eqref{eq:boson-one-loop-k}, and complex conjugated combinations.}, hence, their contributions are suppressed. Based on this argumentation, consider the following integral ($p < \alpha T$):

\begin{align}
I &= \int_0^\infty dq \int_0^T dt_3 \int_0^{t_3} dt_4 F^*(t_3) F(t_4) e^{ip(t_3 - t_4)} \simeq \\
&\simeq \int_0^p dq \Bigg[ \int_0^{\frac{p+q}{2\alpha}} dt_3 \int_0^{t_3} dt_4 e^{2 i p (t_3 - t_4)} + \int_{\frac{p+q}{2\alpha}}^T dt_3 \int_0^{\frac{p+q}{2\alpha}} dt_4 \frac{p}{2 \alpha t_3} e^{i \alpha t_3^2 - 2 i p t_4} + \nonumber \\ &\phantom{ \int_0^p dq \Bigg[}+\int_{\frac{p+q}{2\alpha}}^T dt_3 \int_{\frac{p+q}{2\alpha}}^{t_3} dt_4 \frac{p^2}{4 \alpha^2 t_3 t_4} e^{i \alpha t_3^2 - i \alpha t_4^2} \Bigg] + \\
&+ \int_p^{2 \alpha T - p} dq \Bigg[ \int_0^{\frac{q-p}{2\alpha}} dt_3 \int_0^{t_3} dt_4 \frac{4 p^2 \alpha^2 t_3 t_4}{q^4} e^{i q (t_3 - t_4)}  + \int_{\frac{q-p}{2\alpha}}^{\frac{q+p}{2\alpha}} dt_3 \int_0^{\frac{q-p}{2\alpha}} dt_4 \frac{\sqrt{2} p \alpha t_4}{q^2} e^{i q (t_3 - t_4)} +\nonumber \\ &\phantom{\int_p^{2 \alpha T - p} dq \Bigg[}+ \boxed{\frac{1}{2} \int_{\frac{q-p}{2\alpha}}^{\frac{q+p}{2\alpha}} dt_3 \int_{\frac{q-p}{2\alpha}}^{t_3} dt_4 e^{i q (t_3 - t_4)}} + \int_{\frac{q+p}{2\alpha}}^T dt_3 \int_0^{\frac{q-p}{2\alpha}} dt_4 \frac{q}{2 \alpha t_3} e^{i \alpha t_3^2 - i q t_4} + \nonumber \\ &\phantom{\int_p^{2 \alpha T - p} dq \Bigg[}+ \int_{\frac{q+p}{2\alpha}}^T dt_3 \int_{\frac{q-p}{2\alpha}}^{\frac{q+p}{2\alpha}} dt_4 \frac{q}{2 \sqrt{2} \alpha t_3} e^{i \alpha t_3^2 - i q t_4} + \boxed{\int_{\frac{q+p}{2\alpha}}^T dt_3 \int_{\frac{q+p}{2\alpha}}^{t_3} dt_4 \frac{q^2}{4 \alpha^2 t_3 t_4} e^{i \alpha t_3^2 - i \alpha t_4^2}} \Bigg] +\label{eq:I-main} \\
&+ \int_{2 \alpha T - p}^{2 \alpha T + p} dq \Bigg[ \int_0^{\frac{q-p}{2\alpha}} dt_3 \int_0^{t_3} dt_4 \frac{4 p^2 \alpha^2 t_3 t_4}{q^4} e^{i q (t_3 - t_4)} + \int_{\frac{q-p}{2\alpha}}^T dt_3 \int_0^{\frac{q-p}{2\alpha}} dt_4 \frac{2 p \alpha t_4}{q^2} e^{i q (t_3 - t_4)} + \nonumber \\ &\phantom{\int_{2 \alpha T - p}^{2 \alpha T + p} dq \Bigg[}+ \frac{1}{2} \int_{\frac{q-p}{2\alpha}}^T dt_3 \int_{\frac{q-p}{2\alpha}}^{t_3} dt_4 e^{i q (t_3 - t_4)} \Bigg] \label{eq:I-helper} \\
&+ \int_{2 \alpha T + p}^\infty dq \int_0^T dt_3 \int_0^{t_3} dt_4 \frac{4 p^2 \alpha^2 t_3 t_4}{q^4} e^{i q (t_3 - t_4)}.
\end{align}
In this expression we threw away the subleading terms, i.e. held only leading absolute values and phases of the integrands in the limit in question. However, even this rough estimate shows that there are only two terms which can grow as $T \rightarrow \infty$ (in the above formula these terms are enclosed in the boxes), whereas other contributions give constant or decaying with $T$ corrections:

\begin{align}
\label{eq:Int-1}
I_1 &\equiv \frac{1}{2} \int_p^{2 \alpha T - p} dq \int_{\frac{q-p}{2\alpha}}^{\frac{q+p}{2\alpha}} dt_3 \int_{\frac{q-p}{2\alpha}}^{t_3} dt_4 e^{i q (t_3 - t_4)} \simeq \frac{i p}{4 \alpha} \log \frac{\alpha T}{p} + \mathcal{O}\left(\frac{p}{\alpha}\right), \\
\label{eq:Int-2}
I_2 &\equiv \int_p^{2 \alpha T - p} dq \int_{\frac{q+p}{2\alpha}}^T dt_3 \int_{\frac{q+p}{2\alpha}}^{t_3} dt_4 \frac{q^2}{4 \alpha^2 t_3 t_4} e^{i \alpha t_3^2 - i \alpha t_4^2} \simeq \frac{i}{3} \alpha T + \frac{i p}{2} \log \frac{\alpha T}{p} + \mathcal{O}\left(\frac{p}{\alpha}\right).
\end{align}
Here $\mathcal{O}\left(\frac{p}{\alpha}\right)$ denotes such a function $g(T)$ that $\lambda g(T) = \text{const}$ as $\lambda \rightarrow 0$ and $T \rightarrow \infty$. Now it is obvious that such contributions cannot appear if the integrand contains $F(t_3) F(t_4)$ instead of $F^*(t_3) F(t_4)$, because in this case oscillating terms do not cancel out:

\begin{align}
I_1 &\sim \int_p^{2 \alpha T - p} dq \int dt_3 \int dt_4 e^{i q (t_3 + t_4)} \sim \int_p^{2 \alpha T - p} \frac{dq}{q^2} \sim \frac{1}{p}, \\
I_2 &\sim \int_p^{2 \alpha T - p} dq \int_{\frac{q+p}{2\alpha}}^T dt_3 \int_{\frac{q+p}{2\alpha}}^{t_3} dt_4 \frac{q^2}{4 \alpha^2 t_3 t_4} e^{i \alpha t_3^2 + i \alpha t_4^2} \sim \int_p^{2 \alpha T - p} dq \left(\frac{q^2 e^{i \alpha T^2}}{\alpha^4 T^4} - \frac{q^2 e^{\frac{i (q+p)^2}{4 \alpha}}}{(q+p)^4}\right) \sim \frac{1}{p}.
\end{align}
Also there is no any significant contribution if $p > \alpha T$. In fact, in the latter case the line~\eqref{eq:I-main} is replaced by the line~\eqref{eq:I-helper} which gives leading behavior similar to \eqref{eq:Int-1}. However, this time it is bounded from above:

\beq I \simeq \frac{1}{2} \int_p^{2 \alpha T + p} dq \int_{\frac{q-p}{2\alpha}}^T dt_3 \int_{\frac{q-p}{2\alpha}}^{t_3} dt_4 e^{i q (t_3 - t_4)} + \cdots \simeq \frac{1}{2} \left( \frac{i p}{2\alpha} + i T \right) \log\left(1 + \frac{\alpha T}{p} \right) - iT + \cdots = \mathcal{O}\left(\frac{p}{\alpha}\right). \eeq
Thus, despite the fact that this integral grows at some time intervals, it is suppressed by big external momenta and does not diverge when $T \rightarrow \infty$.

Now let us  combine all the above observations to estimate the expressions \eqref{eq:boson-one-loop-n} and~\eqref{eq:boson-one-loop-k}. Keeping in mind the integrals~\eqref{eq:Int-1} and~\eqref{eq:Int-2}, we consider small external momenta: $p < \alpha T$, neglect the integrands proportional to $F(t_3) F(t_4)$ or $F^*(t_3) F^*(t_4)$, and focus on the interval $p < q < 2 \alpha T - p$, $\frac{q-p}{2\alpha} < t_3 < \frac{q+p}{2\alpha}$, $\frac{q-p}{2\alpha} < t_4 < t_3$. However, this time we calculate the integrals more accurately, i.e. we take into account the next-to-the-leading order terms in the phases of the exponents:

\begin{align}
n_p(T) &\simeq \frac{\lambda^2}{\pi p} \text{Re} \int_p^{2 \alpha T - p} dq \int_{\frac{q-p}{2\alpha}}^{\frac{q+p}{2\alpha}} dt_3 \int_{\frac{q-p}{2\alpha}}^{t_3} dt_4 e^{i (q+p) (t_3 - t_4) + i p (t_3 - t_4) + \frac{1}{2} i \alpha (t_3^2 - t_4^2)} + \nonumber \\ &+ \frac{\lambda^2}{\pi p} \text{Re} \int_p^{2 \alpha T - p} dq \int_{\frac{q+p}{2\alpha}}^T dt_3 \int_{\frac{q+p}{2\alpha}}^{t_3} dt_4 \frac{q^2}{4 \alpha^2 t_3 t_4} e^{i \alpha t_3^2 - i \alpha t_4^2 + ip (t_3 - t_4)} + \frac{\lambda^2}{\pi p} \mathcal{O}\left(\frac{p}{\alpha}\right) \simeq \nonumber \\ &\simeq \frac{\lambda^2}{\pi p} \text{Re} \left[ \frac{i}{3} \alpha T + \frac{i p}{2} \log\frac{\alpha T}{p} + \frac{i p}{\alpha} \log\frac{\alpha T}{p} + \mathcal{O}\left(\frac{p}{\alpha}\right)\right] \sim \nonumber \\ &\sim \lambda^2 \mathcal{O}\left(\frac{p}{\alpha}\right) \rightarrow 0, \quad \text{as} \quad \lambda \rightarrow 0, \; T \rightarrow \infty, \label{eq:boson-n-final} \\
{\rm and} \quad 
\kappa_p(T) &\simeq -\frac{2 \lambda^2}{\pi p} \int_p^{2 \alpha T - p} dq \int_{\frac{q-p}{2\alpha}}^{\frac{q+p}{2\alpha}} dt_3 \int_{\frac{q-p}{2\alpha}}^{t_3} dt_4 e^{-i (q+p) (t_3 - t_4) + \frac{1}{2} i \alpha (t_3^2 - t_4^2)} \cos\left(p(t_3 + t_4)\right) - \nonumber \\ &- \frac{2 \lambda^2}{\pi p} \int_p^{2 \alpha T - p} dq \int_{\frac{q+p}{2\alpha}}^T dt_3 \int_{\frac{q+p}{2\alpha}}^{t_3} dt_4 \frac{q^2}{4 \alpha^2 t_3 t_4} e^{i \alpha t_3^2 - i \alpha t_4^2} \cos\left(p(t_3 + t_4)\right) - \frac{2 \lambda^2}{\pi p} \mathcal{O} \left(\frac{p}{\alpha}\right) \simeq \nonumber \\ &\simeq -\frac{2 \lambda^2}{\pi p} \int_p^{2 \alpha T - p} dq \left[ \frac{2 i \sin\left(\frac{p^2}{\alpha}\right)}{p} \frac{\cos\left(\frac{p q}{\alpha}\right)}{q} + \frac{\sin(2 p T)}{8 p} \frac{q^2}{\alpha^3 T^3} - \frac{\sin\left(\frac{p q}{\alpha}\right)}{p q}\right]  - \frac{2 \lambda^2}{\pi p} \mathcal{O} \left(\frac{p}{\alpha}\right) \simeq \nonumber \\ &\sim \frac{\lambda^2}{p^2} \sin\left(\frac{p^2}{\alpha}\right) \text{Ci}\left(\frac{p^2}{\alpha}\right) + \frac{\lambda^2}{p^2} \sin(2 p T) +\lambda^2 \mathcal{O}\left(\frac{p}{\alpha}\right) \rightarrow 0, \quad \text{as} \quad \lambda \rightarrow 0, \; T \rightarrow \infty, \label{eq:boson-kappa-final}
\end{align}
where $\text{Ci}(x)$ is the cosine integral. In essence, integral~\eqref{eq:boson-n-final} does not grow with $T$ because it is real and the integral~\eqref{eq:boson-kappa-final} does not grow due to the oscillating term $\cos\left(p(t_3 + t_4)\right)$. Thus, both level population and anomalous quantum average do not grow in the limit $T \rightarrow \infty$. They are generated, because the situation is not stationary, but are suppressed by the small $\lambda^2$ factor, which is not accompanied by a growing factor $T^n$, $n\geq 1$. This situation is very different from the case of strong electric and gravitational fields \cite{Krotov, Akhmedov:Et, Akhmedov:Ex, Akhmedov:H, Akhmedov:dS, Bascone}.

The technical reason for the absence of the secular growth in the background scalar field as opposed to its presence e.g. in constant electric field or de Sitter space can be explained as follows. In the constant electric field (de Sitter space) all the quantities depend on the invariant/physical momenta $p_3 - eEt$ ($\left|\vec{p}\right| \, e^{-t/H}$). (Here $p_3$ is the component of the momentum along the external electric field $E$ and $H$ is the Hubble constant in the case of the de Sitter space.) As the result all physical quantities are invariant under the
simultaneous translations $t\to t - a$ and $p_3 \to p_3 - eEa$ ($\left|\vec{p}\right| \to \left|\vec{p}\right| e^{-a/H}$). Due to such symmetries the integrands of $(t_3+t_4)/2$ combination do not depend on it. This fact brings the growing factor of $T^1$. At the same time in the background scalar field under consideration there is no such a symmetry.

Finally, note that Wightman functions $D^{+-}$ and $D^{-+}$ also do not receive growing corrections in the limit $\lambda \rightarrow 0$, $T \rightarrow \infty$ for the same reasons. As we have shown above, these correlation functions can receive growing corrections only from the integrals of the form~\eqref{eq:Int-1} and~\eqref{eq:Int-2}; however, both $D^{+-}$ and $D^{-+}$ contain only the real part of these integrals. This is consistent with our observations above, because imaginary part of such correlation functions is proportional to the retarded propagator, which does not grow in the limit in question.

\subsubsection{One-loop corrections to the fermion Keldysh propagator}
\label{sec:fermion-one-loop}

In this subsubsection we calculate one-loop corrections to the fermion two-point functions (Fig.~\ref{fig:one-loop}). We also work in the same limit for times $T$ and $\tau$ as in the previous subsubsection and set $t_0 = - T$.

For convenience here we restore the mass of the boson field:
\beq
\label{eq:action-mu}
S = \int d^2 x \left[ \frac{1}{2} \partial_{\mu} \phi \partial^{\mu} \phi - \frac{1}{2} \mu^2 \phi^2 + i \bar{\psi} \slashed{\partial} \psi - \lambda \phi \bar{\psi} \psi \right].
\eeq
On one hand, it allows us to avoid uncontrollable infrared divergences in the loop integrals due to massless 2D scalar field. On the other hand, it is a standard textbook exercise to show that the scalar field spontaneously acquires a mass $\mu \sim \lambda$ (see Appendix ~\ref{sec:effective}). We use this estimate to roughly check the self-consistency of the expressions below. 

Obviously, loop corrections to the fermion retarded and advanced propagators do not grow with time. In fact, these propagators have the same causal properties as boson retarded and advanced propagators, and hence the reasoning of the previous subsubsection also works for them. 

First loop correction to the fermionic Keldysh propagator is given by the following expression:

\beq \begin{aligned}
\Delta G_{ab}^K(t_1, t_2; p) = \frac{1}{2} \left[ \Delta G_{ab}^{++}(t_1, t_2; p) + \Delta G_{ab}^{--}(t_1, t_2; p) \right] = \\ = -\frac{\lambda^2}{2} \int dt_3 dt_4 \int \frac{dq}{2\pi} \, \sum_{\sigma_{1,2,3} = \{+,-\}} G_{13}^{\sigma_1\sigma_3}(p) G_{34}^{\sigma_3 \sigma_4}(q) D_{34}^{\sigma_3 \sigma_4}(p-q) G_{42}^{\sigma_4 \sigma_1}(p) \, \text{sgn}(\sigma_3 \sigma_4).
\end{aligned} \eeq
Then we open the brackets, substitute the expressions~\eqref{eq:fermion-correators-def} and~\eqref{eq:boson-correators-def}, take the trace over the external spinor indices and obtain the following leading contributions to the fermion level density and anomalous quantum average:

\begin{align}
n'_p(T) \simeq -2 \lambda^2 \text{Re} \int_{-T}^T dt_3 \int_{-T}^{t_3} dt_4 \int \frac{dq}{2\pi} \frac{e^{i |p-q| (t_3 - t_4)}}{2 |p-q|} \left( \psi_{p,3}^{1*} \psi_{q,3}^{2*} + \psi_{p,3}^{2*} \psi_{q,3}^{1*} \right) \left( \psi_{q,4}^1 \psi_{p,4}^2 + \psi_{q,4}^2 \psi_{p,4}^1 \right), \nonumber \\ \simeq -\lambda^2 \text{Re} \int_0^T dt_3 \int_0^{t_3} dt_4 \int \frac{dq}{2\pi} \Bigg[ \frac{e^{i |p-q| (t_3 - t_4)}}{|p-q|} H^*(t_3) H(t_4) + \sgn q  \frac{e^{i |p-q| (t_3 + t_4)}}{|p-q|} H^*(t_3) H^*(t_4)\Bigg], \label{eq:fermion-one-loop-n} \\
{\rm and} \quad \kappa'_p(T) \simeq 2 \lambda^2 \int_{-T}^T dt_3 \int_{-T}^{t_3} dt_4 \int \frac{dq}{2\pi} \frac{e^{-i |p-q| (t_3 - t_4)}}{2 |p-q|} \left( \psi_{p,3}^{1*} \psi_{q,3}^1 - \psi_{p,3}^{2*} \psi_{q,3}^2 \right) \left( \psi_{q,4}^{1*} \psi_{p,4}^{2*} + \psi_{q,4}^{2*} \psi_{p,4}^{1*} \right) \simeq \nonumber \\ \simeq \lambda^2 \int_0^T dt_3 \int_0^{t_3} dt_4 \int \frac{dq}{2\pi} \Bigg[ \frac{e^{i |p-q| (t_3 - t_4)}}{|p-q|} \left(\tilde{H}(t_3) H^*(t_4) - H(t_3) \tilde{H}^*(t_4) \right) + \nonumber \\ \phantom{\simeq -\lambda^2 \int_0^T dt_3 \int_0^{t_3} dt_4 \int \frac{dq}{2\pi} }+\sgn q  \frac{e^{i |p-q| (t_3 + t_4)}}{|p-q|} \left(\tilde{H}(t_3) H(t_4) + H(t_3) \tilde{H}(t_4) \right) \Bigg], \label{eq:fermion-one-loop-k}
\end{align}
where we introduced functions $H(t)$ and $\tilde{H}(t)$, which are defined as:
\beq \begin{aligned}
H(t) &\equiv \psi_{p,1}^{(+)}(t) \psi_{q,2}^{(+)}(t) + \psi_{p,2}^{(+)}(t) \psi_{q,1}^{(+)}(t), \\
\tilde{H}(t) &\equiv \psi_{p,1}^{(+)*}(t) \psi_{q,1}^{(+)}(t) - \psi_{p,2}^{(+)*}(t) \psi_{q,2}^{(+)}(t).
\end{aligned} \eeq
As in the previous subsubsection, we have divided the area of the integration over $t_3$ and $t_4$ in a specific way and then used the property~\eqref{eq:mirror} of the modes to obtain expressions~\eqref{eq:fermion-one-loop-n} and~\eqref{eq:fermion-one-loop-k}. Also we assumed that $p > 0$. 

For illustrative reasons let us again perform the calculation in the case when there is no any background field, $\phi_{cl} =0$. As in the boson loop calculation (Sec.~\ref{sec:boson-one-loop}), it is straightforward to show that one-loop corrections to the fermion quantum expectation values do not grow with $T$:

\begin{align}
n'_p(T) &\simeq \lambda^2 \int_{t_0}^T dt' \int_{-\infty}^\infty d\tau' \int \frac{dq}{2\pi} \mathcal{M} e^{i(\omega_p+\omega_q+|p-q|) \tau'}\simeq \nonumber \\ &\simeq \lambda^2\int_{t_0}^T dt'\int dq \mathcal{M} \delta(\omega_p+\omega_q+|p-q|) \sim \mathcal{O}(T^0),\\ 
{\rm and} \quad \kappa'_p(T) &\simeq 2\lambda^2 \int\limits_{t_0}^T dt' \int\limits_{0}^{+\infty} d\tau \int \frac{dq}{2\pi} \mathcal{N} e^{2i\omega_p t'}e^{-i(|p-q|+\omega_q)\tau'}= \nonumber \\
&=2 \lambda^2\int dq \mathcal{N} \delta(2\omega_p)\left(\pi\delta(|p-q|+\omega_q)- \mathcal{P} \frac{i}{|p-q|+\omega_q}\right) \sim \mathcal{O}(T^0).
\end{align}
Here we have made the following substitutions: $t'=\frac{t_3+t_4}{2}$, $\tau'=t_3-t_4$, and singled out the time-independent parts of the integrands:
\begin{equation}\label{s6f41}
\begin{aligned}
    \frac{1}{|p-q|} \left( \psi_{p,3}^{1*} \psi_{q,3}^{2*} + \psi_{p,3}^{2*} \psi_{q,3}^{1*} \right) \left( \psi_{q,4}^1 \psi_{p,4}^2 + \psi_{q,4}^2 \psi_{p,4}^1 \right) = \mathcal{M} e^{i(\omega_p+\omega_q)(t_3-t_4)}, \quad \text{where} \\
    \mathcal{M} \equiv \frac{1}{|p-q|} \frac{\left(p(\omega_q-m)+q(\omega_p-m)\right)^2}{4\omega_p\omega_q(\omega_p-m)(\omega_q-m)}, \\
    \frac{1}{|p-q|} \left( \psi_{p,3}^{1*} \psi_{q,3}^1 - \psi_{p,3}^{2*} \psi_{q,3}^2 \right) \left( \psi_{q,4}^{1*} \psi_{p,4}^{2*} + \psi_{q,4}^{2*} \psi_{p,4}^{1*} \right) = \mathcal{N} e^{-i\omega_q(t_3-t_4)+i\omega_p(t_3+t_4)}, \quad \text{where} \\
    \mathcal{N} \equiv \frac{1}{|p-q|} \frac{\left(pq-(\omega_p-m)(\omega_q-m)\right)\left(p(\omega_q-m)+q(\omega_p-m)\right)}{4\omega_p\omega_q(\omega_p-m)(\omega_q-m)}.
\end{aligned}
\end{equation}
As in boson calculation (Sec.~\ref{sec:boson-one-loop}), integrals do not grow due to the delta-functions which ensure the energy conservation law. As the result the two--point functions depend only on the time difference $t_1-t_2$, as it should be in stationary situations.

However, in the strong scalar field background there is no energy conservation. At the same time the integrals~\eqref{eq:fermion-one-loop-n} and~\eqref{eq:fermion-one-loop-k} again cannot be taken exactly. Hence, we estimate them in the limit $T \rightarrow \infty$, $\tau \ll T$. Using expansion~\eqref{eq:modes-bigp} and~\eqref{eq:modes-smallp} one can find the behavior of $H(t)$ and $\tilde{H}(t)$:
\beq H(t) \simeq \begin{cases} \left( \frac{\sgn q + 1}{2} + \frac{\sgn q - 1}{2} \frac{\alpha |t| \left(|q| - p\right)}{2 |q| p} \right) e^{-i \left(p + |q-p|\right) t}, \quad \text{if} \quad t < \min(p,|q|), \\  \frac{\sgn q}{\sqrt{2}} \left(2 \alpha t^2\right)^{\frac{i p^2}{4 \alpha}} e^{-\frac{i \alpha t^2}{2} - i |q| t}, \quad \text{if} \quad p < |q| \quad \text{and} \quad p < \alpha |t| < |q|, \\  \frac{1}{\sqrt{2}} \left(2 \alpha t^2\right)^{\frac{i q^2}{4 \alpha}} e^{-\frac{i \alpha t^2}{2} - i p t}, \quad \text{if} \quad |q| < p \quad \text{and} \quad |q| < \alpha |t| < p, \\  \frac{p+q}{2 \alpha |t|} \left(2 \alpha t^2\right)^{\frac{i(p^2+q^2)}{4 \alpha}} e^{-i \alpha t^2}, \quad \text{if} \quad t > \max(p,|q|), \end{cases} \label{eq:H} \eeq
\beq \tilde{H}(t) \simeq \begin{cases} \left( \frac{1 - \sgn q}{2} + \frac{\sgn q + 1}{2} \frac{\alpha |t| \left(|q| + p\right)}{2 |q| p} \right) e^{-i \left(p - |q-p|\right) t}, \quad \text{if} \quad t < \min(p,|q|), \\  \frac{1}{\sqrt{2}} \left(2 \alpha t^2\right)^{-\frac{i p^2}{4 \alpha}} e^{\frac{i \alpha t^2}{2} - i |q| t}, \quad \text{if} \quad p < |q| \quad \text{and} \quad p < \alpha |t| < |q|, \\  \frac{1}{\sqrt{2}} \left(2 \alpha t^2\right)^{\frac{i q^2}{4 \alpha}} e^{-\frac{i \alpha t^2}{2} + i p t}, \quad \text{if} \quad |q| < p \quad \text{and} \quad |q| < \alpha |t| < p, \\  \left(2 \alpha t^2\right)^{\frac{i(q^2 - p^2)}{4 \alpha}}, \quad \text{if} \quad t > \max(p,|q|). \end{cases} \eeq
Here we showed only the leading terms in the exponents and their prefactors, as in the previous subsubsection.

Note that integrals of $H(t_3) H(t_4)$ and $H(t_3) \tilde{H}(t_4)$ (and similar expressions) are suppressed in comparison with the integral over $H^*(t_3) H(t_4)$, because the former always contain oscillating factors of both $t_3 - t_4$ and $t_3 + t_4$ simultaneously. Hence, due to the same argumentation as in the previous subsubsection, if we would like to single out a growing contribution in the limit $T\to \infty$, it is sufficient to consider the following integral (we assume $p < \alpha T$):
\begin{align}
I &= \int_0^\infty dq \int_0^T dt_3 \int_0^{t_3} dt_4 \left( H^*(t_3) H(t_4) \frac{e^{i |q - p| (t_3 - t_4)}}{|q-p|} + (q \rightarrow -q) \right) = \\
&= \Bigg[ \int_0^{p-\mu} dq \Bigg( \int_0^{\frac{q}{\alpha}} dt_3 \int_0^{t_3} dt_4 +  \int_{\frac{q}{\alpha}}^{\frac{p}{\alpha}} dt_3 \int_0^{\frac{q}{\alpha}} dt_4 +  \int_{\frac{q}{\alpha}}^{\frac{p}{\alpha}} dt_3 \int_{\frac{q}{\alpha}}^{t_3} dt_4 + \\ &\phantom{\Bigg[ \int_0^{p-\mu} dq \Bigg(}+ \int_{\frac{p}{\alpha}}^T dt_3 \int_0^{\frac{q}{\alpha}} dt_4 + \int_{\frac{p}{\alpha}}^T dt_3 \int_{\frac{q}{\alpha}}^{\frac{p}{\alpha}} dt_4 + \int_{\frac{p}{\alpha}}^T dt_3 \int_{\frac{p}{\alpha}}^{t_3} dt_4 \Bigg) + \nonumber \\
&\phantom{\Bigg[}+ \int_{p-\mu}^{p+\mu} dq \Bigg( \int_0^{\frac{p}{\alpha}} dt_3 \int_0^{t_3} dt_4 + \int_{\frac{p}{\alpha}}^T dt_3 \int_0^{\frac{p}{\alpha}} dt_4 + \int_{\frac{p}{\alpha}}^T dt_3 \int_{\frac{p}{\alpha}}^{t_3} dt_4 \Bigg) + \\
&\phantom{\Bigg[}+ \int_{p+\mu}^{\alpha T} dq  \Bigg( \int_0^{\frac{p}{\alpha}} dt_3 \int_0^{t_3} dt_4 +  \boxed{\int_{\frac{p}{\alpha}}^{\frac{q}{\alpha}} dt_3 \int_0^{\frac{p}{\alpha}} dt_4} +  \boxed{\int_{\frac{p}{\alpha}}^{\frac{q}{\alpha}} dt_3 \int_{\frac{p}{\alpha}}^{t_3} dt_4} + \\ &\phantom{\Bigg[ \int_{p+\mu}^{\alpha T} dq \Bigg(}+ \int_{\frac{q}{\alpha}}^T dt_3 \int_0^{\frac{p}{\alpha}} dt_4 + \int_{\frac{q}{\alpha}}^T dt_3 \int_{\frac{p}{\alpha}}^{\frac{q}{\alpha}} dt_4 + \boxed{\int_{\frac{q}{\alpha}}^T dt_3 \int_{\frac{q}{\alpha}}^{t_3} dt_4} \Bigg) + \nonumber \\
&\phantom{\Bigg[}+ \int_{\alpha T}^\infty dq \Bigg( \int_0^{\frac{p}{\alpha}} dt_3 \int_0^{t_3} dt_4 + \int_{\frac{p}{\alpha}}^T dt_3 \int_0^{\frac{p}{\alpha}} dt_4 + \int_{\frac{p}{\alpha}}^T dt_3 \int_{\frac{p}{\alpha}}^{t_3} dt_4 \Bigg) \Bigg] \times \\ &\times \left( H^*(t_3) H(t_4) \frac{e^{i |q - p| (t_3 - t_4)}}{|q-p|} + (q \rightarrow -q) \right). \nonumber
\end{align}
Note that we have restored the mass $\mu \ne 0$, i.e. excluded the integration interval $q \in [p-\mu, p+\mu]$ to get rid of the logarithmic infrared divergencies from the virtual boson. Considering each term in the above sum and using corresponding expansions from~\eqref{eq:H} one finds that the only terms which potentially can grow with $T$ come from the integrals in boxes:

\beq I \simeq \frac{i}{2 \alpha} \log\frac{\alpha T}{p} + \mathcal{O}\left(\frac{1}{\alpha}\right) \cdot \log\frac{p}{\mu} + \mathcal{O}\left(\frac{1}{\alpha}\right). \eeq
Here $\mathcal{O}\left(\frac{1}{\alpha}\right)$ denotes such a function $g(T)$ that $\lambda g(T) = \text{const}$ as $\lambda \rightarrow 0$ and $T \rightarrow \infty$. 
Note that such integrals do not grow if $p > \alpha T$ (in this case they are bounded from above) or if the integrand contains other combinations of $H(t)$, $H^*(t)$, $\tilde{H}(t)$ and $\tilde{H}^*(t)$ (in this case time-oscillating functions reduce the growth rate at least by one power of $T$). Therefore, we get only non-growing with $T$ contributions both in level population and anomalous quantum averages for fermions:

\beq n'_p(T) \sim \kappa'_p(T) \sim \lambda^2 \cdot \log\frac{p}{\mu} \cdot \mathcal{O}\left(\frac{1}{\alpha}\right) \rightarrow 0, \quad \text{as} \quad \lambda \rightarrow 0 \quad \text{and} \quad T \rightarrow \infty. \eeq
This limit holds even if we substitute the mass $\mu \sim \lambda$ expected from the standard equilibrium analysis (Appendix~\ref{sec:effective}). Thus, for the fermions the situation is similar to the one for bosons.

\subsubsection{One-loop corrections to vertexes}
\label{sec:vertex-one-loop}

\begin{figure}[t]
\center{\includegraphics[scale=0.3]{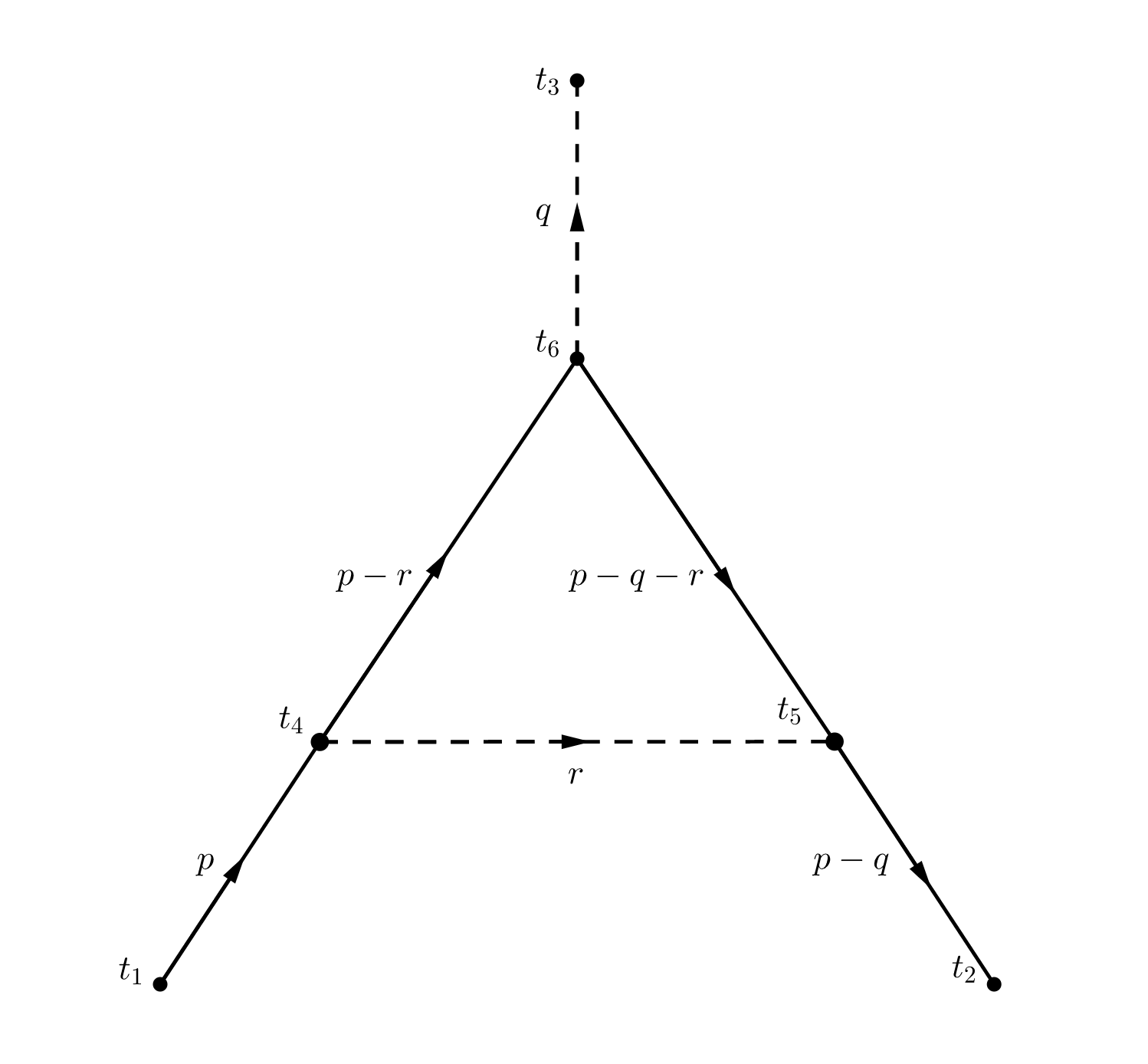}}\caption{One-loop correction to the vertex}\label{fig:vertex-one-loop}
\end{figure}

To make a thorough analysis in this subsubsection we calculate one-loop correction to the three-point correlation function $G_{ab}^{\pm\pm\pm}(x_1, x_2, x_3)$, i.e. to the vertex (Fig.~\ref{fig:vertex-one-loop}). Note that in non--stationary situations in strong background fields vertexes potentially can also show a secular growth \cite{Akhmedov:2019cfd}.

To single out the growing contributions, if any, we consider the limit $|t_i - t_j| \ll T$ and $\frac{1}{3}(t_1 + t_2 + t_3) = T \rightarrow \infty$. For convenience we work before the Keldysh rotation~\eqref{eq:fermion-correators-def},~\eqref{eq:boson-correators-def} and do spatial Fourier transformation.
We set external momenta of the three-point correlation function $|p|, |q| \rightarrow 0$ and consider the virtual momentum in the loop as follows $|r| \gg \alpha T$ (see fig. \ref{fig:vertex-one-loop}). On general physical grounds one can expect that the growing contribution, if any, comes from this region of physical parameters. A generic contribution in this limit has the following form:

\beq
\label{eq:vertex-general}
\begin{aligned} 
&\Delta G^{\pm \pm \pm} \sim \int_{t_0}^T dt_4 dt_5 dt_6 \int_{|r| > M} \frac{dr}{|r|} e^{\pm i |r| (t_4 - t_5) \pm i |r-p-q| (t_5 - t_6) \pm i |r-p| (t_6 - t_4) \pm i |q| t_6 \pm \frac{i \alpha t_4^2}{2\alpha} \pm \frac{i \alpha t_5^2}{2\alpha}} \sim \\ &\sim \int_{t_0}^T dt_4 dt_5 dt_6 \int_{M}^\infty \frac{dr}{r} e^{\pm i |r| (t_4 - t_5) \pm i |r| (t_5 - t_6) \pm i |r| (t_6 - t_4) \pm i |q| t_6 \pm \frac{i \alpha t_4^2}{2} \pm \frac{i \alpha t_5^2}{2}} \cos\left( \pm |p+q|(t_5 - t_6) \pm |p| (t_6 - t_4) \right),
\end{aligned} \eeq
where we took into account different signs of the virtual momentum $r$. Let us estimate the expression~\eqref{eq:vertex-general} for different combinations of signs. For this purpose we need the following integral which is saturated in the vicinity of zero:
\beq \begin{gathered}
\int_0^t e^{\frac{i x^2}{2} + i \rho x} dx = \frac{1+i}{2} \sqrt{\pi} + \mathcal{O}\left(\frac{1}{t}\right) + \mathcal{O}\left(\rho\right), \quad \text{if} \quad \rho \ll 1 \\
\int_0^t e^{\frac{i x^2}{2} + i \rho x} dx = \frac{i}{\rho} + \mathcal{O}\left(\frac{1}{t}\right) + \mathcal{O}\left(\frac{1}{\rho^2}\right) , \quad \text{if} \quad \rho \gg 1.
\end{gathered} \eeq
First, consider the situation when the exponent in the second line of~\eqref{eq:vertex-general} vanishes, i.e. all terms which are proportional to $|r|$ cancel each other. In this case the integral~\eqref{eq:vertex-general} reduces to the following expression:
\beq \Delta G^{\pm \pm \pm} \sim \int_{t_0}^T dt_6 e^{\pm i \left( |p+q| - |p| \pm |q| \right) t_6} \int_M^\Lambda \frac{dr}{r} \lesssim (T - t_0) \log \frac{\Lambda}{\sqrt{\alpha}}. \eeq
Naively one can think that such a term gives growing with $T$ contribution. However, such a term always appears with the following products of theta-functions: $\theta_{45} \theta_{56} \theta_{64}$ or $\theta_{46} \theta_{65} \theta_{54}$, which are identically zero. Hence, this growth does not occur in the vertex.

Second, consider the case when the exponent~\eqref{eq:vertex-general} does not contain the term $i |r| t_6$, but contain terms $\pm i |r| t_4$ and $\pm i |r| t_5$. Then:

\beq \Delta G^{\pm \pm \pm} \sim \int_{t_0}^T dt_6 e^{\pm i \left(|p+q| - |p| \pm |q| \right) t_6} \int_M^\Lambda \frac{dr}{r^3} \lesssim \frac{T - t_0}{M^2} \lesssim \mathcal{O}(T^0). \eeq
Finally, consider a situation when the time $t_6$ does not cancel out in the exponent~\eqref{eq:vertex-general}. Integrating out $t_4$ and $t_5$, one obtains the following expression:

\beq \Delta G^{\pm \pm \pm} \lesssim \int_{t_0}^T dt_6 \int_M^\Lambda \frac{dr}{r} e^{\pm i r t_6} \sim \int_{t_0}^T \frac{dt_6}{\alpha T t_6} \lesssim \mathcal{O}(T^0). \eeq
Our arguments here are generic and, hence, are applicable also to other vertex corrections and to other types of vertexes. Thus, we can conclude that one-loop corrections to the three-point correlation functions also do not grow in the limit $T \rightarrow \infty$. 

\section{Linearly growing in space background scalar field in two dimensions}
\label{sec:2D-Ex}
\setcounter{equation}{0}

In this section we consider the same theory as above~\eqref{eq:action}, but in a different background field. We use the following representation for the Clifford algebra:

\begin{equation}\label{s1f2}
\gamma^0=\begin{pmatrix}
0 & 1 \\ 1 & 0
\end{pmatrix},\;\;\gamma^1=\begin{pmatrix}
-i & 0 \\ 0 & i
\end{pmatrix},
\end{equation}
and consider the background field which linearly grows with space coordinate:
\begin{equation}\label{s1f4}
\phi_{cl}=\frac{m}{\lambda}+E x, \quad \psi_{cl}=0.
\end{equation}
Without loss of generality, we restrict our attention to the case $E > 0$, since the case $E < 0$ is achieved by the inversion $x \rightarrow -x$. Specifically, in the limit $E \rightarrow 0$ this background reproduces free fermion field with the mass $m$. However, note that in the background field this mass can be removed by the translation $x \rightarrow x - \frac{m}{\lambda E}$. In this case the situation is obviously the same as in the time--dependent background above.


\subsection{Modes}\label{s2}

To set up the notations consider again the free massive Dirac field. Unlike the case of the subsection~\ref{sec:modes}, here we have to use the following decomposition for the field:

\begin{equation}\label{s2f1}
\psi(x,t)=\int\limits_{|\omega|>m}\frac{d\omega}{2\pi}\left[a_{\omega} \psi(x,\omega)e^{-i\omega t}+b_{\omega}^\dagger \widetilde{\psi}(x,\omega)e^{i\omega t}\right],
\end{equation}
because in the background that we consider in this section there is time translational invariance rather than the spatial one.

The functions $\psi(x,\omega)e^{-i\omega t}$ and $\widetilde{\psi}(x,\omega)e^{i\omega t}$ solve the free equations of motion~\eqref{eq:free} and creation and annihilation operators $a_\omega$ and $b_\omega$ obey the standard anticommutation relations which are similar to~\eqref{eq:anti-a}.
This fixes the equal-time anticommutation relations~\eqref{eq:anti}. The frequency in this expression runs in the interval $\omega \in (-\infty,-m]\cup[m, \infty)$.

We emphasize that the definite-frequency operators $a_\omega$, $b_\omega$ and definite-momentum operators $a_p$, $b_p$ do not coincide, in particular they act differently on the vacuum state. Namely, there are four non-zero expectation values of the product of two operators:

\beq \label{eq:new-averaging} \begin{gathered}
\langle 0 | a_\omega a_{\omega'}^\dagger | 0 \rangle =  \langle 0 | b_\omega b_{\omega'}^\dagger | 0 \rangle = \theta(\omega) \times 2 \pi \delta(\omega - \omega'), \\ \langle 0 | a_\omega^\dagger a_{\omega'} | 0 \rangle = \langle 0 | b_\omega^\dagger b_{\omega'} | 0 \rangle = \theta(-\omega) \times 2 \pi \delta(\omega - \omega'),
\end{gathered} \eeq
where $|0\rangle$ is the standard vacuum state, which is annihilated by the operators $a_p$ and $b_p$. This is due to the fact that the creation operators with definite positive frequency correspond to the creation operators with definite positive momentum, whereas the creation operators with definite negative frequency correspond to the charge-conjugated annihilation operators with definite negative momentum. For the details see appendix~\ref{sec:definite}.

The form of $\psi(x,\omega)$ and $\widetilde{\psi}(x,\omega)$ spinors is as follows:
\begin{equation}\label{s2f5}
\psi(x,\omega)=\frac{1}{\sqrt{2|\omega| p}}\begin{pmatrix}
\omega \\ m-ip
\end{pmatrix}e^{ipx},\quad \widetilde{\psi}(x,\omega)=\frac{1}{\sqrt{2|\omega| p}}\begin{pmatrix}
\omega \\ -m-ip
\end{pmatrix}e^{-ipx},
\end{equation}
where $p=\sqrt{\omega^2-m^2}$ and we have used the Dirac representation for gamma-matrices~\eqref{s1f2}. 

Now, let us consider the Dirac field on the classical background $\phi_{cl} = \frac{m}{\lambda}+Ex$. In this case we have the analog of the decomposition~\eqref{s2f1}, but with the modes that solve the following equation:

\begin{equation}\label{s2f6}
(i\slashed{\partial}-m-\alpha x)\psi_\omega(x,t)=0,
\end{equation}
where we have defined for short $\alpha=\lambda E$.

Because of the time translational invariance of the equations of motion one can do the time Fourier transformation\footnote{Note that in the subsection~\ref{sec:modes} we did the spatial Fourier transformation.} and obtain the equation for the spatial coordinate dependent part of the modes:
\beq \label{s2f7}
\left[i\gamma^0(-i\omega)+i\gamma^1\partial_x-m-\alpha x\right]\psi(x,\omega)=0. \eeq
As in the time-dependent field case (Sec.~\ref{sec:modes}), one can decouple this system applying the operator $\left[-\gamma^0\omega-i\gamma^1\partial_x-m-\alpha x\right]$ to its left hand side. Then the system reduces to:
\begin{equation}\label{s2f9}
\begin{cases}
\left[\partial_x^2-(m+\alpha x)^2+\omega^2-\alpha\right]\psi_1(x,\omega)=0, \\ \left[\partial_x^2-(m+\alpha x)^2+\omega^2+\alpha\right]\psi_2(x,\omega)=0.
\end{cases}
\end{equation}
The exact solution of this equation can be represented via a sum of two linearly independent parabolic cylinder functions $D_\nu(z)$:

\begin{equation}\label{s2f10}
\begin{aligned}
\psi_1(x,\omega)&=C_1(\omega)D_{\nu-1}\left(z\right)+C_2(\omega)D_{-\nu}\left(i z\right), \\
\psi_2(x,\omega)&=B_1(\omega)D_{\nu}\left(z\right)+B_2(\omega)D_{-\nu-1}\left(i z\right),
\end{aligned}
\end{equation}
where $C_{1,2}$, $B_{1,2}$ are complex constants which we will fix below, and for convenience we define:
\begin{equation}\label{s2f24}
\nu \equiv \frac{\omega^2}{2\alpha}, \quad z(x) \equiv \sqrt{\frac{2}{\alpha}}(m+\alpha x).
\end{equation}
Note that these variables are real unlike the $\phi_{cl} = E t$ case~\eqref{eq:variables-def}.

In order to fix the integration constants $C_{1,2}$, $B_{1,2}$ one should impose additional constraints on the modes~\eqref{s2f10}. To do this, consider the limit $|\omega|\gg\sqrt{\alpha}$ for a fixed $x$. We expect that the modes in the scalar background and without it have similar behavior in such a limit, because high energy modes should not be sensitive to a smooth background field. In other words, the modes~\eqref{s2f10} must behave as plane waves (i.e. as $e^{-i\omega t\pm i|\omega| x}$) for $\omega\rightarrow\infty$. We refer to functions with asymptotic behavior $\sim e^{-i\omega t+i|\omega| x}$ as ``positive frequency modes'' and functions $\sim e^{i\omega t-i|\omega| x}$ as ``negative frequency modes''. As above we choose such modes to have the proper Hadamard behaviour of the propagators. More generic choice of the modes is also possible, as we have discussed at the end of the subsection 3.1.

Note that one obtains the ``negative frequency'' modes from the ``positive frequency'' ones by the following operation:

\begin{equation}\label{s2f12}
    \widetilde{\psi}(x,\omega) = i \gamma^1 \psi^*(x,\omega) \quad \text{or} \quad \widetilde{\psi}(x,\omega)=-\psi^*(x,-\omega).
\end{equation}
Consider the anticommutation relation~\eqref{eq:anti}:
\begin{equation}\label{s2f13}
    \left\{\psi_a(t,x),\psi_b^\dagger(t,y)\right\} = \int\limits_{|\omega|>m} \frac{d\omega}{2\pi} \left[\psi_a(x,\omega) \psi_b^\dagger(y,\omega) + \widetilde{\psi}_a(x,-\omega) \widetilde{\psi}_b^\dagger(y,-\omega) \right] = \delta(x-y) \delta_{ab},
\end{equation}
where $a,b=1,2$ enumerate spinor indices. Using asymptotic normalization method~\cite{Landau:vol3}, the connection between positive and negative frequency modes~\eqref{s2f12} and requirement $\psi(x,\omega)\sim e^{i|\omega|x}$ in the limit $\omega\rightarrow\infty$, we get the following asymptotic behavior at high frequencies:
\begin{equation}\label{s2f14}
\psi_a(x,\omega) \psi_b^\dagger(y,\omega)=\frac{1}{2}e^{i|\omega|(x-y)} \delta_{ab}.
\end{equation}
Hence, the asymptotic behavior of $\psi_1(x,\omega)$ for $\omega\rightarrow\infty$ is as follows:

\begin{equation}\label{s2f15}
\psi_1(x,\omega)=\frac{1}{\sqrt{2}} e^{i|\omega|x + i\varphi(\omega)},
\end{equation}
where $\varphi(\omega)$ is some coordinate independent phase.

Now, using the asymptotics of parabolic cylinder functions for large values of their parameter~\cite{Bateman-2,Whittaker,Olver}, we choose the coefficients $C_{1,2}(\omega)$ in~\eqref{s2f10} in order to get the exponent: $\psi_1(x,\omega)=\frac{1}{\sqrt{2}} e^{i|\omega|x}$ in the limit $|\omega|\gg\sqrt{\alpha}$, $|m+\alpha x|\ll|\omega|$ due to~\eqref{s2f15}. Thus, we obtain the first component of the positive-frequency mode $\psi_1(x,\omega)$:

\begin{equation}\label{s2f16}
\psi_1(x,\omega)=\frac{1}{2}e^{\frac{i\pi\omega^2}{4\alpha}-i|\omega|\frac{m}{\alpha}}\Bigg\{e^{\frac{i\pi\omega^2}{4\alpha}}e^{-\frac{\omega^2}{4\alpha}+\frac{\omega^2}{4\alpha}\log\frac{\omega^2}{2\alpha}}D_{-\nu}(iz)
-\frac{i|\omega|}{\sqrt{2\alpha}}e^{\frac{\omega^2}{4\alpha}-\frac{\omega^2}{4\alpha}\log\frac{\omega^2}{2\alpha}}D_{\nu-1}(z)\Bigg\}.
\end{equation}
We can get rid of the phase factor due to its arbitrariness:

\begin{equation}\label{s2f17}
\psi_1(x,\omega)=\frac{1}{2}\Bigg\{e^{\frac{i\pi\omega^2}{4\alpha}}e^{-\frac{\omega^2}{4\alpha}+\frac{\omega^2}{4\alpha}\log\frac{\omega^2}{2\alpha}}D_{-\nu}\left(iz\right)-\frac{i|\omega|}{\sqrt{2\alpha}}e^{\frac{\omega^2}{4\alpha}-\frac{\omega^2}{4\alpha}\log\frac{\omega^2}{2\alpha}}D_{\nu-1}(z)\Bigg\}.
\end{equation}
Then $\psi_2(x,\omega)$ can be found from the system of equations \eqref{s2f7}:

\beq\label{s2f18}
\psi_2(x,\omega)=\frac{1}{\omega}(m+\alpha x-\partial_x)\psi_1(x,\omega)
=\frac{i}{2}\Bigg\{e^{\frac{i\pi\omega^2}{4\alpha}}e^{-\frac{\omega^2}{4\alpha}+\frac{\omega^2}{4\alpha}\log\frac{\omega^2}{2\alpha}}\frac{|\omega|}{\sqrt{2\alpha}}D_{-\nu-1}(iz)-e^{\frac{\omega^2}{4\alpha}-\frac{\omega^2}{4\alpha}\log\frac{\omega^2}{2\alpha}}D_{\nu}(z)\Bigg\}\text{sgn}(\omega).
\eeq
Here we have used the relations~\eqref{s2f19}. The expressions for the negative frequency modes are obtained using the relation~\eqref{s2f12}.

\subsection{Tree-level scalar current}
\label{s4}


According to the operator equations of motion~\eqref{eq:opEOM}, one needs to calculate the classical current $j_{cl}(x)\equiv \langle \hat{\bar{\psi}}\hat{\psi} \rangle$ to find the response of the classical field $\phi_{cl}=\langle \hat{\phi}\rangle$. We expand the fermion field over the modes~\eqref{s2f1}, substitute the expectation values~\eqref{eq:new-averaging} and use the symmetry~\eqref{s2f12} to find the expression for this current:

\begin{equation}\label{s4f2}
\left\langle0\right|\bar{\psi}\psi\left|0\right\rangle = 2\int\limits_m^\Lambda \frac{d\omega}{2\pi}\left(\widetilde{\psi}_1\widetilde{\psi}_2^*+\widetilde{\psi}_1^*\widetilde{\psi}_2\right),
\end{equation}
where $|0\rangle$ is the state, which is annihilated by the definite-momentum annihilation operators (see Appendix C for the notations). Note that the integration is performed only over the positive frequencies.

For the same reason as in the subsection~\ref{sec:current} we expect the following dependence for the current~\eqref{eq:current-proposal} on the $\phi_{cl} = \frac{m}{\lambda} + E x$ background:

\begin{equation}\label{prop}
   \left<\bar{\psi}\psi\right>\simeq\frac{\lambda\phi_{cl}}{\pi}\log\frac{\lambda\phi_{cl}}{2\Lambda}.
\end{equation}
Note that in this case the analog of the mass parameter is 
\beq M(x) = \lambda \phi_{cl} = m + \lambda E x.\eeq

Let us check this conjecture and calculate the integral~\eqref{s4f2}. Note again that $M(x)=m+\alpha x$ indefinitely grows with $x$-coordinate, so it can overcome an arbitrarily large fixed scale $\Lambda$. However, the case of $M>\Lambda$ is not realistic, because the infinitely growing field $\phi_{cl}$ is not a physically meaningful situation, as we have already mentioned several times. Below we assume that $M^2(x) \gg \alpha$ to single out the leading contributions.

In the case $M < \Lambda$ we divide the region of integration into two segments: $[m,\Lambda]=[m,M]+[M,\Lambda]$. The asymptotics~\eqref{eq:D-bigz} for the parabolic cylinder functions is valid over the integration interval $[m,M]$:

\begin{equation}\label{s4f6}
\left\langle\bar{\psi}\psi\right\rangle_{1} \simeq  \int\limits_{m^2}^{M^2} \frac{d\omega^2}{2\pi} \frac{1}{4 M} \cosh\left[\frac{2 M^2 - \omega^2}{2\alpha} + \frac{\omega^2}{2\alpha} \log\frac{\omega^2}{4 M^2}\right] \sim \frac{\alpha}{M}\, e^{\frac{M^2}{\alpha}}.
\end{equation}
In the interval $[M,\Lambda]$ we use the following asymptotic form of the function 

$$
U(A,z) \equiv D_{-A-\frac{1}{2}}(z),
$$
which works for $A\rightarrow -\infty$, $-2\sqrt{-A}<|z|<2\sqrt{-A}$, $-\frac{\pi}{2}<\arg z<\frac{\pi}{2}$,~\cite{Bateman-2,Whittaker,Olver}:
\begin{equation}\label{s4f7}
U\left(-\frac{1}{2}\mu^2,\mu \tau\sqrt{2}\right)\simeq \frac{2g(\mu)}{(1-\tau^2)^{1/4}}\left(\cos\kappa\sum\limits_{s=0}^{\infty}(-1)^s\frac{\tilde{A}_{2s}(\tau)}{\mu^{4s}}-\sin\kappa\sum\limits_{s=0}^{\infty}(-1)^s\frac{\tilde{A}_{2s+1}(\tau)}{\mu^{4s+2}}\right),
\end{equation}
where
\beqs
\begin{aligned}
g(\mu)&\simeq h(\mu)\left(1+\frac{1}{2}\sum\limits_{s=1}^{\infty}\frac{\gamma_s}{\left(\frac{1}{2}\mu^2\right)^s}\right),\quad h(\mu)=2^{-\frac{1}{4}\mu^2-\frac{1}{4}}e^{-\frac{1}{4}\mu^2}\mu^{\frac{1}{2}\mu^2-\frac{1}{2}},\\
\kappa&=\mu^2\eta-\frac{\pi}{4},\quad\eta=\frac{1}{2}\arccos \tau-\frac{1}{2}\tau\sqrt{1-\tau^2},\quad\tilde{A}_s(\tau)=\frac{u_s(\tau)}{(1-\tau^2)^{\frac{3s}{2}}},
\end{aligned}
\eeqs
$u_s(\tau)$ are polynomials of $\tau$, $\gamma_s$ are numbers depending on $s$; all that matters is that $u_0(\tau)=1$. In our case \beq \mu^2=\frac{\omega^2}{\alpha}-1, \quad \mu \tau=\frac{m+\alpha x}{\sqrt{\alpha}}. \eeq
Taking limits $\mu^2\rightarrow +\infty$, $\tau\rightarrow 0$, we leave the first term from the asymptotic expansion~\eqref{s4f7}:
\begin{equation}\label{s4f11}
U\left(-\frac{1}{2}\mu^2,\mu\tau\sqrt{2}\right)\simeq \frac{2h(\mu)}{(1-\tau^2)^{1/4}}\cos\left(\mu^2\tau-\frac{\pi\mu^2}{4}+\frac{\pi}{4}\right)
\end{equation}
where we used that $g(\mu) \simeq h(\mu)$, $\kappa \simeq \frac{\pi}{4}\left(\mu^2 - 1\right) - \mu^2 \tau$, $\eta \simeq \frac{\pi}{4} - \tau$. Then we rotate the variable $\mu \rightarrow i\mu$ and obtain:
\begin{equation}\label{s4f12}
U\left(\frac{1}{2}\mu^2,i\mu\tau\sqrt{2}\right)\simeq 2\,\frac{e^{-\frac{i\pi}{4}\mu^2-\frac{i\pi}{4}}}{\sqrt{2}\mu h(\mu)(1-\tau^2)^{1/4}}\cos\left(\mu^2\tau-\frac{\pi\mu^2}{4}-\frac{\pi}{4}\right).
\end{equation}
Using these formulas and multiplying by an $x$-independent phase, we find the asymptotic behavior of the components of the Dirac field:
\begin{equation}\label{s4f13}
\begin{aligned}
\psi_1(x,\omega)&\simeq\sqrt{\frac{\omega}{2}}\frac{e^{i|\omega|x}}{(\omega^2-M^2(x))^{1/4}}, & \widetilde{\psi}_1(x,\omega)&\simeq\sqrt{\frac{\omega}{2}}\frac{e^{-i|\omega|x}}{(\omega^2-M^2(x))^{1/4}}, \\
\psi_2(x,\omega)&\simeq\frac{M(x)-i|\omega|}{\sqrt{2\omega}}\frac{e^{i|\omega|x}}{(\omega^2-M^2(x))^{1/4}}, & \widetilde{\psi}_2(x,\omega)&\simeq-\frac{M(x)+i|\omega|}{\sqrt{2\omega}}\frac{e^{-i|\omega|x}}{(\omega^2-M^2(x))^{1/4}}.
\end{aligned}
\end{equation}
Then the integrand for the scalar current acquires the following form:
\begin{equation}\label{s4f14}
\widetilde{\psi}_1\widetilde{\psi}_2^*+\widetilde{\psi}_1^*\widetilde{\psi}_2=-\frac{M(x)}{\sqrt{\omega^2-M^2(x)}}+\cdots,
\end{equation}
where we denoted the subleading (in the limit in question) contribution by ellipsis. 

Finally, we obtain the following expression for the scalar current:
\begin{equation}\label{s4f15}
\langle\bar{\psi}\psi\rangle\simeq\langle\bar{\psi}\psi\rangle_{2}\simeq-2\int\limits_M^\Lambda\frac{d\omega}{2\pi}\frac{M(x)}{\sqrt{\omega^2-M^2(x)}}=-\frac{1}{\pi}M\log\frac{\Lambda+\sqrt{\Lambda^2-M^2}}{M} \simeq\frac{M(x)}{\pi}\log\frac{M(x)}{2\Lambda},
\end{equation}
where we neglected the subleading contributions in the limit $\sqrt{\lambda E} \ll \lambda \phi_{cl}(x) \ll \Lambda$. We also neglected $\langle\bar{\psi}\psi\rangle_{1}$ supposing that $\Lambda$ is very big: $\frac{M^2}{\alpha}\log\frac{\Lambda}{M}\gg e^{\frac{M^2}{\alpha}}$. The obtained result (\ref{s4f15}) coincides with the proposal \eqref{prop}. 

Thus, again we obtain a peculiar behaviour of the scalar current for the large and slowly changing background field, which agrees with the result of \cite{Diatlyk} and of the previous section. We explain such a dependence of the scalar current on the background field in the Appendix B and in the Concluding section.

\subsection{Loop corrections}\label{s4.3}

We do the time Fourier transformation of the two dimensional analog of~\eqref{eq:fermion-correators-def}:
\begin{equation}
    G_{ab}^{\pm \pm}(\underline{x}_1, \underline{x}_2) = \int\limits_{|\omega|>m} \frac{d\omega}{2\pi} G_{ab}^{\pm \pm}(x_1, x_2; \omega) e^{-i \omega (t_1 - t_2)},
\end{equation}
where we denoted $\underline{x}=(t,x)$. Then:
\begin{equation}\label{s6f20}
\begin{aligned}
    G_{ab}^{+-}(x_1,x_2;\omega)&=2\text{Re}\,\psi_{\omega1}^a\psi_{\omega2}^{c*}(\gamma^0)_{cb}\,\theta(\omega) = 2 \text{Re}\,\begin{pmatrix} \psi_{\omega1}^1\psi_{\omega2}^{2*} & \psi_{\omega1}^1 \psi_{\omega2}^{1*} \\ \psi_{\omega1}^2\psi_{\omega2}^{2*} & \psi_{\omega1}^2\psi_{\omega2}^{1*} \end{pmatrix}\theta(\omega)\equiv G_{12}(\omega)\theta(\omega), \\ 
    G_{ab}^{-+}(x_1,x_2;\omega)&=-2\text{Re}\,\tilde{\psi}_{\omega1}^a\tilde{\psi}_{\omega2}^{c*}(\gamma^0)_{cb}\,\theta(-\omega) = -2\text{Re}\,\begin{pmatrix} \psi_{\omega1}^{1*}\psi_{\omega2}^{2} & \psi_{\omega1}^{1*} \psi_{\omega2}^{1} \\ \psi_{\omega1}^{2*}\psi_{\omega2}^{2} & \psi_{\omega1}^{2*}\psi_{\omega2}^{1} \end{pmatrix}\theta(-\omega)=-G_{12}(\omega)\theta(-\omega),
\end{aligned}
\end{equation}
where we use the notations $\psi_{a}(\omega,x_\alpha)=\psi_{\omega\alpha}^a$, $\widetilde{\psi}_{a}(-\omega,x_\alpha)=\tilde{\psi}_{\omega\alpha}^a$. We also use the representation for gamma matrices~\eqref{s1f2}, decomposition~\eqref{s2f1} and relation~\eqref{s2f12}. 

At the same time, using the expansion for the boson field:

\begin{equation}\label{s6f23}
    \phi(t,x)=\int\limits_{-\infty}^{+\infty}\frac{d\omega}{2\pi}\left[\alpha_\omega f_\omega(x)e^{-i\omega t}+\alpha_\omega^\dagger f_\omega^*(x)e^{i\omega t}\right],
\end{equation}
where $f_\omega(x)=\frac{1}{\sqrt{2|\omega|}}e^{i|\omega|x}$, and doing the time Fourier transformation of the two dimensional analog of~\eqref{eq:boson-correators-def}:
\begin{equation}
    D^{\pm \pm}(\underline{x}_1, \underline{x}_2) = \int\limits_{-\infty}^{+\infty} \frac{d\omega}{2\pi} D^{\pm \pm}(x_1, x_2; \omega) e^{-i \omega (t_1 - t_2)},
\end{equation}
we obtain (see appendix~\ref{sec:definite} for the details on definite-frequency operators $\alpha_\omega$ and $\alpha_\omega^\dagger$):
\begin{equation}\label{s6f24}
\begin{aligned}
    D^{+-}(x_1,x_2;\omega)&=\left[f_\omega(x_1)f^*_\omega(x_2)+h.c.\right]\theta(\omega)=\frac{\cos\{|\omega|(x_1-x_2)\}}{|\omega|}\theta(\omega)\equiv D(x_1-x_2;|\omega|)\,\theta(\omega), \\
    D^{-+}(x_1,x_2;\omega)&=\left[f_\omega(x_1)f^*_\omega(x_2)+h.c.\right]\theta(-\omega)=\frac{\cos\{|\omega|(x_1-x_2)\}}{|\omega|}\theta(-\omega)=D(x_1-x_2;|\omega|)\,\theta(-\omega).
\end{aligned}
\end{equation}
It is convenient to do the Keldysh rotation~\eqref{1s2f28} and keep in mind that if one does the quantum average over an arbitrary state $\left|\chi\right>$, the Keldysh propagators acquire the following form:

\begin{equation}\label{s6f25}
\begin{aligned}
    D^{K}(\underline{x}_1,\underline{x}_2)&=\int\frac{d\omega}{2\pi}\int\frac{d\omega'}{2\pi}\bigg[\left(n_{\omega\omega'}+\frac{1}{2}2\pi\delta(\omega-\omega')\right)f_\omega(x_1)f^*_{\omega'}(x_2)+\kappa_{\omega\omega'} f_\omega(x_1)f_{\omega'}(x_2)+h.c.\bigg]e^{-i\omega t_1+i\omega' t_2},\\
    \text{tr}\,G_{ab}^K(\underline{x}_1,\underline{x}_2)&=\int\frac{d\omega}{2\pi}\int\frac{d\omega'}{2\pi}\bigg[\left(\frac{1}{2}2\pi\delta(\omega-\omega')-n'_{\omega\omega'}\right)\left(\psi_{\omega1}^1\psi_{\omega'2}^{2*}+\psi_{\omega1}^2\psi_{\omega'2}^{1*}\right)+\\
    &+\kappa'_{\omega\omega'}\left(\psi_{\omega1}^1\psi_{\omega'2}^2+\psi_{\omega1}^2\psi_{\omega'2}^1\right)+\left( c.c, \, p.c, \, h.c. \right)\bigg]e^{-i\omega t_1+i\omega' t_2}.
\end{aligned}
\end{equation}
Here we have introduced the following notations for the quantum averages. First, the bosonic Keldysh propagator contains $\left<\chi\right|\alpha_\omega^\dagger\alpha_{\omega'}\left|\chi\right>\equiv n_{\omega\omega'}$ and $\left<\chi\right|\alpha_\omega\alpha^\dagger_{\omega'}\left|\chi\right>\equiv \tilde{n}_{\omega\omega'}$, anomalous quantum averages\\ $\left<\chi\right|\alpha_\omega\alpha_{-\omega'}\left|\chi\right>\equiv\kappa_{\omega\omega'}$ and $\left<\chi\right|\alpha^\dagger_\omega\alpha^\dagger_{-\omega'}\left|\chi\right>\equiv\tilde{\kappa}_{\omega\omega'}$. Second, the trace of the fermionic Keldysh propagator contains $\left<\chi\right|a_\omega^\dagger a_{\omega'}\left|\chi\right>\equiv n'_{\omega\omega'}$, $\left<\chi\right|b_{\omega}b^\dagger_{\omega'}\left|\chi\right>\equiv\tilde{n}'_{\omega\omega'}$, anomalous quantum averages $\left<\chi\right|a_\omega b_{-\omega'}\left|\chi\right>\equiv\kappa'_{\omega\omega'}$ and $\left<\chi\right|a^\dagger_\omega b^\dagger_{-\omega'}\left|\chi\right>\equiv\tilde{\kappa}'_{\omega\omega'}$.

\subsubsection{One-loop corrections to the boson propagators}\label{s6.3.1}

Similarly to the time-dependent background field, one can show that loop corrections to the retarded and advanced propagators do not grow, when $|t_1-t_2|\ll\frac{t_1+t_2}{2}=T\rightarrow\infty$. Let us now  calculate the one--loop correction to the Keldysh propagator:

\begin{equation}\label{s6f49}
\begin{aligned}
&\Delta D^K(\underline{x}_1, \underline{x}_2) = \frac{1}{2} \biggl[ \Delta D^{++}(\underline{x}_1, \underline{x}_2) + \Delta D^{--}(\underline{x}_1, \underline{x}_2) \biggr] = \\ &= -\frac{\lambda^2}{2} \int d^2x_3 d^2x_4 \sum_{\sigma_{1,3,4} = \{+,-\}} D^{\sigma_1\sigma_3}(\underline{x}_1,\underline{x}_3) G_{ab}^{\sigma_3 \sigma_4}(\underline{x}_3,\underline{x}_4) G_{ba}^{\sigma_4 \sigma_3}(\underline{x}_4,\underline{x}_3) D^{\sigma_4 \sigma_1}(\underline{x}_4,\underline{x}_2) \, \text{sgn}(\sigma_3 \sigma_4)\simeq \\ &\simeq \frac{\lambda^2}{2}\int dx_3 dx_4\left(\int\limits_{0}^{+\infty}\frac{d\omega}{2\pi}\int\limits_{m}^{\omega}\frac{d\omega'}{2\pi}+\int\limits_{-\infty}^{0}\frac{d\omega}{2\pi}\int\limits_{\omega}^{-m}\frac{d\omega'}{2\pi}\right) D(x_1-x_3;|\omega|)J(\omega',\omega'-\omega)D(x_4-x_2;|\omega|)e^{-i\omega(t_1-t_2)}\simeq \\ &\simeq \frac{\lambda^2}{2}\int dx_3 dx_4\left(\int\limits_{0}^{+\infty}\frac{d\omega}{2\pi}\int\limits_{m}^{\omega}\frac{d\omega'}{2\pi}+\int\limits_{-\infty}^{0}\frac{d\omega}{2\pi}\int\limits_{\omega}^{-m}\frac{d\omega'}{2\pi}\right)\frac{\cos\{|\omega|(x_1-x_3)\}\cdot\cos\{|\omega|(x_4-x_2)\}}{\omega^2}J(\omega',\omega'-\omega)e^{-i\omega(t_1-t_2)},  
\end{aligned}
\end{equation}
where we have introduced the following notation:

\begin{equation}
\begin{aligned}
     J(\omega_1,\omega_2)&\equiv G_{34}(\omega_1) G_{43}(\omega_2)=\left[\psi_{\omega_2}^{1*}(x_3)\psi_{\omega_1}^{2*}(x_3)+\psi_{\omega_2}^{2*}(x_3)\psi_{\omega_1}^{1*}(x_3)\right]\left[\psi_{\omega_2}^{1}(x_4)\psi_{\omega_1}^{2}(x_4)+\psi_{\omega_1}^{1}(x_4)\psi_{\omega_2}^{2}(x_4)\right]+\\
     &+\left[\psi_{\omega_1}^{2}(x_3)\psi_{\omega_2}^{1*}(x_3)+\psi_{\omega_2}^{2*}(x_3)\psi_{\omega_1}^{1}(x_3)\right]\left[\psi_{\omega_2}^{1}(x_4)\psi_{\omega_1}^{2*}(x_4)+\psi_{\omega_1}^{1*}(x_4)\psi_{\omega_2}^{2}(x_4)\right]+h.c.
\end{aligned}
\end{equation}
Thus, one can see that the loop corrections are finite in the limit $(T-t_0)\rightarrow\infty$ because $\Delta D^K(\underline{x}_1, \underline{x}_2)$ depends only on $(t_1-t_2)$.

\begin{figure}[t]
\center{\includegraphics[scale=0.23]{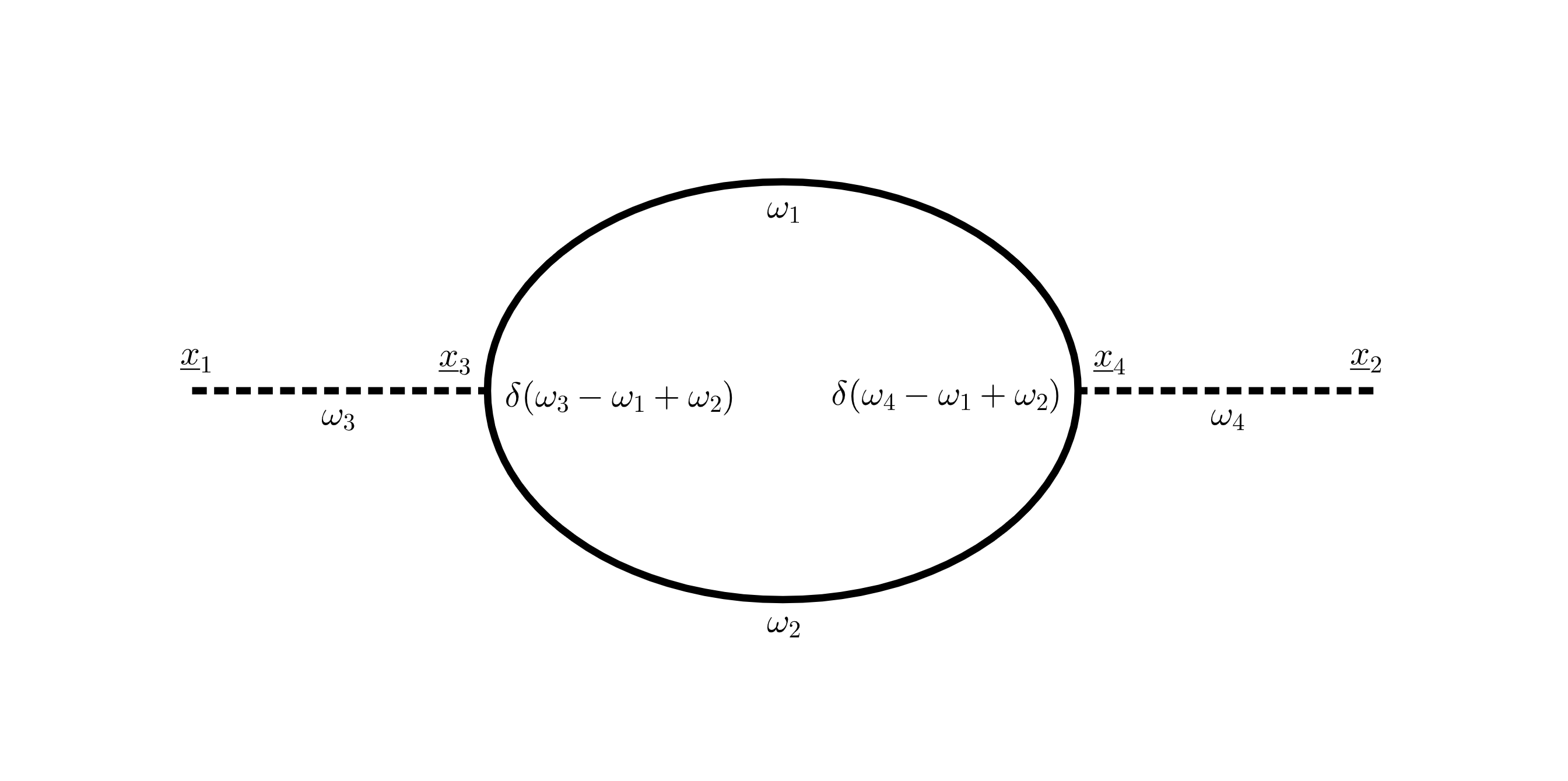}}\caption{One-loop correction to the boson two-point function with energy conservation laws}\label{fig:ECL}
\end{figure}

\subsubsection{One-loop corrections to the fermion propagators}\label{s6.3.2}

Again it can be similarly shown that the loop corrections to the retarded and advanced propagators do not grow with time. Let us then calculate the first loop correction to the Keldysh propagator:
\begin{equation}\label{s6f39}
\begin{aligned}
&\Delta G_{ab}^K(\underline{x}_1, \underline{x}_2) = \frac{1}{2} \left( \Delta G_{ab}^{++}(\underline{x}_1, \underline{x}_2) + \Delta G_{ab}^{--}(\underline{x}_1, \underline{x}_2) \right) = \\ &= -\frac{\lambda^2}{2} \int d^2x_3 d^2x_4 \, \sum_{\sigma_{1,3,4} = \{+,-\}} G_{ac}^{\sigma_1\sigma_3}(\underline{x}_1,\underline{x}_3) G_{cd}^{\sigma_3 \sigma_4}(\underline{x}_3,\underline{x}_4) D^{\sigma_3 \sigma_4}(\underline{x}_3,\underline{x}_4) G_{db}^{\sigma_4 \sigma_1}(\underline{x}_4,\underline{x}_2) \, \text{sgn}(\sigma_3 \sigma_4)\simeq \\ &\simeq \frac{\lambda^2}{2}\int dx_3 dx_4\left(\int\limits_{m}^{+\infty}\frac{d\omega}{2\pi}\int\limits_{m}^{\omega}\frac{d\omega'}{2\pi}-\int\limits_{-\infty}^{-m}\frac{d\omega}{2\pi}\int\limits_{\omega}^{-m}\frac{d\omega'}{2\pi}\right) G_{13}(\omega) G_{34}(\omega') D(x_3-x_4;|\omega-\omega'|) G_{42}(\omega) e^{-i\omega(t_1-t_2)}.
\end{aligned}
\end{equation}
Again we see that the loop corrections are finite in the limit $(T-t_0)\rightarrow\infty$ because $\Delta G^K_{ab}(\underline{x}_1, \underline{x}_2)$ depends only on $(t_1-t_2)$.

\subsubsection{One-loop corrections to vertexes}
In this subsubsection we will show that one-loop corrections to the vertexes (Fig.~\ref{fig:vertex-one-loop}) do not grow with time.

\begin{figure}[t]
\center{\includegraphics[scale=0.47]{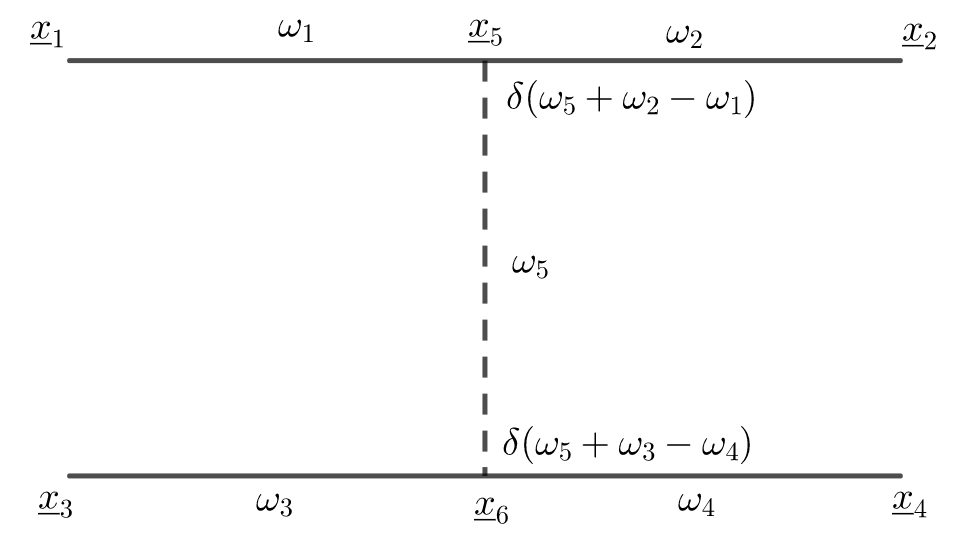}}\caption{Correction to the four-point correlation function}\label{fig:ladder}
\end{figure} 

Let us consider the four-point correlation function $G^{\sigma_1 \sigma_2 \sigma_3 \sigma_4}$, (Fig.~\ref{fig:ladder}). Note that it is a part of the three-point correlation function $G^{\pm\pm\pm}_{ab}$. To single out the growing contributions, if any, we consider the limit $|t_i-t_j|\ll T$ and $\frac{1}{4}(t_1+t_2+t_3+t_4)=T\rightarrow\infty$. A generic contribution in this limit has the following form:
\begin{equation}
\begin{aligned}
    \Delta G^{\sigma_1\sigma_2\sigma_3\sigma_4}&\sim \lambda^2\int d^2x_5 d^2x_6\,D^{+-}_{56}\sum\limits_{\sigma_{5,6}=\{+,-\}}G^{\sigma_1\sigma_5}_{15}G^{\sigma_5\sigma_2}_{52}G^{\sigma_3\sigma_6}_{36}G^{\sigma_6\sigma_4}_{64}\,\text{sgn}(\sigma_5\sigma_6)=\\&=\lambda^2\int dx_5 dx_6\int dt_5 dt_6\int\limits_{m}^{+\infty}\frac{d\omega_1}{2\pi}\int\limits_{m}^{+\infty}\frac{d\omega_2}{2\pi}\int\limits_{m}^{+\infty}\frac{d\omega_3}{2\pi}\int\limits_{m}^{+\infty}\frac{d\omega_4}{2\pi}\int\limits_{0}^{+\infty}\frac{d\omega_5}{2\pi}\,e^{-it_5(\omega_5+\omega_2-\omega_1)}e^{it_6(\omega_5+\omega_3-\omega_4)}\times\\&\times e^{-i\omega_1 t_1}e^{i\omega_2 t_2}e^{-i\omega_3 t_3}e^{i\omega_4 t_4} D^{+-}_{56}(\omega_5)\sum\limits_{\sigma_{5,6}=\{+,-\}}G^{\sigma_1\sigma_5}_{15}(\omega_1)G^{\sigma_5\sigma_2}_{52}(\omega_2)G^{\sigma_3\sigma_6}_{36}(\omega_3)G^{\sigma_6\sigma_4}_{64}(\omega_4)\,\text{sgn}(\sigma_5\sigma_6)=\\&=\lambda^2\int dx_5 dx_6\int\limits_{m}^{+\infty}\frac{d\omega'}{2\pi}\int\limits_{m}^{+\infty}\frac{d\omega''}{2\pi}\int\limits_{0}^{+\infty}\frac{d\omega}{2\pi}e^{i\omega(t_3-t_2)}e^{i\omega'(t_2-t_1)}e^{i\omega''(t_4-t_3)}D^{+-}_{56}(\omega)\times\\&\times\sum\limits_{\sigma_{5,6}=\{+,-\}}G^{\sigma_1\sigma_5}_{15}(\omega')G^{\sigma_5\sigma_2}_{52}(\omega'-\omega)G^{\sigma_3\sigma_6}_{36}(\omega''-\omega)G^{\sigma_6\sigma_4}_{64}(\omega'')\,\text{sgn}(\sigma_5\sigma_6).
\end{aligned}
\end{equation}
Thus, we see that the last integral depends only on $(t_i-t_j)$ and does not grow in the limit in question.

\section{Coherent state}\label{s5}
\label{sec:coherent}

In the previous section we have considered $\phi_{cl}(x)=Ex+\frac{m}{\lambda}$ and found the exact modes $\psi(t,x)$ in such a background. Such an approach means that the background field $\phi_{cl}(x)$ is set by a brutal external force to be the same for all times. Such an approach can work only when the backreaction on the background is weak.

The situation, which we consider in this section, corresponds to a different set up. Namely, at some point in time there was formed a state $|\phi_{cl}\rangle$ which corresponds to the presence of the external field $\phi_{cl}(x)$ in the sense that we will see in a moment. And then this state is released to evolve freely. Our goal is to find out how it will be changing in time. 

To start with, we define the coherent state $\left|\phi_{cl}\right>$ as follows:
\begin{equation}\label{s5f1}
    \left<\phi_{cl}\right|\hat{\phi}(y)\left|\phi_{cl}\right>=\phi_{cl}(y).
\end{equation}
In appendix~\ref{pa} it is shown that one can represent the state as follows:

\begin{equation}\label{s5f2}
\left|\phi_{cl}\right>=e^{-i\int\phi_{cl}\hat{\pi}_\phi dx}\left|0\right>, \quad \text{where} \quad a_p\left|0\right>=0.
\end{equation}
In what follows we want to calculate the following expectation value:

\begin{equation}\label{s5f3}
\langle\phi\rangle(t,x)=\left\langle\phi_{cl}\right|\hat{U}^\dagger(t,t_0)\hat{\phi}_I(t,x)\hat{U}(t,t_0)\left|\phi_{cl}\right\rangle,
\end{equation}
where $\hat{U}(t,t_0)$ is the evolution operator and we use the interaction picture and the modes are ordinary plane waves:

\begin{equation}\label{s5f5}
\begin{aligned}\phi_I(t,x)&=\int\frac{dp}{2\pi}\frac{1}{\sqrt{2|p|}}\left(\alpha_p e^{ipx-i|p|t}+\alpha_p^\dagger e^{-ipx+i|p|t}\right),\\
\psi_I(t,x)&=\int\frac{dp}{2\pi}\frac{1}{\sqrt{2E_p}}\left(a_p u_pe^{ipx-iE_pt}+b_p^\dagger v_pe^{-ipx+iE_pt}\right),
\end{aligned}
\end{equation}
unlike the case of the previous section.
Here $u_p$ and $v_p$ are modes of the two-dimensional free massive Dirac field~\eqref{eq:free-modes}.

One can find the equation for $\langle\phi\rangle(t,x)$:

\begin{equation}\label{s5f7}
\Box \langle\phi\rangle(t,x)=-\lambda\left<\phi_{cl}\right|\hat{U}^\dagger(t,t_0)\hat{\bar{\psi}}_I(t,x)\hat{\psi}_I(t,x)\hat{U}(t,t_0)\left|\phi_{cl}\right>.
\end{equation}
Now let us transform the right hand side of this equation. We commute $\hat{U}$ with the exponent in the definition of the coherent state (\ref{s5f2}). Let us denote:

\begin{equation}\label{s5f8}
    \hat{X}=-i\int d^2x \lambda\hat{\phi}_I\hat{\bar{\psi}}_I\hat{\psi}_I,\quad {\rm and} \quad \hat{Y}=-i\int dy\phi_{cl}\hat{\pi}_\phi,
\end{equation}
and use that
\begin{equation}\label{s5f9}
    \phi_I(x,t)=e^{iH_0 t}\phi_I(x)e^{-iH_0 t}.
\end{equation}
Then 
\begin{multline}\label{s5f10}
    [X,Y]=-\int d^2x dy \lambda\phi_{cl}(y)\bar{\psi}_I(x,t)\psi_I(x,t)[\phi_I(x,t),\pi_\phi(y)]= \\ =-\int d^2x dy \lambda\phi_{cl}(y)\bar{\psi}_I(x,t)\psi_I(x,t)\left(e^{iH_0t}\phi_I(x)[e^{-iH_0t},\pi_\phi(y)]+i\delta(x-y)+[e^{iH_0t},\pi_\phi(y)]\phi_I(x)e^{-iH_0t}\right).
\end{multline}
To simplify the last expression we denote

\begin{equation}\label{s6f32}
    A=-iH_0t=-it\int dx\left(\frac{1}{2}\pi_\phi^2+\frac{1}{2}\phi_I'^2\right),\quad {\rm and} \quad B=-i\int\phi_{\text{cl}}\pi_\phi dx,
\end{equation}
and check that the commutator of $A$ and $B$ is vanishing in our case:
\begin{multline}\label{s6f33}
    [A,B]=-\frac{t}{2}\int dx\int dy[\phi_I'^2(y),\pi_\phi(x)]\phi_{cl}(x)= \\ =-t\int dx\int dy\, \phi_I'(y)\phi_{cl}(x)\partial_y[\phi_I(y),\pi_\phi(x)]=it\int dx\,\phi_I''(x)\phi_{cl}(x)=0,
\end{multline}
because the background field that we consider here is the linear function of $x$: $\phi_{cl}=\frac{m}{\lambda}+Ex$ and $\phi_I(y)$ does vanish at infinity.

Hence, due to \eqref{s6f33} the first and the third terms in~\eqref{s5f10} are vanishing. Therefore

\begin{equation}\label{s5f11}
    [X,Y]=-i\int d^2x\lambda\phi_{cl}\bar{\psi}_I\psi_I,
\end{equation}
and then

\begin{equation}\label{s5f12}
    \hat{U}e^{-i\int\phi_{cl}\hat{\pi}_\phi dx}=e^{-i\int d^2x\lambda\phi_{cl}\bar{\psi}_I\psi_I}e^{-i\int\phi_{cl}\hat{\pi}_\phi dx}\hat{U}.
\end{equation}
Thus, we obtain that:

\begin{equation}\label{s5f13}
\begin{aligned}
    &\left<\phi_{cl}\right|\hat{U}^\dagger(t,t_0)\hat{\bar{\psi}}_I(t,x)\hat{\psi}_I(t,x)\hat{U}(t,t_0)\left|\phi_{cl}\right>=\\
    &=\left<0\right|\hat{U}^\dagger(t,t_0)e^{i\int d^2x\lambda\phi_{cl}\bar{\psi}_I\psi_I}\hat{\bar{\psi}}_I(t,x)\hat{\psi}_I(t,x)e^{-i\int d^2x\lambda\phi_{cl}\bar{\psi}_I\psi_I}\hat{U}(t,t_0)\left|0\right>.
\end{aligned}
\end{equation}
We will use these relations below.

Meanwhile to find the relation between the problem of the previous section to the one here, note that:

$$
e^{i\int\phi_{cl}\hat{\pi}_\phi dx} {\cal O}(\hat{\phi}) e^{-i\int\phi_{cl}\hat{\pi}_\phi dx} = {\cal O}(\phi_{cl}+\hat{\phi}),
$$
for any operator ${\cal O}(\hat{\phi})$ in the theory.
Using this relation on the right hand side of the eq. (\ref{s5f7}) one can assume that we obtain here the same scalar current as in the previous section. However, note that in the previous section the average in the correlation function was done with respect to the ground Fock space state corresponding to the exact fermionic modes in the $\phi_{cl}(x)$ background, while in (\ref{s5f7}) the expectation value is taken with respect to the ordinary Poincare invariant state for fermions.

\subsection{Loop corrections (coherent state) }\label{s6}

In this subsection we calculate the right hand side of the equation~\eqref{s5f7}. The tree--level result for the scalar current in the present case is obviously trivial (the same as in empty space). To restore the tree--level result of the previous section within the present settings one has to sum up infinite number of terms, as is explained in the footnote~\ref{footnote:loops} in appendix~\ref{sec:effective}. 

In what follows we consider the corrections of the order $\lambda^2$ to the propagators (Fig.~\ref{fig:one-loop}) in the limit

\begin{equation}\label{s6f27}
    \tau=t_1-t_2=\text{const},\;\; T=\frac{1}{2}(t_1+t_2)\rightarrow+\infty,\;\; t_0\rightarrow -\infty,
\end{equation}
where $t_1$ and $t_2$ are arguments of the two-point functions, or, more specifically:

\begin{equation}\label{s6f28}
    |T|,|t_0|\gg \frac{1}{\sqrt{\alpha}},\;\;|T|\gg|\tau|.
\end{equation}
Note that the variant with the averaging over the coherent state allows one not to specify the form of $\phi_{cl}$, therefore it allows to get more general result. It is also interesting to compare the answers in these two problems (sections~\ref{s4.3} and~\ref{s6}) and to find out if the expressions for the first loop corrections show a different behaviour.

To do the calculation in question, we need to find

\begin{equation}\label{s6f29}
    \phi(x,t)\left|\phi_{cl}\right>=e^{iH_0t}\phi(x)e^{-iH_0t}e^{-i\int\phi_{\text{cl}}\hat{\pi}_\phi dx}\left|0\right>,
\end{equation}
where

\begin{equation}\label{s6f30}
    H_0=\int dx\left(\frac{1}{2}\pi_\phi^2+\frac{1}{2}\phi'^2\right).
\end{equation}
Taking into account the Becker-Hausdorff formula and the result of~\eqref{s6f33}, we obtain that:

\begin{equation}\label{s6f34}
\begin{aligned}
    \phi(x,t)\left|\phi_\text{cl}\right>&=e^{iH_0t}\phi(x)e^{-i\int\phi_{\text{cl}}\hat{\pi}_\phi dx}e^{-iH_0t}\left|0\right>=e^{-iE_0t}e^{iH_0t}\phi(x)\left|\phi_\text{cl}\right>= \\ 
    &=e^{-iE_0t}\phi_{cl}(x)e^{iH_0t}e^{-i\int\phi_{\text{cl}}\hat{\pi}_\phi dx}\left|0\right>+e^{-iE_0t}e^{iH_0t}e^{-i\int\phi_{\text{cl}}\hat{\pi}_\phi dx}\phi(x)\left|0\right>= \\
     &=\phi_{cl}(x)\left|\phi_\text{cl}\right>+e^{-iE_0t}e^{-i\int\phi_{\text{cl}}\hat{\pi}_\phi dx}e^{iH_0t}\phi(x)\left|0\right>=\phi_{cl}(x)\left|\phi_\text{cl}\right>+e^{-i\int\phi_{\text{cl}}\hat{\pi}_\phi dx}\phi(x,t)\left|0\right>.
\end{aligned}
\end{equation}
Thus, if we consider quantum averages over the state $\left|\phi_{cl}\right>$, then according to~\eqref{s6f34} instead of~\eqref{eq:boson-propagator} we have that:

\begin{equation}\label{s6f35}
\begin{aligned}
     D^{+-}(\underline{x}_1,\underline{x}_2)&=\phi_{cl}(x_1)\phi_{cl}(x_2)+\left<0\right|\phi(\underline{x}_1)\phi(\underline{x}_2)\left|0\right>=\phi_{cl}(x_1)\phi_{cl}(x_2)+D(\underline{x}_1-\underline{x}_2), \\
     D^{-+}(\underline{x}_1,\underline{x}_2)&=\phi_{cl}(x_1)\phi_{cl}(x_2)+D(\underline{x}_2-\underline{x}_1), \\
     D^{--}(\underline{x}_1,\underline{x}_2)&=\phi_{cl}(x_1)\phi_{cl}(x_2)+\theta(t_1-t_2)D(\underline{x}_1-\underline{x}_2)+\theta(t_2-t_1)D(\underline{x}_2-\underline{x}_1), \\
     D^{++}(\underline{x}_1,\underline{x}_2)&=\phi_{cl}(x_1)\phi_{cl}(x_2)+\theta(t_1-t_2)D(\underline{x}_2-\underline{x}_1)+\theta(t_2-t_1)D(\underline{x}_1-\underline{x}_2),
\end{aligned}
\end{equation}
where we denoted $\underline{x}=(t,x)$ and
\begin{equation}\label{s6f36}
   D(\underline{x}_1-\underline{x}_2)=\int\frac{dp}{2\pi}\frac{1}{2|p|}e^{-i|p|(t_1-t_2)+ip(x_1-x_2)}
\end{equation}
is just the empty space scalar propagator. The fermion propagators in the situation under consideration are the same as in the theory without background field.

\subsection{One-loop corrections to the fermion propagators}

We continue with the loop corrections to the boson and fermion correlation functions. For the retarded and advanced propagators we have the usual story as was described in the previous sections. Hence, below we concentrate on the calculations of the loop corrections to the Keldysh propagators. 

We start with the calculation of the first loop correction to the fermion Keldysh propagator~\eqref{s6f39}. Due to the fact that in the one-loop correction to the fermion propagator we have only one tree-level bosonic Green function and in the free case fermion Green functions does not receive growing with time corrections (see subsubsection~\ref{sec:fermion-one-loop}), we can use the tree-level bosonic propagator in the following form:
\begin{equation}\label{s6f43}
    D^{+-}(\underline{x}_1,\underline{x}_2)=D^{-+}(\underline{x}_1,\underline{x}_2)=D^{++}(\underline{x}_1,\underline{x}_2)=D^{--}(\underline{x}_1,\underline{x}_2)=\phi_{cl}(x_1)\phi_{cl}(x_2)\equiv D(x_1,x_2),
\end{equation}
instead of~\eqref{s6f35}, and define the following expressions:
\begin{equation}
\begin{aligned}
    H(p,q,p')&\equiv\left( \psi_{p,3}^{1*} \psi_{q,3}^{2*} + \psi_{p,3}^{2*} \psi_{q,3}^{1*} \right)\left( \psi_{q,4}^1 \psi_{p',4}^2 + \psi_{q,4}^2 \psi_{p',4}^1 \right)\left( \psi_{p,1}^1 \psi_{p',2}^{1*} - \psi_{p,1}^2 \psi_{p',2}^{2*} \right),\\
    K(p,q,p')&\equiv\left( \psi_{p,1}^1 \psi_{p',2}^2 + \psi_{p,1}^2 \psi_{p',2}^1 \right)\left( \psi_{p',3}^{1*} \psi_{q,3}^1 - \psi_{p',3}^{2*} \psi_{q,3}^2 \right) \left( \psi_{q,4}^{1*} \psi_{p,4}^{2*} + \psi_{q,4}^{2*} \psi_{p,4}^{1*} \right).
\end{aligned}
\end{equation}
Then, taking into account the Fourier representation of $\phi_{cl}(x)$:

\begin{equation}\label{s6f46}
    \tilde{\phi}_{cl}(p)=\int dx \, \phi_{cl}(x) \, e^{-ipx} = 2\pi \, \left(\alpha\delta(p)+i\beta\delta'(p)\right),
\end{equation}
we obtain that the one loop correction to the fermion Keldysh propagator is as follows:

\begin{equation}\label{s6f44}
\begin{aligned}
    \Delta G_{ab}^K(\underline{x}_1, \underline{x}_2)  =&-\frac{\lambda^2}{4} \int d^2x_3 d^2x_4 D(x_3,x_4)\sum_{\sigma_{1,3,4} = \{+,-\}} G_{ac}^{\sigma_1\sigma_3}(\underline{x}_1,\underline{x}_3) G_{cd}^{\sigma_3 \sigma_4}(\underline{x}_3,\underline{x}_4) G_{db}^{\sigma_4 \sigma_1}(\underline{x}_4,\underline{x}_2) \text{sgn}(\sigma_3 \sigma_4)\simeq \\
    \simeq &-\frac{\lambda^2}{2}\int_{t_0}^T dt_3 \int_{t_0}^T dt_4 \int\frac{dp}{2\pi}\frac{dq}{2\pi}\frac{dp'}{2\pi}\tilde{\phi}_{cl}(p-q)\tilde{\phi}_{cl}(q-p')e^{ipx_1-ip'x_2}\big[H(p,q,p')+h.c.\big]+\\
    &+\frac{\lambda^2}{2}\int_{t_0}^T dt_3 \int_{t_0}^{t_3} dt_4 \int\frac{dp}{2\pi}\frac{dq}{2\pi}\frac{dp'}{2\pi}\tilde{\phi}_{cl}(p-q)\tilde{\phi}_{cl}(q-p')e^{ipx_1-ip'x_2}\big[K(p,q,p')+K(p',q,p)+h.c\big]=\\
    =&-\frac{\lambda^2}{2}\int\frac{dp}{2\pi}e^{ip(x_1-x_2)}\int_{t_0}^T dt_3\int_{t_0}^T dt_4\bigg[H(p,p,p)\bigg\{\alpha^2+\alpha\beta (x_1+x_2) -\alpha\beta\frac{p}{\omega_p} (t_1+t_2-t_3-t_4)+\\&+\beta^2\bigg(i\frac{p}{\omega_p}(t_2-t_4)-ix_2+\frac{2m^2-p^2}{4\omega_p^2p}\bigg)\bigg(i\frac{p}{\omega_p}(t_3-t_1)+ix_1+\frac{2m^2-p^2}{4\omega_p^2p}\bigg) \bigg\}+h.c.\bigg]+ \\
    +&\lambda^2\int\frac{dp}{2\pi}e^{ip(x_1-x_2)}\int_{t_0}^T dt_3 \int_{t_0}^{t_3} dt_4\bigg[K(p,p,p)\bigg\{\alpha^2+\alpha\beta(x_1+x_2)-\alpha\beta\frac{p}{\omega_p}(t_1-t_2)+\\&+\beta^2\bigg(i\frac{p}{\omega_p}\frac{t_3+t_4-2t_2}{2}-ix_2+\frac{3m^2-p^2}{4\omega_p^2p}\bigg)\bigg(i\frac{p}{\omega_p}\frac{t_3+t_4-2t_1}{2}+ix_1+\frac{3m^2-p^2}{4\omega_p^2p}\bigg)\bigg\}+h.c.\bigg]=\\
    &=\mathcal{O}(T^0),
\end{aligned}
\end{equation}
where we have used that

\begin{equation}\label{s6f45}
\begin{aligned}
    \left( \psi_{p,3}^{1*} \psi_{p,3}^{2*} + \psi_{p,3}^{2*} \psi_{p,3}^{1*} \right) \left( \psi_{p,4}^1 \psi_{p,4}^2 + \psi_{p,4}^2 \psi_{p,4}^1 \right)&=\frac{p^2}{\omega_p^2}e^{2i\omega_p(t_3-t_4)}, \\ 
    \left( \psi_{p,3}^{1*} \psi_{p,3}^1 - \psi_{p,3}^{2*} \psi_{p,3}^2 \right) \left( \psi_{p,4}^{1*} \psi_{p,4}^{2*} + \psi_{p,4}^{2*} \psi_{p,4}^{1*} \right)&=\frac{pm}{\omega_p^2}e^{-i\omega_p(t_3-t_4)}e^{i\omega_p(t_3+t_4)}.
\end{aligned}
\end{equation}
and the following expressions for the derivatives of the fermion field components:

\begin{equation}\label{s6f47}
    \partial_p\psi_{p,\alpha}^1=-\frac{m(\omega_p-m)}{2\omega_p^2 p}\psi_{p,\alpha}^1-it_\alpha\frac{p}{\omega_p}\psi_{p,\alpha}^1,\;\;\partial_p\psi_{p,\alpha}^2=\frac{mp}{2\omega_p^2 (\omega_p-m)}\psi_{p,\alpha}^2-it_\alpha\frac{p}{\omega_p}\psi_{p,\alpha}^2.
\end{equation}
Thus, as follows follows from (\ref{s6f44}) one-loop correction to the fermion propagator does not grow as $T\rightarrow\infty$.

\subsection{One-loop corrections to the boson propagators}\label{s6.2.2}

The one-loop correction to the free bosonic Keldysh propagator does not grow with time (see subsubsection~\ref{sec:boson-one-loop}).
To show that in the present case let us denote:

\begin{equation}
    F(p,q)\equiv\left( \psi_{q,3}^{1*} \psi_{p,3}^{2*} + \psi_{p,3}^{1*} \psi_{q,3}^{2*} \right)\left( \psi_{q,4}^1 \psi_{p,4}^2 + \psi_{p,4}^1 \psi_{q,4}^2 \right) = \frac{p^2}{\omega_p^2}e^{2i\omega_p(t_3-t_4)}.
\end{equation}
Therefore, keeping in mind eq.~\eqref{s6f46}, we obtain:

\begin{equation}\label{s6f50}
\begin{aligned}
   \Delta D^K(\underline{x}_1, \underline{x}_2)&\simeq -\lambda^2\int_{t_0}^T dt_3 \int_{t_0}^T dt_4 \int\frac{dp}{2\pi}\frac{dq}{2\pi}\bigg[\bigg(\phi_{cl}(x_1)\tilde{\phi}_{cl}(q-p)\phi_{cl}(x_2)\tilde{\phi}_{cl}(p-q)+\phi_{cl}(x_1)\tilde{\phi}_{cl}(q-p)\times\\
   &\times\frac{e^{-i|p-q|(t_4-t_2)}}{2|p-q|}e^{-i(p-q)x_2}+\phi_{cl}(x_2)\tilde{\phi}_{cl}(p-q)\frac{e^{-i|p-q|(t_1-t_3)}}{2|p-q|}e^{i(p-q)x_1}\bigg)F(p,q)+h.c.\bigg]+\\
   &+\lambda^2\int_{t_0}^T dt_3 \int_{t_0}^{t_3} dt_4 \int\frac{dp}{2\pi}\frac{dq}{2\pi} \bigg[\bigg(2\phi_{cl}(x_1)\tilde{\phi}_{cl}(q-p)\phi_{cl}(x_2)\tilde{\phi}_{cl}(p-q)+\phi_{cl}(x_1)\tilde{\phi}_{cl}(q-p)\times\\
    &\times \frac{e^{-i|p-q|(t_2-t_4)}}{2\sqrt{|p-q|^2 + \mu^2}} e^{-i(p-q)x_2} + \phi_{cl}(x_2) \tilde{\phi}_{cl}(p-q) \frac{e^{-i|p-q|(t_1-t_3)}}{2\sqrt{|p-q|^2 + \mu^2}} e^{i(p-q)x_1} + \phi_{cl}(x_2) \tilde{\phi}_{cl}(q-p) \times \\
    & \times\frac{e^{-i|p-q|(t_1-t_4)}}{2\sqrt{|p-q|^2 + \mu^2}} e^{-i(p-q)x_1} + \phi_{cl}(x_1) \tilde{\phi}_{cl}(p-q) \frac{e^{-i|p-q|(t_2-t_3)}}{2\sqrt{|p-q|^2 + \mu^2}} e^{i(p-q)x_2}\bigg)F^*(p,q) + h.c.\bigg]\simeq \\
    &\simeq -\frac{\lambda^2}{\mu}(T-t_0)\int_{-\infty}^{+\infty} d\tau' \int\frac{dp}{2\pi}\bigg[\alpha(\phi_{cl}(x_1)+\phi_{cl}(x_2)) +2\beta(\phi_{cl}(x_1)x_2-\phi_{cl}(x_2)x_1)\bigg]\frac{p^2}{\omega_p^2}e^{-2i\omega_p\tau'}=\\
    &=\mathcal{O}(T^0),
\end{aligned}
\end{equation}  
where we have restored the spontaneously acquired mass of the boson field $\mu\sim\lambda$ (see appendix~\ref{sec:effective}) to eliminate the singularity $\frac{1}{|p-q|}$ in the denominator. Note that the growing factor $(T-t_0)$ is multiplied by $\delta(\omega_p)$ which is never zero. This situation is similar to the free cases from subsections~\ref{sec:boson-one-loop} and~\ref{sec:fermion-one-loop}.

\section{Conclusions}
\label{sec:discussion}
\setcounter{equation}{0}

We consider one of the simplest examples of nontrivial quantum field theory out of equilibrium --- the Yukawa model in strong scalar field backgrounds in $(0+1)$ and $(1+1)$ dimensions. Our main interest is in the response of the dynamical scalar and fermion fields to such a background. To find this response, we calculate the tree--level scalar current $\langle \bar{\psi} \psi \rangle$ (i.e. the fermion propagator at coincident points) and loop corrections to both fermion and boson correlation functions. To take into account possible non-equilibrium effects we use Schwniger--Keldysh diagrammatic technique instead of the Feynman one. In this section we summarize our results and explain their physical meaning.

\textbf{1.} In $(0+1)$ dimensions the dynamics of fermion and scalar fields is nearly trivial. First of all, due to the properties of one--dimensional fermions the scalar current can be exactly calculated from the very beginning. Then the corrections to the two-point correlation functions of the scalar field basically reduce to the disconnected corrections to the one-point functions --- so-called ``tadpoles''. We show this fact both in operator formalism and diagrammatic approach. Moreover, it is not difficult to generalize this result to arbitrary orders of the perturbation theory and arbitrary $n$-point functions, because ``tadpoles'' do not receive any loop corrections in one dimension. This result means that no external scalar perturbation can change the initial state of the theory.

\textbf{2.} The dynamics in $(1+1)$ dimensions is more interesting. First, in the case of indefinitely growing scalar field, in particular, $\phi_{cl} = \frac{m}{\lambda} + Et$ and $\phi_{cl} = \frac{m}{\lambda} + Ex$, one should accurately choose the exact modes. Namely, one should demand a correct UV behavior of the modes, because at the space-time infinities they do not tend to the plane waves. Such a correct UV behaviour is necessary to have the same UV renormalization in the background field as is in its absence, which is meaningful on general physical grounds.

Second, in the leading order (when $\phi_{cl}'$ is small, while $\phi_{cl}$ itself is large) the scalar current on the backgrounds $\phi_{cl} = \frac{m}{\lambda} + Et$ and $\phi_{cl} = \frac{m}{\lambda} + Ex$ coincides with the current in the theory of free fermions with the time-dependent mass $m(t) = \lambda \phi_{cl}(t)$:

\beq \langle \bar{\psi} \psi \rangle \simeq \frac{\lambda \phi_{cl}}{\pi} \log \frac{\lambda \phi_{cl}}{\Lambda}, \eeq
where $\Lambda$ is the UV cut-off. 

The last equation is an analog of the causal equations that have been derived in, e.g., \cite{Cooper:1994hr} for the scalar and electromagnetic fields. In the expression under consideration we have explicitly calculated the right hand side (the scalar current) for the given background fields in the tree--level approximation and for large and slowly changing backgrounds.
This result indicates that the leading expressions in the strong scalar fields are insensitive to the choice of the initial state. Note that subleading corrections to the scalar current do depend on the choice of the initial state \cite{Diatlyk}. 

Third, neither level population nor anomalous quantum average of the scalar and fermion fields do grow with time. Hence, in the limit of small coupling constant time--dependent corrections to the tree-level correlation functions (including scalar current) are negligible despite the strength of the background. This type of behavior does not resemble the one in strong electric~\cite{Akhmedov:Et, Akhmedov:Ex} or gravitational~\cite{Akhmedov:H, Akhmedov:dS} fields, in which loop corrections to these quantities do grow with time\footnote{Note that by calculating corrections to the Keldysh propagator (to the level population and anomalous quantum averages) we examine if there are contributions to the scalar current which grow with time, but do not check if there are corrections which are large, when $\phi_{cl}$ is large, but does not change the state of the system.}. 

This is a very strange phenomenon for the case of time dependent scalar background. In fact, it seems that the secular growth under discussion is forbidden due to a specific behaviour of the exact modes in the background. In particular, it means that even if one starts with any non-stationary state (e.g. non-plankian initial distribution) there will not be any substantial change of the level population and of the anomalous averages, if the mass of a particle changes in time, $M(t) = \lambda \, \phi(t)$. This we find as quite a non--trivial observation, which should be compared to the time-dependent gauge and gravitational backgrounds.  

The technical reason why there is no secular growth in the background scalar field as opposed to its presence e.g. in constant electric field or de Sitter space can be explained as follows. In the constant electric field (de Sitter space) all the quantities depend on the invariant/physical momenta $p_3 - eEt$ ($\left|\vec{p}\right| \, e^{-t/H}$). (Here $p_3$ is the component of the momentum along the external electric field $E$ and $H$ is the Hubble constant in the case of the de Sitter space.) As the result all physical quantities are invariant under the simultaneous translations $t\to t - a$ and $p_3 \to p_3 - eEa$ ($\left|\vec{p}\right| \to \left|\vec{p}\right| e^{-a/H}$). Furthermore, in the field theory without background field, but with an initial non--stationary (non--plankian) distribution there is the time translational invariance at the tree--level. That is the reason why there is secular growth in the loops in all the listed in this paragraph situations.  Meanwhile in the background fields $\phi_{cl} = Et$ there is no time translational invariance. 

At the same time in the background $\phi_{cl} = \frac{m}{\lambda} + Ex$ the simple explanation for the absence of the secular growth is not yet clear to us.
What remains to be checked now if there is a secular growth for any other states of the type (\ref{abgest}) for the spatial coordinate dependent background.

We should probably stress here that for finite coupling constants the corrections to the quantum averages, $\langle a^+ a \rangle$ and $\langle aa \rangle$, are non-zero, i.e. the theory~\eqref{eq:action} is indeed non-stationary. Also let us emphasize that the scalar currents calculated in different ground states (e.g. currents~\eqref{eq:Et-current} and~\eqref{s4f15}) coincide only in the leading order, whereas subleading corrections to these quantities depend on the state. One can find more examples in~\cite{Diatlyk}.

Finally, usually one studies the behavior of the fields on fixed backgrounds, e.g. fixed electric field or gravitational field of collapsing matter. In most of our article we also follow this approach. However, in addition to such a standard setup in the last section we consider dynamics of the ``coherent state'':

$$
\left\langle \phi_{cl}\left| \hat{\phi}(t=0,x)\right|\phi_{cl}\right\rangle = \phi_{cl}(x),
$$
i.e. a self-guided dynamics of a freely evolving in time initially set up ``coherent state''. Namely, we were attempting to calculate $\left\langle \phi_{cl}\left| \hat{\phi}(t,x)\right|\phi_{cl}\right\rangle$ for arbitrary $t$ in the full theory.
We have found that the behavior of the correlation functions in this case is qualitatively the same as the one previously found for the strong fixed scalar backgrounds.

\textbf{3.} As soon as the dynamics in the strong scalar field (when $\phi_{cl}'$ is small, while $\phi_{cl}$ itself is large) is weakly sensitive to the choice of the ground state, we can estimate its effective action in the equilibrium approach. It is a standard textbook exercise to show that in this approach the effective action for the scalar field (the action one obtains after the integration over the fermion degrees of freedom) in the leading order looks as follows:

\beq \label{eq:D1} S_{eff} = \int d^2x \left[ \frac{1}{2} (\partial_\mu \phi)^2 - V_{eff}[\phi] \right], \quad \text{where} \quad V_{eff}[\phi] \simeq \frac{(\lambda \phi)^2}{2 \pi} \log \frac{\phi}{\langle \phi \rangle_{GS}} - \frac{(\lambda \phi)^2}{4 \pi} \eeq
and $\langle \phi \rangle_{GS}$ is the minimum of the renormalized effective potential $V_{eff}[\phi]$. Note that scalar field acquires non-zero mass $\mu = \frac{\lambda}{\sqrt{\pi}}$ at the bottom of the potential due to the quantum fluctuations. Also we remind that the derivation of~\eqref{eq:D1} assumes that the scalar field is non-dynamical, large and slowly changing, $\left| \slashed{\partial} \phi \right| \ll \lambda \phi^2$. We review the derivation of this expression in appendix~\ref{sec:effective}. 

The equation of motion that follows from the action~\eqref{eq:D1}:

\beq \label{eq:D2} \partial^2 \phi_{cl} + \frac{\lambda^2 \phi_{cl}}{\pi} \log \frac{\phi_{cl}}{\langle \phi \rangle_{GS}} = 0, \eeq
obviously reproduces the results of sections~\ref{sec:current} and~\ref{s4} with classical backgrounds $\phi_{cl} = Et$ and $\phi_{cl} = \frac{m}{\lambda} + Ex$. In fact, it generalizes these results to arbitrary large,  but slowly changing scalar field backgrounds. So it is not surprising that the calculations in the strong scalar wave background~\cite{Diatlyk} result in the same answer for the scalar current. However, we emphasize again that this result is correct only in the leading order, whereas the subleading corrections can be different for different choices of initial states.

\textbf{4.} The ``universality'' of the leading order approximation to the effective action can be interpreted as follows. First, note that fermion modes with high enough momenta behave as plane waves. The critical scale is roughly $p \sim \lambda \phi$. Such a behavior is necessary for the proper treatment of UV divergences, as we have already mentioned above in this section.

Second, the main contribution to the scalar current and effective potential comes from exactly such high-momenta modes (e.g. see equations~\eqref{eq:current-calc-1} and~\eqref{eq:current-calc-2}). This is due to the spatiotemporal oscillations of nearly-zero momenta modes, which are significantly faster than oscillations of higher-momenta modes. In fact, compare asymptotic behaviors~\eqref{eq:modes-smallp}, $\psi(t) \sim e^{i \alpha t^2}$, and~\eqref{eq:modes-bigp}, $\psi(t) \sim e^{i |p| t}$.

Third, when calculating the contribution from such modes, the variations of the background scalar field can be neglected. Roughly speaking, plane waves with large momenta ($p \gtrsim \lambda \phi$) dominate in regions with small spatial and temporal size. Hence, at each moment the background plays the role of a fixed mass of the fermion field. Therefore, one can just substitute $m \to \lambda \phi_{cl}(t,x) \simeq \text{const}$ in the expressions for the free case~\eqref{eq:current-proposal}.

In summary, one expects that the effective action coincides for arbitrary strong scalar fields because such fields are not sensitive, at the leading order, to the properties of the low lying initial state. In the next orders this sensitivity does manifest itself \cite{Diatlyk}.

\textbf{5.} Thus, the calculation with the use of the Feynman approach shows that zero point fluctuations of the fermion field polarize the vacuum and deform the classical scalar field background\footnote{Note that the Feymnan diagrammatic technique takes care of only zero point fluctuations. To see excitations of higher levels one has to apply the Schwinger--Keldysh technique. That is the reason why we calculate loops in the latter technique.}. However, we remind that this calculation is valid only if $\left| \slashed{\partial} \phi \right| \ll \lambda \phi^2$ and $\lambda \rightarrow 0$ (in the opposite case loop corrections to the level density and anomalous quantum average are non-zero). Both of these conditions hold in the limit $\lambda \rightarrow 0$, $t \rightarrow \infty$ for $\phi_{cl} = Et$ or $\lambda \rightarrow 0$, $x \rightarrow \infty$ for $\phi_{cl} = \frac{m}{\lambda} + Ex$. Obviously, they also hold near the minimum of the effective potential. Therefore, in this limit the scalar field just classically rolls down to the minimum of such a potential.

It would be interesting to calculate loop corrections to the quantum averages on top of such a rolling classical solution. In principle such corrections can change the situation under consideration~\cite{Krotov, Akhmedov:dS, Bascone, Akhmedov:Et, Akhmedov:Ex, Akhmedov:H, Alexeev, Astrakhantsev, Trunin}. If one considers a coherent state decay, most likely this would not happen: of course, strong initial perturbation can induce complex dynamics for a while, but one expects that eventually the field falls on the classical trajectory described by the equation~\eqref{eq:D2} for large and slowly changing values of $\phi_{cl}$. However, if one pumps energy into the system, i.e., maintains a strong field with substantial derivatives, loop corrections potentially can grow. In this case the choice of the initial state is important, the leading approximation~\eqref{eq:D1} is not valid anymore, and the dynamics of the field is less predictable. This case will be studied elsewhere.

\section{Acknowledgments}

We would like to thank Artem Alexandrov, Francesco Bascone, Kirill Gubarev, Olexander Diatlyk, Andrew Semenov, Ugo Moschella, Fedor Popov and Daniil Sherstnev for useful comments and discussions.

We would like to thank Hermann Nicolai and Stefan Theisen for the hospitality at the Albert Einstein Institute, Golm, where the work on this project was partly done. AET would like to thank Ugo Moschella for the hospitality during his stay at the INFN, Sez di Milano and the University of Insubria at Como, where the work on this project was partly done. AET also gratefully acknowledges support from the Simons Center for Geometry and Physics, Stony Brook University at which some of the research for this paper was performed.

The work of DAT and ENL was supported by the Russian Ministry of Science and Education, project number 3.9911.2017/BasePart. The work of AET was supported by the Russian Ministry of Science and Education, project number 3.9904.2017/BasePart. The work of ETA and DAT was supported by the grant from the Foundation for the Advancement of Theoretical Physics and Mathematics ``BASIS''. The work of ETA was partially supported by the RFBR grant 19-02-00815.

\appendix

\section{Asymptotic behavior of parabolic cylinder function for large order}
\label{sec:asymptotics}
\setcounter{equation}{0}

The asymptotic behavior of the parabolic cylinder functions has been widely studied in the literature (e.g., see~\cite{Bateman-2, Whittaker, Olver, Crothers}). But the only asymptotic expansion for an arbitrary complex $|\nu| \gg 1$ we have found in the literature is as follows~\cite{Bateman-2}:

\beq
\label{eq:Cherry-app}
D_\nu(z) = \frac{1}{\sqrt{2}} \exp\left[ \frac{1}{2} \nu \log(-\nu) - \frac{1}{2} \nu - \sqrt{-\nu} z \right] \left[ 1 + \mathcal{O} \left(\frac{1}{\sqrt{|\nu|}}\right) \right],
\eeq
where $|\arg (-\nu)| \le \frac{\pi}{2}$ and $|z|$ is bounded. The error of this expansion is too large for our purposes: e.g. when one integrates $D_{-\frac{i p^2}{2 \alpha}}(z)$ over $dp$, due to terms of the order $\mathcal{O}\left(\frac{1}{\sqrt{|\nu|}}\right) \sim \mathcal{O}\left(\frac{1}{p}\right)$ the integral can diverge. Thus, we have to obtain a more accurate asymptotic expansion. 

\begin{figure}[t]
\center{\includegraphics[scale=0.3]{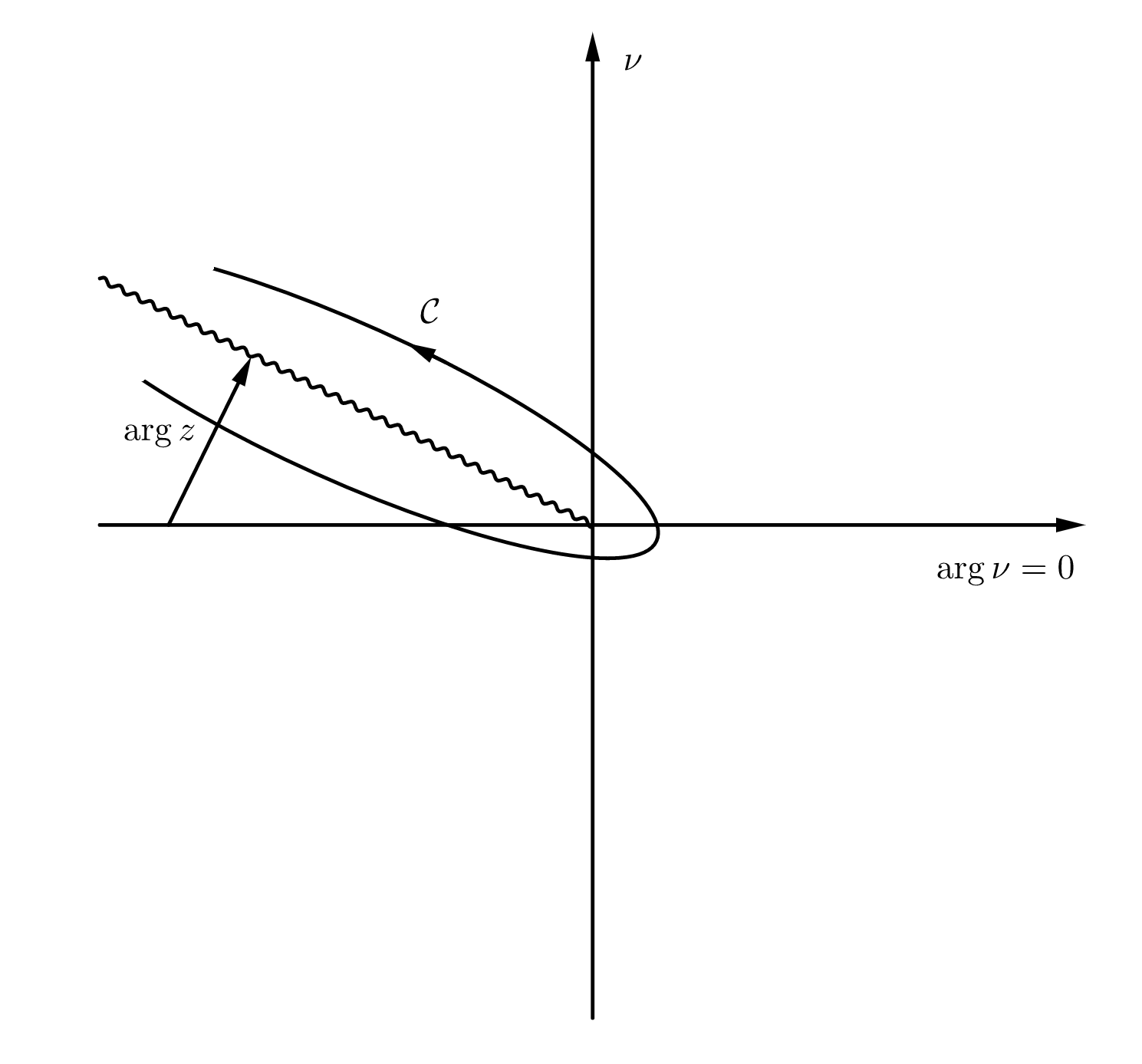}}\caption{The integration contour}\label{fig:contour-app}
\end{figure}

Following~\cite{Crothers}, we start with the integral representation of the parabolic cylinder function:

\beq \label{A22}
D_\nu(z) = \frac{\Gamma(1+\nu)}{2 \pi i} e^{-\frac{1}{4}z^2}z \int_{\mathcal{C}} \exp\left[z^2 \, \left(v-\frac{1}{2}v^2\right) - (1+\nu) \, \log(zv)\right] \, dv, \eeq
where the integration contour $\mathcal{C}$ is depicted on the Fig.~\ref{fig:contour-app}. One can check that this expression indeed solves the differential equation for parabolic cylinder function. For the case $\phi_{cl} = E t$ we have

\beq \label{eq:app-notation} \nu = -\frac{ip^2}{2\alpha}, \quad z = e^{i\pi/4}\sqrt{\frac{2}{\alpha}}M(t), \quad M(t) = \alpha t, \quad \alpha = \lambda E. \eeq
Using the saddle-point approximation in (\ref{A22}), one obtains its decomposition as follows:

\beq
\label{eq:saddleintegral-app}
D_\nu(z) \simeq \frac{\Gamma(1+\nu)}{i \sqrt{2 \pi}} z e^{-\frac{1}{4}z^2} \sum_{j = 0, 1} \frac{\exp(i\alpha_j + f(v_j))}{|f''(v_j)|^{\frac{1}{2}}} \left[1 + \sum_{l = 2}^\infty \frac{(2l - 1)!! \exp(2 i l \alpha_j)}{v_j^{2l} |f''(v_j)|^{l}} \sum_{\lambda_n} \prod_{n = 3}^{2l} \frac{[(1+\nu)/n]^{\lambda_n}}{\lambda_n!} \right],
\eeq
where
\beqs \begin{aligned}
f(v) &= z^2(v-\frac{1}{2}v^2) - (1+\nu)\log(zv), \\
\alpha_j &= \frac{1}{2}\pi - \frac{1}{2}\arg(f''(v_j)),
\end{aligned} \eeqs
and we denoted the critical points of the function $f(v)$ as $v_{0, 1} = -\frac{1}{2} \pm \frac{1}{2} \left(1 - \frac{4(1+\nu)}{z^2} \right)^{\frac{1}{2}}$. The innermost sum in eq.~\eqref{eq:saddleintegral-app} is taken over all distinct partitions of $2l$ given by non-negative integer solutions ${\lambda_n}$ such that $\sum_{n = 3}^{2l} n \lambda_n = 2l$. Let us estimate this sum. The $l$-th term in it contains the $l$-th power of the following expression:

\beqs \frac{1}{v_{0,1}^2 f''(v_{0,1})} = \frac{1}{2(1+\nu)} \left[1 \mp \left(1 - \frac{4(1 + \nu)}{z^2} \right)^{-\frac{1}{2}} \right], \eeqs
and not greater than the $\lfloor \frac{2l}{3} \rfloor$-th power of $[(1+\nu)/n]$. In the case $|\nu| \gg 1$ the square root in the brackets is small, and $\frac{1}{v_{0,1}^2 f''(v_{0,1})} \sim \frac{1}{2(1 + \nu)} = \mathcal{O}\left(\frac{1}{\nu}\right)$ for both signs ``$\mp$''. This means that the innermost sum is $\mathcal{O}\left(\frac{1+\nu}{[v_{0,1}^2 f''(v_{0,1})]^2}\right) = \mathcal{O}\left(\frac{1}{\nu}\right)$, so we neglect it in the integrals over $dp$. Substituting the values of the saddle points into the decomposition~\eqref{eq:saddleintegral-app} we obtain:
\begin{align} 
D_\nu(z) &= \frac{1}{\sqrt{2}}\exp\left[\frac{1 - \nu}{2} + \left(\nu + \frac{1}{2}\right) \log\nu - \left(\frac{\nu}{2} + \frac{1}{2}\right) \log\frac{z^2}{4} - \frac{1}{4} \log\left(1 - \frac{4(1 + \nu)}{z^2}\right)\right] \times \nonum &\times \sum_\pm \exp\left[\left( -\frac{1}{2} - \nu \right) \log\left(1 \pm \left(1 - \frac{4(1 + \nu)}{z^2}\right)^{\frac{1}{2}} \right) \pm \frac{z^2}{4}\left(1 - \frac{4(1 + \nu)}{z^2}\right)^{\frac{1}{2}} \right] \left[1 + \mathcal{O}\left(\frac{1}{\nu}\right)\right].
\end{align}
Then we note that in the notations of~\eqref{eq:app-notation} and limit $|\nu| \gg 1$ (i.e. $p^2 \gg \alpha$) or $|z| \gg 1$ (i.e. $M^2 \gg \alpha$), we have that

\beqs \left(1 - \frac{4(1 + \nu)}{z^2}\right)^{\frac{1}{2}} = \frac{\sqrt{M^2 + p^2}}{M} \left[ 1 + \frac{i \alpha}{M^2 + p^2} + \mathcal{O}\left(\frac{\alpha}{M^2 + p^2}\right)^2 \right]. \eeqs
Hence, denoting $V = \sqrt{M^2 + p^2}$ for short, we obtain:

\beq 
\label{eq:D_bignu-app}
D_{-\frac{i p^2}{2 \alpha}}\left(\frac{1+i}{\sqrt{\alpha}} M\right) \simeq \frac{e^{\frac{\pi p^2}{8 \alpha}}}{\sqrt{2}} \left(\frac{M}{V} + 1\right)^{\frac{1}{2}} e^{\frac{i p^2}{4 \alpha} - \frac{i p^2}{4 \alpha} \log \frac{(V+M)^2}{2 \alpha} - \frac{i M V}{2 \alpha}} \left[1 + \mathcal{O}\left(\frac{\alpha}{V^2}\right) \right].
\eeq
Then for the squared module of the parabolic cylinder function we get:

\beq
\label{eq:D_bignu-2-app}
\left| D_{-\frac{i p^2}{2 \alpha}}\left(\frac{1+i}{\sqrt{\alpha}} M\right) \right|^2 \simeq \frac{1}{2} e^{\frac{\pi p^2}{4 \alpha}} \left(\frac{M}{\sqrt{M^2 + p^2}} + 1 \right) \left[1 + \mathcal{O}\left(\frac{\alpha}{M^2 + p^2}\right) \right].
\eeq
Here we neglected the second term in the sum because it contains the factor of $e^{-\frac{\pi p^2}{2 \alpha}}$. Note that we have chosen the sheet on the complex plane in which $-1 = e^{-i \pi}$.
One can check that~\eqref{eq:D_bignu-app} coincides with~\eqref{eq:Cherry-app} up to $\mathcal{O}\left(\frac{1}{p}\right)$. But the new equation also contains the next term of the asymptotic expansion.

We emphasize that eqs.~\eqref{eq:D_bignu-app} and~\eqref{eq:D_bignu-2-app} work for arbitrary values $M^2 \gg \alpha$. However, they simplify in extremal cases. For instance,
\beq D_{-\frac{i p^2}{2 \alpha}}\left(\frac{1 + i}{\sqrt{\alpha}} M\right) \simeq \frac{1}{\sqrt{2}} e^{\frac{\pi p^2}{8 \alpha} +\frac{i p^2}{4 \alpha} - \frac{i p^2}{4 \alpha} \log \frac{p^2}{2 \alpha}} e^{\frac{-i |p| M}{\alpha} - \frac{i M^3}{6 |p| \alpha} + \frac{M}{2|p|}} \left[ 1 + \mathcal{O}\left(\frac{M^2 + \alpha}{p^2}\right)\right], \eeq
if $M^2 \ll p^2$, and
\beq D_{-\frac{i p^2}{2 \alpha}}\left(\frac{1 + i}{\sqrt{\alpha}} M\right) \simeq \begin{cases} \left( 1  - \frac{p^2}{8 M^2} \right) e^{\frac{\pi p^2}{8 \alpha} - \frac{i M^2}{2 \alpha} - \frac{i p^2}{4 \alpha} \log \frac{2 M^2}{\alpha}} \left[ 1 + \mathcal{O}\left(\frac{p^2 + \alpha}{M^2}\right)\right], \quad M > 0, \\ \frac{p}{2 |M|} e^{\frac{\pi p^2}{8 \alpha} + \frac{i M^2}{2 \alpha} + \frac{i p^2}{4 \alpha} \log \frac{2 M^2}{\alpha} -  \frac{i p^2}{2 \alpha} \log \frac{p^2}{2 \alpha}} \left[ 1 + \mathcal{O}\left(\frac{p^2 + \alpha}{M^2}\right)\right], \quad M < 0 \end{cases} \eeq
if $M^2 \gg p^2$.

In the opposite case $|\nu| \ll 1$ one should exactly calculate the innermost sum in~\eqref{eq:saddleintegral-app}, because $\frac{1}{1+\nu} \sim 1$ (at least for the ``$+$'' sign). Furthermore, in the case $|\nu| \ll 1 \ll |z|$ (i.e., $p^2 \ll \alpha \ll M^2$) we can use the following decomposition, which can be obtained from another integral representation for parabolic cylinder function~\cite{Bateman-2}: 
\beq \begin{aligned}
& D_{\nu}(z)= z^{\nu}e^{-\frac{z^2}{4}}\left[\sum_{n=0}^N \frac{\left(-\frac{\nu}{2} \right)_n\left(\frac{1}{2}-\frac{\nu}{2} \right)_n}{n!\left(-\frac{z^2}{2} \right)^n}+\mathcal{O}\left(\left|z^2 \right|^{-N-1}\right) \right], \\ & \left(\gamma\right)_0=1, \quad \left( \gamma\right)_{n \neq 0}=\gamma\left(\gamma+1 \right) \cdots \left(\gamma+n-1 \right).
\end{aligned} \eeq
Hence, we find that:
\beq
\label{eq:D_smallnu-app}
D_{-\frac{i p^2}{2 \alpha}}\left(\frac{1+i}{\sqrt{\alpha}} M\right) \simeq e^{\frac{\pi p^2}{8 \alpha} - \frac{i M^2}{2 \alpha} - \frac{i p^2}{4 \alpha} \log \frac{2 M^2}{\alpha}} \left(1 - \frac{p^2}{8 M^2} \right) \left[ 1 + \mathcal{O}\left(\frac{\alpha^2}{M^4}\right) \right],
\eeq
and for the squared module:
\beq
\left| D_{-\frac{i p^2}{2 \alpha}}\left(\frac{1+i}{\sqrt{\alpha}} M\right) \right|^2 \simeq e^{\frac{\pi p^2}{4 \alpha}} \left[ 1 - \frac{p^2}{4 M^2} + \mathcal{O}\left(\frac{\alpha^2}{M^4}\right) \right].
\eeq
Note that expressions~\eqref{eq:D_bignu-app} and~\eqref{eq:D_smallnu-app} approximately coincide if $|\nu| \ll |z|$, $|\nu| \gg 1$, as it should be.
\setcounter{equation}{0}

\section{Effective action}
\label{sec:effective}
\setcounter{equation}{0}

\subsection{Path integral calculation}

In sections~\ref{sec:2D-Et} and~\ref{sec:2D-Ex} we have shown that the leading behavior of the fermion current does not depend on the ground state of the theory (see also~\cite{Diatlyk}). Moreover, in the limit of small coupling constants  loop corrections to the scalar and fermion propagators do not grow. Therefore, if $\phi$ is large and slowly changing function we can estimate the effective action using standard equilibrium technique, assuming that the field $\phi$ is not dynamical. In this appendix we review the textbook calculation of the Feynman effective action~\cite{Coleman, Zinn-Justin, Malbouisson, Mosel} for the theory~\eqref{eq:action}.

To find the effective action for scalars, we integrate out the fermionic degrees of freedom in the functional integral:
\beq
\label{eq:path-integral}
e^{i S_{eff}[\phi]} = \frac{\int \mathcal{D}\bar{\psi} \mathcal{D}\psi e^{ i \int d^2 x \left( \frac{1}{2} (\partial_\mu \phi)^2 +  \bar{\psi}(i\slashed{\partial} - \lambda \phi) \psi \right)}}{\int \mathcal{D}\bar{\psi} \mathcal{D}\psi e^{ i \int d^2 x \bar{\psi} i \slashed{\partial} \psi }} = \exp \left[ i \int d^2 x \, \frac{1}{2} (\partial_\mu \phi)^2 + \tr \log \frac{i\slashed{\partial} - \lambda \phi}{i\slashed{\partial}} \right],
\eeq
which we normalize to the partition function of a free massless fermion for the correct definition of the operator determinant.

As we have just mentioned, in this section we consider the situation, when the scalar field is non-dynamical. At the same time in (\ref{eq:path-integral}) we calculate the time--ordered Feynman effective action rather than Schwinger--Keldysh one. Note that this approximation in general is not valid if one takes into account the quantum fluctuations of the scalar field. In this calculation it is implicitly assumed that the state of the theory does not change in time. However, we have seen in the sections~\ref{sec:loops} and~\ref{s4.3} that both of these approximations are good enough if we work in the limit of large and slowly changing background scalar field. 

Let us evaluate the determinant in~\eqref{eq:path-integral}. For simplicity we consider scalar fields smaller than the UV cut-off: $\lambda \phi \ll \Lambda$ (these fields still can be strong: $\phi \gg 1$). This relation allows us to expand the logarithm and separate the operators which are local in $x$ and $p$~\cite{Mosel}. Using the reflection symmetry, i.e multiplying the expression by $1 = (\gamma^5)^2$, and anti-commuting $\gamma^5$ and $\gamma^\mu$, one obtains:

\beq \tr \log \frac{i\slashed{\partial} - \lambda \phi}{i\slashed{\partial}} = \frac{1}{2} \tr \log \frac{(i\slashed{\partial} - \lambda \phi)(-i\slashed{\partial} - \lambda \phi)}{(i\slashed{\partial})(-i\slashed{\partial})} = \frac{1}{2} \tr \log \frac{\partial^2 + (\lambda \phi)^2 - i \lambda \slashed{\partial} \phi}{\partial^2} \simeq \tr \log \Big(1 + \frac{(\lambda \phi)^2}{\partial^2} \Big), \eeq
where we took the trace over the spinor indices and neglected the derivatives $\partial_t \phi \ll \lambda \phi^2$ and $\partial_x \phi \ll \lambda \phi^2$. E.g. for $\phi_{cl} = E t$ we have exactly such situation when $t \gg \frac{1}{\sqrt{\lambda E}}$ and for $\phi_{cl} =\frac{m}{\lambda}+E x$ when $x \gg \frac{\left|\sqrt{\lambda E} - m\right|}{\lambda E}$.

To evaluate the $\tr \log$ we do Wick rotation into the Euclidean space~\cite{Peskin}. One can actually do such a transformation, which is not valid in non--stationary situation, in the approximation that we are adopting here. Then we expand the logarithm:

\beq
\label{eq:det-calc}
\tr \log \left(1 + \frac{(\lambda \phi)^2}{\partial^2} \right) = \int d^2 x \int \frac{i \, d^2p}{(2 \pi)^2} \log \left( 1 + \frac{(\lambda \phi)^2}{p^2} \right) \simeq i \int \frac{d^2 x}{4 \pi} \left[ (\lambda \phi)^2 \log \frac{\Lambda^2}{(\lambda \phi)^2} + (\lambda \phi)^2 \right].
\eeq
For the last equality we neglected the terms of the order $\frac{(\lambda \phi)^4}{\Lambda^2}$ and smaller. Thus, in the leading order for large $\phi$ and small derivatives of $\phi$ the effective action has the following form\footnote{\label{footnote:loops} Let us recall that the calculation of the effective action corresponds to the summation of the Feynman diagrams (e.g. see~\cite{Peskin,Coleman}). Indeed, consider the soft bosonic corrections to the free fermion propagator:

\beqs G(p) = \frac{i}{\slashed{p} + i \epsilon} + (-i \lambda \phi) \left(\frac{i}{\slashed{p} + i \epsilon}\right)^2 + \frac{2!}{2!} (-i \lambda \phi)^2 \left(\frac{i}{\slashed{p} + i \epsilon}\right)^3 + \cdots = \frac{i}{\slashed{p} + i \epsilon} \frac{1}{1 - \frac{\lambda \phi}{\slashed{p} + i \epsilon}} = \frac{i}{\slashed{p} - \lambda \phi + i \epsilon}. \eeqs
Such corrections take into account the interaction between the fermion field and fixed scalar field background, so it is not surprising that we have obtained the inverse operator of the second equation in the system~\eqref{eq:system} in almost constant $\phi_{cl}$ background. The fermion current corresponds to the exact propagator with the coincident initial and end points, i.e. to the sum of the closed fermionic loops with an even number of external legs (diagrams with an odd number of legs are zero due to Furry's theorem~\cite{Peskin}). Hence, the summation of such diagrams should reproduce the result~\eqref{eq:det-calc} in the limit that we consider in this section.}:

\beq
\label{eq:effective-action}
S_{eff} \simeq \int d^2 x \left[ \frac{1}{2}\partial_\mu \phi \partial_\mu \phi - V_{eff}[\phi] \right], \quad \text{where} \quad V_{eff}[\phi] \simeq \frac{(\lambda \phi)^2}{2 \pi} \log \frac{\lambda \phi}{\Lambda} - \frac{(\lambda \phi)^2}{4 \pi}.
\eeq
The partition function $Z = \int \mathcal{D}\phi \, e^{i S_{eff}[\phi]}$ is predominantly gained on the functions which solve the classical equation of motion. Hence:
\beq
\label{eq:effective-current}
\partial^2 \langle\phi\rangle + \lambda \langle \bar{\psi} \psi \rangle \approx \partial^2 \langle\phi\rangle + \frac{\lambda^2 \langle\phi\rangle}{\pi} \log \frac{\lambda \langle\phi\rangle}{\Lambda} = 0.
\eeq
This expression is consistent with the values of the scalar currents~\eqref{eq:Et-current} and~\eqref{s4f15} for $\phi_{cl} = E t$ and $\phi_{cl} =\frac{m}{\lambda}+E x$, respectively, which were obtained in the main body of the text. However, \eqref{eq:effective-current}~works for strong, but slowly changing classical backgrounds (see also \cite{Diatlyk}). Note that subleading corrections to the scalar current (and, hence, to the effective action) do depend on the state with respect to which the averaging is done in the correlation functions \cite{Diatlyk}. The corrections should be calculated with the use of the Schwinger--Keldysh technique.

Now the classical fields $\phi_{cl} = E t$ and $\phi_{cl} =\frac{m}{\lambda}+E x$ do not solve the corrected equation of motion~\eqref{eq:effective-current}, although they do solve the free equation~\eqref{eq:system}. This basically means that such classical fields have to decay due to quantum fluctuations of the fermions. This resembles the decay of strong constant electric field~\cite{Akhmedov:Et,Akhmedov:Ex}. However, in contrast to the strong electric field in this case loop corrections to boson and fermion level populations do not grow, as we have shown in the main body of the text. 

\subsection{Renormalization}

One can see that expressions~\eqref{eq:effective-action} and~\eqref{eq:effective-current} explicitly depend on the UV cut-off, i.e. they are seemingly not invariant with respect to renormalization group. Of course, this dependence has no physical sense, because observables must be renormalization group invariant. To resolve the issue we restore the mass of the scalar field and take into account UV counterterms (we recall that Yukawa theory in two dimensions is renormalizable, since coupling constant $\lambda$ has positive mass dimension):
\beq S_{eff} = \int d^2x \left[ \frac{1}{2} \left(\partial_\mu \phi\right)^2 - \frac{1}{2} \mu_0^2 \phi^2 - V_{eff}[\phi] + \frac{1}{2} A \left(\partial_\mu \phi\right)^2 - \frac{1}{2} B \phi^2 \right]. \eeq
Usually, one defines the renormalized mass as the value of the inverse propagator at zero momentum:
\beq \mu^2 = \frac{\partial^2 V}{\partial \phi^2} \Bigg|_0, \eeq
where $V$ includes both the effective potential, mass term and counterterms. However, in the present case this definition is meaningless: the second derivative of $V$ at the origin does not exist due to the logarithmic singularity. Due to this reason we define the mass at an arbitrary but non-zero value $M_R$:

\beq \mu^2 = \frac{\partial^2 V}{\partial \phi^2} \Bigg|_{M_R}. \eeq
This implies the following expression for the counterterm $B$:
\beq B = -\frac{\lambda^2}{\pi} \log \frac{\lambda M_R}{\Lambda} - \frac{\lambda^2}{\pi}. \eeq
and for the renormalized potential:
\beq V = \frac{1}{2} \mu_0^2 \phi^2 + \frac{(\lambda \phi)^2}{2 \pi} \log \frac{\lambda \phi}{M_R} - \frac{3 (\lambda \phi)^2}{4 \pi}. \eeq
It is easy to check that this expression is invariant under the change of renormalization scale. Also one can note that the effective potential has the minimum, which is not $\phi = 0$. This situation is obviously similar to the well-known Coleman--Weinberg potential~\cite{Coleman, Peskin}.

Finally, we set $\mu_0 = 0$, replace an arbitrary parameter $M_R$ by the ground state expectation value of the scalar field which minimizes the renormalized potential (we emphasize that this value differs from the average over the original state):
\beq M_R = \frac{1}{e} \lambda \langle \phi \rangle_{GS}, \eeq
where $e$ is the Euler's constant, and obtain the following renormalization group invariant expression for the effective potential:
\beq
\label{eq:effective-renorm}
V_{eff} = \frac{(\lambda \phi)^2}{2 \pi} \log \frac{\phi}{\langle \phi \rangle_{GS}} - \frac{(\lambda \phi)^2}{4 \pi}.
\eeq
The expansion of this potential near the minimum $\phi = \langle \phi \rangle_{GS} + \tilde{\phi}$ has the following form:
\beq V_{eff} \simeq - \frac{\lambda^2 \langle \phi \rangle_{GS}^2}{4 \pi} + \frac{\lambda^2}{2 \pi} \tilde{\phi}^2 + \cdots, \eeq
i.e. the field spontaneously acquires the mass $\mu^2 = \frac{\lambda^2}{\pi}$.

Note that eqs.~\eqref{eq:effective-current} and~\eqref{eq:effective-renorm} were obtained in the approximation $\lambda \phi \ll \Lambda$ which is obviously not satisfied near the minimum of the potential. However, higher loops corrections do not change the form of the potential near $\phi = 0$. Therefore, loop corrections cannot shift the minimum of the effective potential to zero, although they can affect its absolute value~\cite{Coleman}. I.e. the expression~\eqref{eq:effective-renorm} provides a good qualitative description of the situation.

\section{Definite-frequency and definite-momentum operators}
\label{sec:definite}
\setcounter{equation}{0}

In this appendix we find the relation between the definite-frequency and definite-momentum creation and annihilation operators. In order to do this let us consider the standard mode decomposition for the free massive fermion field:

\small \beq \begin{aligned}
\psi(t,x) & = \iint_{-\infty}^\infty \frac{d\omega dp}{2\pi} \, \delta(\omega^2 - p^2 - m^2) \, \left[ a_p \psi_{p} e^{- i \omega t + i p x} + b_p^\dagger \tilde{\psi}_{p} e^{i \omega t - i p x} \right] = \\ &= \int_{-\infty}^\infty \frac{dp}{2\pi} \left[ a_p \psi_{p,\omega_p} e^{- i \omega_p t + i p x} + b_p^\dagger \tilde{\psi}_{p,\omega_p} e^{i \omega_p t - i p x} \right] = \\
&= \int_0^\infty \frac{dp}{2\pi} \left[ a_p \psi_{p,\omega_p}  e^{- i \omega_p t + i p x} + b_p^\dagger \tilde{\psi}_{p,\omega_p}  e^{i \omega_p t - i p x} + a_{-p} \psi_{-p,\omega_p}  e^{- i \omega_{-p} t - i p x} + b_{-p}^\dagger \tilde{\psi}_{-p,\omega_p}  e^{i \omega_{-p} t + i p x} \right] = \\
&= \int_m^\infty \frac{d\omega}{2\pi} \left[ a_{p_\omega} \psi_{p_\omega,\omega} e^{- i \omega t + i p_\omega x} + b_{p_\omega}^\dagger \tilde{\psi}_{p_\omega,\omega} e^{i \omega t - i p_\omega x} +a_{-p_\omega} \psi_{-p_\omega,\omega} e^{- i \omega t - i p_\omega x} + b_{-p_\omega}^\dagger \tilde{\psi}_{-p_\omega,\omega} e^{i \omega t + i p_\omega x} \right] = \\
&= \int_{|\omega| > m} \frac{d\omega}{2\pi} \Big[ \left( \theta(\omega) a_{p_\omega} \psi_{p_\omega,\omega} + \theta(-\omega) b_{-p_\omega}^\dagger \tilde{\psi}_{-p_\omega,-\omega} \right) e^{- i \omega t + i p_\omega x} + \\ &\phantom{\int_{|\omega| > m} \frac{d\omega}{2\pi} \Big[}+ \left( \theta(\omega) b_{p_\omega}^\dagger \tilde{\psi}_{p_\omega,\omega} + \theta(-\omega) a_{-p_\omega} \psi_{-p_\omega,-\omega} \right) e^{i \omega t - i p_\omega x} \Big] = \\
&=  \int_{|\omega| > m} \frac{d\omega}{2\pi} \left[ a_\omega \chi_\omega e^{- i \omega t + i p_\omega x} + b_\omega^\dagger \tilde{\chi}_\omega e^{i \omega t - i p_\omega x} \right].
\end{aligned} \eeq \normalsize
Here $\omega_p = \sqrt{p^2 + m^2}$ and $p_\omega = \sqrt{\omega^2 - m^2}$, both functions are even and always positive; $\psi_{p,\omega}$ and $\tilde{\psi}_{p,\omega}$ correspond to fermion and antifermion spinors. The creation and annihilation operators with definite frequency ($a_\omega$ and $b_\omega$) are expressed via the corresponding operators with definite momentum as follows:

\beq \label{eq:new-operators}
a_\omega = \theta(\omega) a_{p_\omega} + \theta(-\omega) b_{-p_\omega}^\dagger, \quad b_\omega = \theta(\omega) b_{p_\omega} + \theta(-\omega) a_{-p_\omega}^\dagger. \eeq
In other words, for positive frequencies new operators coincide with definite-momentum operators, but for negative frequencies fermion creation operator and antifermion annihilation operators switch places.

The definite-frequency and definite-momentum spinors are connected by similar expressions:

\beq \label{eq:new-modes}
\chi_\omega = \theta(\omega) \psi_{p_\omega,\omega} + \theta(-\omega) \tilde{\psi}_{-p_\omega,-\omega} = \psi_{p_\omega,\omega}, \quad \tilde{\chi}_\omega = \theta(\omega) \tilde{\psi}_{p_\omega,\omega} + \theta(-\omega) \psi_{-p_\omega,-\omega} = \tilde{\psi}_{p_\omega,\omega}, \eeq
where we have used that $\tilde{\psi}_{-p_\omega,-\omega} = \psi_{p_\omega,\omega}$. Note that in sections~\ref{sec:2D-Et} and~\ref{sec:2D-Ex} we use different representations for the Clifford algebra. So the modes~\eqref{s2f5} and~\eqref{eq:free-modes} are related by the expression~\eqref{eq:new-modes} plus the unitary transformation, which maps gamma-matrices~\eqref{eq:gamma-1} into~\eqref{s1f2}.

We emphasize that operators~\eqref{eq:new-operators} obey the standard anticommutation relations:
\beq \left\{ a_\omega, a_{\omega'}^\dagger \right\} = \left\{ b_\omega, b_{\omega'}^\dagger \right\} = 2 \pi \delta(\omega - \omega'), \eeq
if these relations hold for the operators $a_p$ and $b_p$. However, the expectation value for the standard vacuum state, which is annihilated by $a_p$ and $b_p$, is not trivial:
\beq \begin{gathered}
\langle 0 | a_\omega a_{\omega'}^\dagger | 0 \rangle =  \langle 0 | b_\omega b_{\omega'}^\dagger | 0 \rangle = \theta(\omega) \times 2 \pi \delta(\omega - \omega'), \\ \langle 0 | a_\omega^\dagger a_{\omega'} | 0 \rangle = \langle 0 | b_\omega^\dagger b_{\omega'} | 0 \rangle = \theta(-\omega) \times 2 \pi \delta(\omega - \omega').
\end{gathered} \eeq
The operators $a_\omega$ and $b_\omega$ in the decomposition~\eqref{s2f1} are related to the operators $a_p$ and $b_p$ via the relation \eqref{eq:new-modes}.

Similarly one can show that the definite-frequency and definite-momentum boson creation and annihilation operators are related as follows:

\beq \alpha_\omega = \theta(\omega) \alpha_{p_\omega} + \theta(-\omega) \alpha_{-p_\omega}^\dagger, \eeq
which implies that:
\beq \langle 0 | \alpha_\omega \alpha_{\omega'}^\dagger | 0 \rangle = \theta(\omega) \times 2 \pi \delta(\omega - \omega'), \quad \langle 0 | \alpha_\omega^\dagger \alpha_{\omega'} | 0 \rangle = \theta(-\omega) \times 2 \pi \delta(\omega - \omega'). \eeq

\section{Derivation of the coherent state}\label{pa}
\setcounter{equation}{0}

In this appendix we show that the coherent state that we use in the main body of the text has the following form:

\begin{equation}\label{paf1}
\left|\phi_\text{cl}\right>=e^{-i\int\phi_{\text{cl}}\hat{\pi}_\phi dx}\left|0\right>,\quad {\rm where} \quad a_p\left|0\right>=0.
\end{equation}
Let us apply the operator $\hat{\phi}(y)$ to the state $\left|\phi_\text{cl}\right>$:

\begin{multline}\label{paf2}
\hat{\phi}(y)\left|\phi_\text{cl}\right>=\hat{\phi}(y)e^{-i\int\phi_{\text{cl}}\hat{\pi}_\phi dx}\left|0\right>= \\ =\sum\limits_{n=0}^{\infty}\frac{(-i)^n}{n!}\int dx_1\dots dx_n\phi_{\text{cl}}(x_1)\dots\phi_{\text{cl}}(x_n)\hat{\phi}(y)\hat{\pi}_\phi(x_1)\dots\hat{\pi}_\phi(x_n)\left|0\right>.
\end{multline}
Commuting $\hat{\phi}(y)$ with $\hat{\pi}_\phi(x_i)$:
\begin{equation}\label{paf3}
[\hat{\phi}(y),\hat{\pi}_\phi(x_i)]=i\delta(x_i-y),
\end{equation}
we get that:
\begin{multline}\label{paf4}
\hat{\phi}(y)\left|\phi_\text{cl}\right> = \left\{\sum\limits_{n=0}^\infty\frac{(-i)^n}{n!}i\phi_{\text{cl}}(y)n\left(\int\phi_{\text{cl}}(x)\hat{\pi}_\phi(x)dx\right)^{n-1}+e^{-i\int\phi_{\text{cl}}\hat{\pi}_\phi dx}\hat{\phi}(y)\right\}\left|0\right>= \\ =\left\{\phi_{\text{cl}}(y)\sum\limits_{n=1}^\infty\frac{(-i)^{n-1}}{(n-1)!}\left(\int\phi_{\text{cl}}(x)\hat{\pi}_\phi(x)dx\right)^{n-1}+e^{-i\int\phi_{\text{cl}}\hat{\pi}_\phi dx}\hat{\phi}(y)\right\}\left|0\right>= \\ =\phi_{\text{cl}}\left|\phi_\text{cl}\right>+e^{-i\int\phi_{\text{cl}}\hat{\pi}_\phi dx}\hat{\phi}(y)\left|0\right>.
\end{multline}
From the last expression it is straightforward to show that eq.~\eqref{s5f1} is true.

Now let us find the normalization factor. Define

\begin{equation}\label{paf5}
\left|\phi_\text{cl}\right>=C(\phi_{\text{cl}})e^{-i\int\phi_{\text{cl}}\hat{\pi}_\phi dx}\left|0\right>.
\end{equation}
Then
\begin{equation}\label{paf6}
\left<\phi_\text{cl}|\phi_\text{cl}\right>=|C(\phi_{\text{cl}})|^2\left<0\right|e^{i\int\phi_{\text{cl}}\hat{\pi}_\phi dx}e^{-i\int\phi_{\text{cl}}\hat{\pi}_\phi dx}\left|0\right>=|C(\phi_{\text{cl}})|^2=1.
\end{equation}
Hence, $C(\phi_\text{cl})=1$, because $\left<0|0\right>=1$. Thus, we confirm the expression ~\eqref{s5f2} for the coherent state.

\end{document}